\documentclass[review]{elsarticle}

\usepackage{lineno,hyperref}
\modulolinenumbers[5]
\usepackage{graphicx}
\usepackage[caption=false,font=footnotesize]{subfig}
\usepackage{tabu}
\usepackage[table,xcdraw]{xcolor}

\usepackage{comment}

\begin{document}

\begin{frontmatter}

\title{Are Social Networks Watermarking Us or\\ Are We (Unawarely) Watermarking Ourself?}

\author[firstaddress]{Flavio Bertini}
\ead{flavio.bertini2@unibo.it}
\author[secondaddress]{Rajesh Sharma}
\ead{flavio.bertini2@unibo.it}
\author[firstaddress]{Danilo Montesi}
\ead{danilo.montesi@unibo.it}
\address[firstaddress]{Department of Computer Science and Engineering, University of Bologna, Bologna, Italy}
\address[secondaddress]{Institute of Computer Science, University of Tartu, Tartu, Estonia}

\begin{abstract}
In the last decade, Social Networks (SNs) have deeply changed many aspects of society, and one of the most widespread behaviours is the sharing of pictures. However, malicious users often exploit shared pictures to create fake profiles leading to the growth of cybercrime. Thus, keeping in mind this scenario, authorship attribution and verification through image watermarking techniques are becoming more and more important. In this paper, firstly, we investigate how 13 most popular SNs treat the uploaded pictures, in order to identify a possible implementation of image watermarking techniques by respective SNs. Secondly, on these 13 SNs, we test the robustness of several image watermarking algorithms. Finally, we verify whether a method based on the Photo-Response Non-Uniformity (PRNU) technique can be successfully used as a watermarking approach for authorship attribution and verification of pictures on SNs. The proposed method is robust enough in spite of the fact that the pictures get downgraded during the uploading process by SNs. The results of our analysis on a real dataset of 8,400 pictures show that the proposed method is more effective than other watermarking techniques and can help to address serious questions about privacy and security on SNs.
\end{abstract}

\begin{keyword}
Copyright protection \sep Privacy \sep Social network services \sep Watermarking.
\end{keyword}

\end{frontmatter}

\section{Introduction}\label{sec:intro}
In recent years, various Social Networks (SNs) have been introduced in order to cater different needs of the users: social interactions (Facebook), professional interconnections (LinkedIn), photo sharing (Instagram), instant messaging (WhatsApp), to name a few. An important reason for the huge popularity of social platforms among users is the increase in usage of smartphones, which in turn has introduced changes in the user habits with respect to multimedia content on SNs \cite{salehan2013social}. In particular, some social platforms are predominantly used through mobile devices, such as Instagram and WhatsApp.

On the flip side, the illicit actions across these SNs are constantly growing, such as illegal copying, identity impersonation and pedopornography \cite{rathore2017social}. In particular, fake profiles creation is an important problem across these SNs, which have seen a sharp increase in recent times \cite{facebook10k}. In fake profiles which are also known as \emph{impersonating profiles}, a malicious user copies images from other profile and claims to be the person in that profile pictures. A natural solution to hinder the process of fake profile creation is by encouraging authorship attribution and verification through the images.

Broadly, two techniques, namely watermarking and steganography, techniques belonging to \emph{information hiding}, are used for embedding messages in digital content \cite{cox2007digital}. Watermarking is the practice of imperceptibly altering digital content to embed a mark. Watermarking is the most commonly used technique for owner identification, proof of ownership, authorship attribution and verification \cite{cox2007digital} and it is generally used in the domain of digital copyright protection \cite{kahng1998watermarking}. The crux of the idea is to embed an invisible payload (i.e., the mark) into the digital content helping users in proving their ownership and, subsequently, avoiding privacy violations of the shared media. Whereas steganography is the practice of undetectably altering digital content to embed a secret message. Since steganography techniques can be used with a mark, and not only with a secret message, we refer to watermarking approaches both for conventional watermarking and steganography, throughout this work.

\subsection{Problem Statement}
In this study, the scope of digital content is limited to the images, since image sharing is the most commonly observed behaviour on SNs. In particular, this work explores the possibility of watermarking the images being uploaded on SNs, and answers to the following three questions:

\begin{enumerate}
\item \textbf{Social Network Watermarking -} \emph{Are SNs watermarking our images?}\\In Section \ref{sec:invest}, we present the results of our experiments to understand whether SNs mark the images being uploaded on the social platforms. Our findings revealed that in general SNs do not perform any watermarking techniques on images. Out of all the 13 SNs being analysed, we found out that only Facebook changes to some extent some of the metadata associated with images. We performed extensive tests in order to verify if these changes can be imputed to a watermarking function.

\item \textbf{User Explicit Watermarking -} \emph{Can conventional watermarking techniques pass through SNs unaffected?}\\In Section \ref{sec:explicit}, we explore various watermarking algorithms as a tool for reliable marking the images to be uploaded on SNs. From the analysis, we found out that there is not a single watermarking technique that can be successfully used across all the selected SNs.

\item \textbf{User Unaware Watermarking -} \emph{Are we unawarely watermarking our images?}\\ By taking inspiration from previous works where researchers exploited sensor imperfections to extract the fingerprint to identify a smartphone \cite{dey2014accelprint}, \cite{bojinov2014mobile}, \cite{das2014you}, we analysed if the camera of smartphones, through which the images have been taken, can be used for creating a watermark. According to conventional watermarking literature \cite{cox2007digital}, the proposed method is \emph{invisible} and \emph{detectable} and belongs to \emph{fragile} and \emph{blind} categories (see Section \ref{sec:backw}). In practice, the fingerprint of the smartphone camera can be extracted by the images without altering the device. In Section \ref{sec:implicit}, we provide more details about our method. We demonstrate that the proposed method is robust enough despite the uploading and downloading process of the SNs that downgrades the shared images.
\end{enumerate}

\subsection{Contributions}\label{sec:intro_contr}
To the best of our knowledge, this is the first study that investigates the use of several image watermarking approaches on 13 SNs. The results of our analysis showed that no profile-dependent watermark operations are performed by SNs. Our study also reveals that conventional watermarking techniques can only pass unaffected through a subset of the SNs being considered in this study.

\begin{figure}[!h]
	\centering
	\includegraphics[width=3.3in,clip,keepaspectratio]{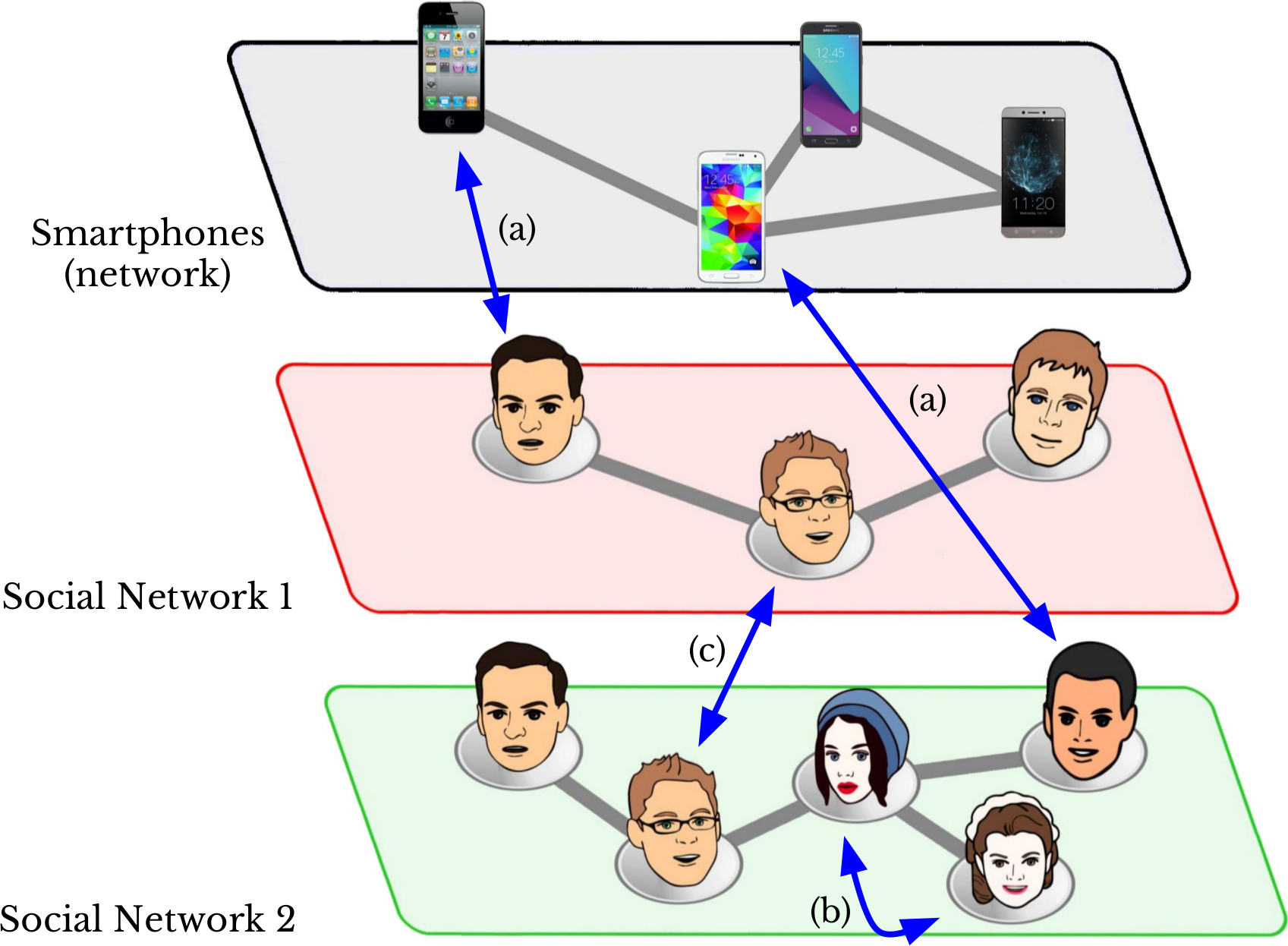}
	\caption{\emph{User unaware watermarking}'s three main tasks: \emph{profile attribution}~(a), \emph{intra-layer user profiles linking} (b), and \emph{inter-layer user profiles linking} (c).}
	\label{fig:multilayer}
\end{figure}

The most important finding of this study is that even if the author of the image does not consciously perform any \emph{user explicit watermarking} technique, the smartphone camera still embeds its characteristic fingerprint in every taken picture. Likewise to conventional watermarking approach, the method allows to identify the rightful owner of the images and can be successfully applied in all the 13 considered SNs. Figure \ref{fig:multilayer} shows users' smartphones and two different SNs where users shared images taken from their smartphones. We used the fingerprints left by the smartphone's camera to deal with the following problems:
\begin{enumerate}
\item \emph{Profile Attribution -} the task to match, through a set of shared images, a user profile to the right smartphone within a set of devices, case (a) in Figure~\ref{fig:multilayer}.
\item \emph{Intra-layer User Profiles Linking -} the task to decide if a restricted set of user profiles, within the same SN, belong to the same user, case (b) in Figure \ref{fig:multilayer}.
\item \emph{Inter-layer User Profiles Linking -} likewise to the previous one, this task tries to match user profiles that belong to different SNs, case (c) in Figure \ref{fig:multilayer}.
\item \emph{Fake Profiles Detection -} the task to identify unauthorized clone of user profiles. This task is a corollary of all the previous tasks since an untrusted/fake profile can be linked to a verified one using the shared images.
\end{enumerate}

It is noteworthy that the two types of \emph{user profiles linking} tasks described above are not always possible to achieve by using conventional watermarking techniques since in some cases it is necessary to have original the image to extract the watermark. The proposed method is robust enough in spite of the fact that the images get downgraded during the uploading and downloading process on the SNs. Moreover, it is possible to address the authorship attribution and verification in SNs without having the original images or altering the device.\\

The remainder of the paper is organized as follows. In Section \ref{sec:backw}, we provide a brief background in watermarking techniques. Section \ref{sec:related} presents literature on three different domains, that is image watermarking, smartphone fingerprinting and user profiles linking in SNs. The SNs used in our work are briefly described in Section \ref{sec:expSet}. The investigation activities regarding possible built-in watermarking techniques adopted by SNs are presented in Section \ref{sec:invest}. In Section \ref{sec:explicit}, \emph{user explicit watermarking} techniques on SNs are discussed, and our methodology for \emph{user unaware watermarking} is presented in Section \ref{sec:implicit}. Concluding remarks are made in Section \ref{sec:conc}.

\section{Background in Watermarking}\label{sec:backw}
Depending on the digital contents, different watermarking techniques are used: syntactic, semantic, and structural transformation for text \cite{kaur2015existential}; domain transformation for images \cite{potdar2005survey} and video \cite{bhattacharya2006survey}. In accordance with the literature \cite{cox2007digital}, watermarking methods can be categorized as follows:
\begin{itemize}
\item \emph{Readable or Detectable} - If the user can read the watermark it is called readable, while if the user can only check whether the watermark is embedded or not, it is called detectable.
\item \emph{Visible or Invisible} - A visible watermarking is visually perceptible by the user.
\item \emph{Blind or Non-Blind} - The watermarking is blind if the extraction process does not need the original non-marked digital content.
\item \emph{Zero Watermarking} - The embedding process does not modify the digital content and exploits its characteristics.
\item \emph{Simple or Multiple} - A multiple watermarking can be applied more than one time without affecting the previous watermark embedding steps.
\item \emph{Fragile, Semi-Fragile, Robust} - A fragile watermarking can be altered or erased and it is used for integrity authentication. A robust watermarking cannot be easily erased and is most suitable for copyright protection, while semi-fragile watermarking is suited for content authentication.
\end{itemize}
The following features are usually required for watermarking methods \cite{nyeem2014digital}:
\begin{itemize}
\item \emph{Robustness} is the ability to resist to processing operations and attacks.
\item \emph{Imperceptibility} represents the ability to be detected through special processing of watermark detector.
\item \emph{Security} is the capacity to not be altered or removed without having full knowledge of the watermarking algorithm.
\item \emph{Data Payload} denotes the maximum number of extra information that can be embedded in the digital content.
\item \emph{Computational Cost} is the cost required in the embedding and extraction process.
\item \emph{Non-inevitability} represents the possibility to not extract a watermark from a non-marked digital content.
\end{itemize}
All this categorization can also be applied to image watermarking techniques.

\section{Related Works}\label{sec:related}
In this section, we describe literature from three domains, all relevant for a full comprehension of our work. Firstly, we focus on conventional image watermarking techniques in SNs. Next, we present various approaches for uniquely identifying smartphones. Finally, we explain methods for associating user profiles in SNs, which is the main outcome of this work.

\subsection{Image Watermarking in Social Networks}
Image watermarking techniques embed a mark in a visually imperceptible way for authentication and copyright protection tasks \cite{naskar2014reversible}. According to the embedding domain \cite{potdar2005survey}, these techniques can be classified in \emph{spatial} and \emph{transform domain watermarking}.\\
The methods in \emph{spatial domain watermarking} class directly modify the image pixel acting on the bit value. The simplest approach embeds the watermark in the Least Significant Bits (LSB) \cite{bamatraf2010digital}. Intermediate Significant Bit (ISB) \cite{zeki2009novel} improves the LSB method and defines the watermarked location according to the range of each bit-plane. Patchwork \cite{bender1996techniques} is another spatial domain approach where the watermark is embedded into the image by changing the pixels' brightness. In general, the simple implementation of the spatial domain methods implies less robustness to affine transformations and image processing attacks \cite{mishra2014optimized}.\\
The methods in \emph{transform domain watermarking} class apply a transformation to the original image and exploit the transformed coefficients to embed the watermark. There are four main techniques in transform domain: Discrete Cosine Transform (DCT), Discrete Wavelet Transform (DWT), Discrete Fourier Transform (DFT), and Singular Value Decomposition (SVD). Typically, DCT algorithms segment the image into blocks and modify a set of selected coefficients \cite{cox1997secure}. Similarly, in DWT techniques the original image is decomposed into three spatial directions (i.e., horizontal, vertical, and diagonal), then the watermark is embedded in the wavelet coefficients \cite{kundur1998digital}. DWT algorithms are computationally efficient and the visual artefacts introduced are less evident compared to DCT. The DFT algorithms employ the Fourier transform since it offers robustness against geometric attacks  \cite{o1997rotation}. DFT decomposes the original image in phase and magnitude representation. Then, the mark is embedded into the magnitude representation. SVD is one of the most powerful numeric analysis techniques \cite{chang2005svd}. In particular, SVD allows embedding of the mark into the singular matrix of the frequency domain or spatial domain coefficients with very less loss of information \cite{saravanan2016digital}. Moreover, in order to exploit the various properties, SVD is combined with the other techniques: DWT-SDV \cite{lai2010digital}, DFT-SVD \cite{zhang2006geometric}, and DWT-DFT-SVD \cite{ansari2012robust}. Since the transform domain methods lead to robust watermarking \cite{olanrewaju2011development}, in our tests we used methods that belong to the \emph{transform domain watermarking} class.\\
Recently, researchers investigated various watermarking techniques to be used in SNs and also the potential attacks and corresponding solutions \cite{chomphoosang2011survey}. In \cite{zigomitros2012social}, the researchers proposed a dual watermarking scheme for Facebook and Google+ by partially redesigning the SN uploading service which may not be feasible. In \cite{hiney2015using} authors considered Facebook as a closed system and tested different steganography methods. A watermarking method based on wavelet decomposition proposed in \cite{banos2015novel} was successfully tested by simulating different attacks that can be performed on SNs. In \cite{thongkor2013image} the authors proposed a method based on DWT coefficients, however, the tests were performed on images with low resolution that are not compressed during the uploading process on the selected four SNs. Similarly, in \cite{lyu2017reversible} a method based on DCT transformation was proposed for Facebook using images that the SN does not resize. A method based on backpropagation neural network was proposed in \cite{singh2016multiple}, however, the authors did not perform any test on real SNs and used the images with very low resolution in comparison to the resolution of current smartphone cameras. An orthogonal approach was presented in \cite{szczypiorski2016steghash}, where the author proposed to use SN hashtags (i.e., labels containing a word starting with the ``\#'' symbol) to hide information through the images uploaded on Twitter and Instagram. To the best of our knowledge, this is the first work where images with high resolution and thirteen SNs are used to investigate image watermarking techniques in SNs.

\subsection{Fingerprinting the Smartphone Devices}
Smartphones are becoming more and more pervasive in daily activities. Recently, researchers proposed methods for identifying and fingerprinting the smartphone by exploiting personalized configurations \cite{kurtz2016fingerprinting}, touchscreen interaction \cite{velten2015user}, and on-board camera \cite{huang2015camera}. Today the smartphones are equipped with a bunch of sensors. These sensors are produced according to industrial standards, which makes them ideally identical, however, each sensor has an imperfection that makes it unique and identifiable \cite{baldini2017survey}. Microphones and speakers, through playback and recording of audio samples, were exploited in \cite{das2014you}. In \cite{dey2014accelprint} the authors proposed a technique using the integrated accelerometers for identifying mobile phones. An improvement compared to \cite{dey2014accelprint} was proposed in \cite{bojinov2014mobile}, where the authors combined speakerphone-microphone along with the accelerometer.\\
Among all the sensors described above, the most investigated sensor in the field of digital forensic is camera \cite{van2007survey}. The reason is that the camera offers several components inside it which can be used to identify the source camera. The chromatic aberration introduced by the lens was exploited in \cite{van2007identifying} whereas in \cite{bayram2005source} the camera identification was achieved through the colour filter array (CFA), that can also be used to recognize a counterfeit image in \cite{popescu2005exposing}. In \cite{lukas2006digital} the authors proposed a sensor-based method exploiting the Photo-Response Non-Uniformity (PRNU) for successfully distinguishing cameras of the same model. A PRNU-based method able to operate with different image sizes was presented in \cite{goljan2008camera}. In \cite{castiglione2011forensic}, Castiglione et al. classified all the various changes made by the SNs on uploaded images, that commonly causes a loss of effectiveness of the camera fingerprinting methods. Later in \cite{castiglione2013experimentations}, they demonstrated the robustness of the PRNU-based method on the images downloaded from 6 SNs and online photo-sharing platforms. Since the effectiveness of PRNU-based approach has been widely demonstrated in \cite{kang2012enhancing}, including on large scale image dataset \cite{goljan2009large}, we decided to use it for source camera fingerprinting on SNs.

\subsection{User Profiles Linking in Social Networks}
Trust is a fundamental ingredient in SNs \cite{agreste2015trust} and a significant increase in the number of impersonating fake profiles has given rise to various solutions for deciding whether a given user profile is an unauthorized clone of another one \cite{kontaxis2011detecting}. In \cite{fire2014friend}, the authors proposed a three layered tool for Facebook which was able to \emph{i}) identify suspicious users, \emph{ii}) expand basic privacy settings, \emph{iii}) warn the user against malicious applications. Similarly, a graph-based framework for detecting fake profile attacks was proposed in \cite{conti2012fakebook}.\\
The more generic user profile linking task allows matching different user profiles belonging to the same user, which is analogous to missing data problem in multilayer networks \cite{sharma2015investigating}. An invasive and device-dependent solution exploits the log information stored on the device's internal memory during the use of the SNs' application \cite{al2012forensic}. The more effective solutions for user profile linking exploit the information and multimedia contents that transit on SNs. A framework for user profile linking, based on the profile's attributes, was proposed in \cite{raad2010user}. While in \cite{iofciu2011identifying} the authors combined tags and user ID to match users' profile across social tagging systems, the solutions proposed in \cite{vosecky2009user} and \cite{liang2013fully} match user profiles by using information about users' identities without compromising their privacy. In \cite{bartunov2012joint}, the authors presented a method that combines profile attributes and social linkage to outperform common attribute-based approach, whereas in \cite{jain2013seek} network attributes and profile attributes were combined to improve a traditional identity search algorithms. A weighted ontology-based user profile linking technique was proposed in \cite{cortis2013ontology}. In order to improve the performance of the attribute-based approach proposed in \cite{goga2015reliability} and \cite{peled2013entity} the authors proposed machine learning techniques for matching user profiles across multiple SNs. Typically, these approaches fail if the malicious user falsifies the information stored into the fake profile, as it usually happens.

\section{Experimental Settings} \label{sec:expSet}
In this section, we describe the social networks and the images' characteristics that were used in our analysis.

\subsection{Social Networks} \label{sec:social}
Thirteen most popular SNs shown in Table \ref{tab:sn} were used for our investigation. The selection of these SNs was based on two different criteria: the number of user accounts and the different features offered by these social platform. All the selected SNs in total count more than 100 million users and cover different needs, like social interactions (Facebook), photo sharing (Instagram), blogging (Tumblr), instant messaging (WhatsApp) to name a few. Moreover, we decided to include SNs developed outside the United States and European Union, like Telegram, VK (originally VKontakte), and WeChat. Since we are interested in image watermarking algorithms, in Table \ref{tab:sn}, we also specify the default pixel resolution accepted by each social platform.

\begin{table}[!h]
	\renewcommand{\arraystretch}{0.85}
	\centering
	\caption{Social Networks used for the investigations activities and the default pixel resolution accepted by each platform.}\label{tab:sn}
	\begin{tabular}{|c|l||c|}
		\hline
		\textbf{ID} &\textbf{Social Networks} & \textbf{Image sizes}\\ \hline \hline
        SN01 & Facebook & 2048$\times$1152\\ \hline
        SN02 & Flickr & 2048$\times$1152\\ \hline
        SN03 & Google+ & 2048$\times$1152\\ \hline
        SN04 & Instagram & 1080$\times$1080\\ \hline
        SN05 & LinkedIn & 2048$\times$1152\\ \hline
        SN06 & Pinterest & 2048$\times$1152\\ \hline
        SN07 & Telegram & 1280$\times$720\\ \hline
        SN08 & Tumblr & 1280$\times$720\\ \hline
        SN09 & Twitter & 2048$\times$1152\\ \hline
        SN10 & Viber & 1280$\times$720\\ \hline
        SN11 & VK & 2560$\times$1440\\ \hline
        SN12 & WeChat & 1280$\times$720\\ \hline
        SN13 & WhatsApp & 1600$\times$1200\\ \hline
	\end{tabular}
\end{table}

\subsection{Images' Characteristics} \label{sec:img}
All the SNs accept Joint Photographic Experts Group (JPEG or JPG) images with standard pixel resolution, that is default image sizes (see Column III in Table \ref{tab:sn}), beyond which the image is automatically scaled to the default resolution. For this reason, we carried out tests for \emph{social network watermarking} and \emph{user explicit watermarking} with three different resolutions: the standard, that matches the images sizes of the SN; the larger than the standard (4128$\times$2322) and smaller than the standard (640$\times$480). We used 10 different images for each resolution class: standard, large, and small. Moreover, both in \emph{social network watermarking} and in \emph{user explicit watermarking}, we created two different user profiles for each of the SNs (i.e., \emph{P}$_1$ and \emph{P}$_2$). This had twofold advantages. Firstly, it allowed us to compare the original image with the downloaded one, that is ``Original vs Shared'' case. Secondly, we were able to compare the same image being uploaded twice on two different user profiles, that is ``Shared \emph{P}$_1$ vs Shared \emph{P}$_2$'' case. In \emph{user unaware watermarking}, we used a real dataset of 8,400 pictures collected using 6 different smartphones and for each smartphone, a different user profile on each SN was created (for more details see Section \ref{sec:implicit}).

\section{Social Networks Watermarking}\label{sec:invest}
The first direction of investigation is about \emph{Are SNs watermarking our images?} The context of watermarking can be broadened in terms of changes performed by a particular SN, such as different name and metadata associated with the image after being uploaded on the SN. Thus, different comparisons can be performed among uploaded and downloaded images. Firstly, we performed some preliminary analysis about the compression of the images. Next, we compared names, contents (i.e., the image without metadata), and metadata of the uploaded and downloaded images on each SN. Finally, we executed some extra analysis on Facebook metadata, including the Content Delivery Network (CDN) that is an intermediate layer of proxy servers distributed globally. Typically, a CDN is widely used by high traffic websites, since it allows to provide digital content with high availability and performance. In the following subsections, we describe all the tests about \emph{social networks watermarking}.

\subsection{Preliminary Analysis}\label{sec:pan}
We observed that when an image is uploaded to a SN, it is compressed by the JPG compression algorithm adopted by the platform for preserving the good quality of the image. The compression is due to an optimization of the image made by the SN. In particular, the quantization matrix coefficients in the JPG compression algorithm control the compression ratio. This means that any uploaded image may incur different file sizes, compared to the original, after downloading. To the best of our knowledge, SNs do not publish any information about the image processing algorithms being used inside their platform and thus for us, they are like a ``black box''. For this reason, we investigated the behaviour of the selected SNs for particular compression algorithms by describing the outcomes of our analysis. Table \ref{tab:compr} provides information about the average compression results for each SNs and each resolution class, namely standard, large (4128$\times$2322), and small (640$\times$480).

\begin{table}[!h]
	\renewcommand{\arraystretch}{0.85}
	\centering
	\caption{Average compression results obtained from each Social Networks and each resolution class.}\label{tab:compr}
	\begin{tabular}{|l||c|c|c|c|}
		\hline
		\textbf{Social Networks} & \textbf{Standard} & \textbf{Large} & \textbf{Small} \\ \hline \hline
		Facebook & 66.54\% & 91.30\% & 76.25\% \\ \hline
		Flickr & \textbf{0.00\%} & \textbf{0.00\%} & \textbf{0.00\%} \\ \hline
		Google+ & \textbf{0.00\%} & \textbf{0.00\%} & \textbf{0.00\%} \\ \hline
		Instagram & 31.94\% & 94.14\% & 64.32\% \\ \hline
		LinkedIn & 68.12\% & 67.39\% & 74.94\% \\ \hline
		Pinterest & 46.04\% & 83.82\% & 52.96\% \\ \hline
		Telegram & 62.91\% & 95.55\% & 70.32\% \\ \hline
		Tumblr & 30.42\% & 82.83\% & 35.37\% \\ \hline
		Twitter & 53.27\% & 88.41\% & 57.12\% \\ \hline
		Viber & 59.72\% & 94.50\% & -46.50\% \\ \hline
		VK & 2.33\% & 79.17\% & 62.43\% \\ \hline
		WeChat & 65.97\% & 96.07\% & 55.97\% \\ \hline
		WhatsApp & 58.49\% & 93.60\% & 70.59\% \\ \hline
	\end{tabular}
\end{table}

Our findings pointed that for each resolution class, Flickr and Google+ do not apply any compression and preserve the original image size. On an average, VK applies a very low compression to standard images and all remaining ten SNs compress the standard images more than 54.34\%, on an average. Generally, large resolution images are strongly compressed by all SNs (except Flickr and Google+), that is in the range of 67\% to 96\% and the pixel resolution drops to default one (see Table \ref{tab:sn}). It is interesting to note that Viber reduces the compression rate for images with low resolution turning the JPG quality from 96 to 100, which resulted in a negative value of \mbox{-46.50\%}. In only four cases the pixel resolution does not match the default size in Table \ref{tab:sn} (Column III): Instagram, Pinterest, Tumblr and WhatsApp large images were scaled to 1350$\times$1080, 2064$\times$1161, 1920$\times$1080, and 1600$\times$900, respectively.

\subsection{Images Comparison}
In this section,  we  present results of four different kinds of comparison on images before and after uploading of images on SNs to investigate if these SNs perform any watermarking activity on the uploaded (shared) images.
\begin{itemize}
\item \emph{Name comparison} - This test allowed us to understand if SNs change the name of the uploaded images.
\item \emph{Full comparison} - The full comparison was performed by exploiting the \emph{Secure Hash Algorithm} version 1 (SHA-1), to find the difference between pair of images. SHA-1 takes an image as input and produces a unique 160-bit message digest. Any change in content and metadata of the image implies a different digest in output.
\item \emph{Content comparison} - This test was performed by using a bit by bit comparison of the image's content excluding the image's metadata. The test compared two images in binary representation highlighting the differences between pixels.
\item \emph{Metadata  comparison} - Metadata is defined as the data providing extra information about one or more aspects of the file. The images' metadata is specified by the \emph{Exchangeable image file format} (Exif) standard that includes information like time, location, camera settings, descriptions, and copyright information.
\end{itemize}
Firstly, the 30 images for each resolution class described in Section \ref{sec:expSet} were uploaded and downloaded on both \emph{P}$_1$ and \emph{P}$_2$ profiles on each SNs. Then, for each SN, we performed two different kinds of test. In the first one, we compared the original images with the downloaded one, ``Original vs Shared'' case. In the second test, we compared the shared images on different profiles, ``Shared \emph{P}$_1$ vs Shared \emph{P}$_2$'' case. We wanted to examine for any SN, whether the changes reverberate in the same way across different profiles or it is unique for each profile. In the following subsections, we present the results for each of the four kinds of comparison.

\subsubsection{Name Comparison}
We investigated how the SNs change the image's name and if the same image receives different names if shared on different user profiles. The name of the uploaded images was in the following format: 20161028\_085447. Out of all SNs, Google+ does not change the original name even when the same image is shared on two different profiles. All remaining SNs use a specific encoding for the image's name, as shown in Table \ref{tab:name} (the name of the Viber file is quite long, so the variable part of 65 characters is defined with the regex \mbox{[a-z0-9]$\{$65$\}$}).

\begin{table}[!h]
	\renewcommand{\arraystretch}{0.85}
	\centering
	\caption{Name format of the downloaded image. The original uploaded name was same for all SNs.}\label{tab:name}
	\begin{tabular}{|l||c|}
		\hline
		\textbf{Social Networks} & \textbf{Downloaded image name} \\ \hline
		Facebook & 14633305\_13935419006590\_2203780186632661\_o \\ \hline
		Flickr & 30319899670\_77e6fd4bed\_o \\ \hline
		Google+ & \textbf{unchanged} \\ \hline
		Instagram & 14533468\_7761667291914\_4275777725718855\_n \\ \hline
		LinkedIn & 9f86293d-bdaf-4bce-b5ce-c5610e2cd9b8-original \\ \hline
		Pinterest & 2ecd9963cac22479edbd03d65b43dd2a \\ \hline
		Telegram & IMG\_20171029\_184428 \\ \hline
		Tumblr & tumblr\_ofswzjXl7K1vjbnv5o1\_1280 \\ \hline
		Twitter & Cv3ahPvXEAAo6CR.jpg-large \\ \hline
		Viber & image-0-02-05-[a-z0-9]$\{$65$\}$-V \\ \hline
		VK & 0ysdRR9cIVc \\ \hline
		WeChat & mmexport1477761254890 \\ \hline
		WhatsApp & IMG-20171029-WA0019 \\ \hline
	\end{tabular}
\end{table}

Instagram uses the same name format of Facebook, while WhatsApp uses a different one even though they both belong to Facebook. In some cases, like Telegram and WhatsApp, the name is created by using the downloading date and time. The Pinterest name format is the most interesting. The original name is not preserved and the image receives the same name when shared on different profiles, irrespective of the resolution class. This means that the name might be a good candidate for watermarking the images, however, it is quite short and the watermark could be easily removed or changed.

\subsubsection{Full Comparison}
For this comparison, we used the SHA-1 algorithm to check if the integrity of the original image was preserved. If SHA-1 produces different digests, then it means that the same input image has undergone a few changes during the uploading and downloading process. 

\begin{table}[!h]
	\renewcommand{\arraystretch}{0.85}
	\centering
	\caption{The SHA-1 results for both the original and shared images and the shared images on different profiles. All the resolution classes are grouped together.}\label{tab:sha}
	\begin{tabular}{|l||c|c|}
		\hline
		\textbf{Social Networks} & \textbf{Original vs Shared} & \textbf{Shared \emph{P}$_1$ vs Shared \emph{P}$_2$} \\ \hline \hline
		Facebook & All differ & \textbf{All differ} \\ \hline
		Flickr & \textbf{All equals} & All equals \\ \hline
		Google+ & \textbf{All equals} & All equals \\ \hline
		Instagram & All differ & All equals \\ \hline
		LinkedIn & All differ & \textbf{All differ} \\ \hline
		Pinterest & All differ & All equals \\ \hline
		Telegram & All differ & All equals \\ \hline
		Tumblr & All differ & All equals \\ \hline
		Twitter & All differ & All equals \\ \hline
		Viber & All differ & All equals \\ \hline
		VK & All differ & All equals \\ \hline
		WeChat & All differ & All equals \\ \hline
		WhatsApp & All differ & All equals \\ \hline
	\end{tabular}
\end{table}

In Table \ref{tab:sha}, we provide collective results for all the resolution classes. The term ``All differ'' (or ``All equals'') means that for a given SN and a given test case we obtain different (or the same) results in all classes. The ``Original vs Shared'' column is consistent with the results in Table \ref{tab:compr}. The SHA-1 algorithm produces equal digests only for the two SNs that do not apply any compression, namely Flickr and Google+. Moreover, it means that even the metadata is not changed during the sharing process. The compression applied by the other SNs changes the images, thus the produced digests are different among original and shared ones.\\
The most interesting results were obtained for Facebook and LinkedIn in ``Shared \emph{P}$_1$ vs Shared \emph{P}$_2$'' case. The two SNs return different digests for all resolution classes when the images are shared on different profiles. With the next two comparisons, we will go more deeply to investigate if Facebook's and LinkedIn's unusual results are due to content or metadata changes.

\subsubsection{Content Comparison}
In this test, we compared images through a bit by bit different operation. Only if the two images have the same content the difference produces a full-zero matrix. The \emph{content comparison} narrows the field to the pixels' value excluding the metadata. Grouped results for all the resolution classes are shown in Table \ref{tab:bit}.

\begin{table}[!t]
	\renewcommand{\arraystretch}{0.85}
	\centering
	\caption{The bit by bit results for both the original and shared images and the shared images on different profiles. All the resolution classes are grouped together.}\label{tab:bit}
	\begin{tabular}{|l||c|c|}
		\hline
		\textbf{Social Networks} & \textbf{Original vs Shared} & \textbf{Shared \emph{P}$_1$ vs Shared \emph{P}$_2$} \\ \hline \hline
		Facebook & All differ & All equals \\ \hline
		Flickr & \textbf{All equals} & All equals \\ \hline
		Google+ & \textbf{All equals} & All equals \\ \hline
		Instagram & All differ & All equals \\ \hline
		LinkedIn & All differ & All equals \\ \hline
		Pinterest & All differ & All equals \\ \hline
		Telegram & All differ & All equals \\ \hline
		Tumblr & All differ & All equals \\ \hline
		Twitter & All differ & All equals \\ \hline
		Viber & All differ & All equals \\ \hline
		VK & All differ & All equals \\ \hline
		WeChat & All differ & All equals \\ \hline
		WhatsApp & All differ & All equals \\ \hline
	\end{tabular}
\end{table}

The ``Original vs Shared'' results are consistent with the results in Table \ref{tab:compr}. If the SN does not apply any compression, such as Flickr and Google+, the content of the original image matches bit by bit with the content of the shared one, while the compression of the other SNs has its obvious effects on bit-level. Moreover, we used an open-source framework \cite{zampoglou2017large} to test different algorithms to detect double JPG compression. The results showed no evidence of watermark based on double JPG compression. In ``Shared \emph{P}$_1$ vs Shared \emph{P}$_2$'' case, we have our first important result. Since in all SNs the image shared on \emph{P}$_1$ matches bit by bit with the same image shared on \emph{P}$_2$, we can infer that all the selected SNs do not apply any watermark into images' content \cite{naskar2014reversible}.

\subsubsection{Metadata Comparison}
Metadata summarizes basic information about the associated file. \emph{Exif} standard defines the metadata attributes for digital images \cite{exif}. The full attributes list can be categorized in: date and time information, static and dynamic camera settings, general descriptions, copyright information, and the thumbnail (i.e., a smaller version of the image for indexing and previewing).

The \emph{metadata comparison} produced exactly the same \emph{full comparison} results as in Table \ref{tab:sha}. The analysis of each single attributes list for each SN gave rise to some interesting results.\\
Flickr and Google+ preserve all the original attributes list, whereas, Twitter erases all of them. Tumblr removes only the thumbnails related fields, while all remaining SNs (i.e., Instagram, Pinterest, Telegram, Viber, VK, WeChat, and WhatsApp) preserve only a few subsets of the original attributes list concerning general descriptions and static camera settings. In most of the cases, this subset does not include any GPS information. However, in all these SNs the attributes list is the same when the same image is shared on two different profiles.

We performed further investigations on LinkedIn and Facebook as the metadata changes when the same image is shared through different user profiles. In particular, LinkedIn uses the subset of the same attributes of the previous SNs and changes some non-significant attributes, like ``Exif Byte Order'' and the ``Resolution Unit''. However, these changes can not be related to a watermark since they do not produce a unique identifier, but rather a standard string for those attributes. In comparison to LinkedIn, Facebook, substitutes a set of attributes by using the \emph{Information Interchange Model} (IIM), that is a set of metadata attributes defined by the International Press Telecommunications Council (IPTC). Three of the new attributes, ``Special Instructions'', ``Current IPTC Digest'' and ``Original Transmission Reference'' receive very unusual values, that is three alphanumeric-characters codes. Moreover, the ``Special Instructions'' does not change if the same image is shared on two different profiles, as the other two are strictly related to the user profile that shares the image.\\
According to the standard, the ``Special Instructions'' is a text field that can include special restrictions, additional permissions, and credits required when publishing. The ``Current IPTC Digest'' is the hash digest of the other IPTC data. Finally, ``Original Transmission Reference'' is typically used to improve the transmission and routing purposes of the image. Facebook does not release any details about these fields' encoding. Even if the last two fields change when the same image is shared on different profiles, we can not claim that they are used for watermarking purposes.\\ \indent
In order to investigate the Facebook metadata case, we performed some additional tests. In particular, we wanted to study the following three crucial aspects:
\begin{itemize}
\item \emph{Time test} - The images were uploaded and downloaded twice on the same profile, after a certain time. The aim was to determine whether the metadata was time-dependent.
\item \emph{Sharing test} - The aim was to determine whether the metadata changed when the image was shared across profiles. In practice, the images were uploaded on profile \emph{P}$_1$ and downloaded three times: from \emph{P}$_1$, from \emph{P}$_2$ that had visited \emph{P}$_1$, and from \emph{P}$_2$ that had shared the images on his/her wall (i.e., the web page where others users, like friends and fans, can post their thoughts, images and video). The aim was to determine whether the metadata was sharing-dependent
\item \emph{Location test} - CDN provides digital content from locations closer to the user. Since it is unknown which node of the CDN serves our request, we used a VPN to repeat \emph{sharing test} forcing one of the two profiles to be located in the following countries: Russia, China, United States and United Kingdom. The aim was to determine whether the metadata was location-dependent.
\end{itemize}
In all these three new tests, the previous \emph{name comparison}, \emph{full comparison}, and \emph{content comparison} produced the same results. The \emph{metadata comparison} outcome was much more interesting. The attributes list was the same as the previous test, but the three new fields ``Special Instructions'', ``Current IPTC Digest'', and ``Original Transmission Reference'' had different behaviour. In \emph{time test}, the ``Special Instructions'' was the same even after 24 hours, as ``Current IPTC Digest'' and ``Original Transmission Reference'' received different values just after few seconds. In \emph{sharing test} and \emph{location test} all the three attributes preserved the same values. This means that the metadata is surely profile-dependent and time-dependent, but not sharing and location dependent.\\

At the time of writing, the \emph{social network watermarking} test reveals that none of the selected SNs applies any profile-dependent watermark visible outside the network. Facebook introduces some suspicious values in three new attributes. However, the metadata can be easily erased or modified and cannot be considered a good solution for authorship attribution and verification purpose.

\section{User Explicit Watermarking}\label{sec:explicit}
In this section, we present the results of some conventional image watermarking techniques on SNs we considered in our study. The goal is to figure out if these approaches can reliably be used on SNs for marking the images to be uploaded. In order to carry out our tests, we selected several free \emph{invisible} watermarking algorithms. We performed our analysis on the set of SNs and images described in Section \ref{sec:expSet}.

There are a number of watermarking tools, however, we selected open-source or freeware tools that work with the image file formats accepted by SNs (i.e., jpg and png). These tools can be categorized into \emph{spatial} and \emph{transform domain watermarking} algorithms. The first category includes easy to implement and low complexity methods. On the flip side, these watermarking tools present weaknesses like weak robustness and low security. The methods in \emph{transform domain watermarking} class are widely applied and they can reach a good balance between robustness and imperceptibility. For our tests we selected the following thirteen algorithms:

\begin{itemize}
	\item[{\footnotesize A1)}] \emph{BlindHide} replaces the least significant bits (LSB) of each pixel with the watermark \cite{hempstalk2006hiding}.
	\item[{\footnotesize A2)}] \emph{HideSeek} tries to get around the security issues in BlindHide by distributing the watermark across the image with a different pixel order \cite{hempstalk2006hiding}.
	\item[{\footnotesize A3)}] \emph{FilterFirst} adopts an edge-detecting filter (Laplace) to check the area with value-homogeneous pixels \cite{hempstalk2006hiding}.
	\item[{\footnotesize A4)}] \emph{BattleSteg} is based on the Battleships game and identifies the best area of the image to embed the watermark \cite{hempstalk2006hiding}.
	\item[{\footnotesize A5)}] \emph{Dynamic FilterFirst} improves the FilerFirst by using dynamic programming \cite{hempstalk2006hiding}.
	\item[{\footnotesize A6)}] \emph{Dynamic BattleSteg}, like the \emph{Dynamic FilterFirst}, uses dynamic programming to improve BattleSteg \cite{hempstalk2006hiding}.
	\item[{\footnotesize A7)}] \emph{F5} implements matrix encoding during the JPEG compression process \cite{steganalysis2001f5}.
	\item[{\footnotesize A8)}] \emph{OpenPuff} applies multiple layers of protection and data decorrelation in order to embed the watermark \cite{sloan2015steganalysis}.
	\item[{\footnotesize A9)}] \emph{OpenStego} implements Dugad's algorithm \cite{Dugad}, a wavelet-based method, for watermarking the image \cite{openStego}.
	\item[{\footnotesize A10)}] \emph{Secretbook} is specifically designed for performing steganography on Facebook by exploiting the quality factor and the quantisation matrix \cite{secretbook}.
	\item[{\footnotesize A11)}] \emph{SilentEye} combines LSB and Advanced Encryption Standard (AES) technique to hide the watermark \cite{silenteye}.
	\item[{\footnotesize A12)}] \emph{SteganPEG} performs data compression and decompression before watermarking the image \cite{reddy2011steganpeg}.
	\item[{\footnotesize A13)}] \emph{Steghide} uses a graph-theoretic algorithm to find pairs of positions of the image to be swapped in order to embed the watermark \cite{steghide}.
\end{itemize}

\begin{table}[!b]
	\renewcommand{\arraystretch}{0.95}
	\centering
	\caption{The watermark extraction results for each selected algorithms for all SNs. The success of the extraction process is shown with a grey cell following this sequence standard, large (4128$\times$2322), and small (640$\times$480) resolution images.}\label{tab:col}
\resizebox{12.2cm}{!}{
	\begin{tabu}{|l||l|l|l|[1.2pt]l|l|l|[1.2pt]l|l|l|[1.2pt]l|l|l|[1.2pt]l|l|l|[1.2pt]l|l|l|[1.2pt]l|l|l|[1.2pt]l|l|l|[1.2pt]l|l|l|[1.2pt]l|l|l|[1.2pt]l|l|l|[1.2pt]l|l|l|[1.2pt]l|l|l|}
		\hline
		\textbf{SNs} & \multicolumn{3}{c|}{\textbf{A1}} & \multicolumn{3}{c|}{\textbf{A2}} & \multicolumn{3}{c|}{\textbf{A3}} & \multicolumn{3}{c|}{\textbf{A4}} & \multicolumn{3}{c|}{\textbf{A5}} & \multicolumn{3}{c|}{\textbf{A6}} & \multicolumn{3}{c|}{\textbf{A7}} & \multicolumn{3}{c|}{\textbf{A8}} & \multicolumn{3}{c|}{\textbf{A9}} & \multicolumn{3}{c|}{\textbf{A10}} & \multicolumn{3}{c|}{\textbf{A11}} & \multicolumn{3}{c|}{\textbf{A12}} & \multicolumn{3}{c|}{\textbf{A13}} \\ \hline \hline
		Facebook & & & & & & & & & & & & & & & & & & & & & & & & & & & & \cellcolor[HTML]{9B9B9B} & \cellcolor[HTML]{9B9B9B} & \cellcolor[HTML]{9B9B9B} & \cellcolor[HTML]{9B9B9B} & & & & & & & & \\ \hline
		Flickr & \cellcolor[HTML]{9B9B9B} & \cellcolor[HTML]{9B9B9B} & \cellcolor[HTML]{9B9B9B} & \cellcolor[HTML]{9B9B9B} & \cellcolor[HTML]{9B9B9B} & \cellcolor[HTML]{9B9B9B} & \cellcolor[HTML]{9B9B9B} & \cellcolor[HTML]{9B9B9B} & \cellcolor[HTML]{9B9B9B} & \cellcolor[HTML]{9B9B9B} & \cellcolor[HTML]{9B9B9B} & \cellcolor[HTML]{9B9B9B} & \cellcolor[HTML]{9B9B9B} & \cellcolor[HTML]{9B9B9B} & \cellcolor[HTML]{9B9B9B} & \cellcolor[HTML]{9B9B9B} & \cellcolor[HTML]{9B9B9B} & \cellcolor[HTML]{9B9B9B} & \cellcolor[HTML]{9B9B9B} & \cellcolor[HTML]{9B9B9B} & \cellcolor[HTML]{9B9B9B} & \cellcolor[HTML]{9B9B9B} & \cellcolor[HTML]{9B9B9B} & \cellcolor[HTML]{9B9B9B} & \cellcolor[HTML]{9B9B9B} & \cellcolor[HTML]{9B9B9B} & \cellcolor[HTML]{9B9B9B} & \cellcolor[HTML]{9B9B9B} & \cellcolor[HTML]{9B9B9B} & \cellcolor[HTML]{9B9B9B} & \cellcolor[HTML]{9B9B9B} & & \cellcolor[HTML]{9B9B9B} & \cellcolor[HTML]{9B9B9B} & \cellcolor[HTML]{9B9B9B} & & \cellcolor[HTML]{9B9B9B} & & \cellcolor[HTML]{9B9B9B} \\ \hline
		Google+ & \cellcolor[HTML]{9B9B9B} & \cellcolor[HTML]{9B9B9B} & \cellcolor[HTML]{9B9B9B} & \cellcolor[HTML]{9B9B9B} & \cellcolor[HTML]{9B9B9B} & \cellcolor[HTML]{9B9B9B} & \cellcolor[HTML]{9B9B9B} & \cellcolor[HTML]{9B9B9B} & \cellcolor[HTML]{9B9B9B} & \cellcolor[HTML]{9B9B9B} & \cellcolor[HTML]{9B9B9B} & \cellcolor[HTML]{9B9B9B} & \cellcolor[HTML]{9B9B9B} & \cellcolor[HTML]{9B9B9B} & \cellcolor[HTML]{9B9B9B} & \cellcolor[HTML]{9B9B9B} & \cellcolor[HTML]{9B9B9B} & \cellcolor[HTML]{9B9B9B} & \cellcolor[HTML]{9B9B9B} & \cellcolor[HTML]{9B9B9B} & \cellcolor[HTML]{9B9B9B} & \cellcolor[HTML]{9B9B9B} & \cellcolor[HTML]{9B9B9B} & \cellcolor[HTML]{9B9B9B} & \cellcolor[HTML]{9B9B9B} & \cellcolor[HTML]{9B9B9B} & \cellcolor[HTML]{9B9B9B} & \cellcolor[HTML]{9B9B9B} & \cellcolor[HTML]{9B9B9B} & \cellcolor[HTML]{9B9B9B} & \cellcolor[HTML]{9B9B9B} & & \cellcolor[HTML]{9B9B9B} & \cellcolor[HTML]{9B9B9B} & \cellcolor[HTML]{9B9B9B} & & \cellcolor[HTML]{9B9B9B} & & \cellcolor[HTML]{9B9B9B} \\ \hline
		Instagram & & & & & & & & & & & & & & & & & & & & & & & & & & & & & & & & & & & & & & & \\ \hline
		LinkedIn & & & & & & & & & & & & & & & & & & & & & & & & & & & & & & & & & & & & & & & \\ \hline
		Pinterest & & & & & & & & & & & & & & & & & & & & & & & & & & & & & & & \cellcolor[HTML]{9B9B9B} & & \cellcolor[HTML]{9B9B9B} & & & & & & \\ \hline
		Telegram & & & & & & & & & & & & & & & & & & & \cellcolor[HTML]{9B9B9B} & & \cellcolor[HTML]{9B9B9B} & & & & & & & & & & \cellcolor[HTML]{9B9B9B} & & \cellcolor[HTML]{9B9B9B} & & & & & & \\ \hline
		Tumblr & & & \cellcolor[HTML]{9B9B9B} & & & \cellcolor[HTML]{9B9B9B} & & & \cellcolor[HTML]{9B9B9B} & & & \cellcolor[HTML]{9B9B9B} & & & \cellcolor[HTML]{9B9B9B} & & & \cellcolor[HTML]{9B9B9B} & \cellcolor[HTML]{9B9B9B} & & \cellcolor[HTML]{9B9B9B} & \cellcolor[HTML]{9B9B9B} & \cellcolor[HTML]{9B9B9B} & \cellcolor[HTML]{9B9B9B} & \cellcolor[HTML]{9B9B9B} & \cellcolor[HTML]{9B9B9B} & \cellcolor[HTML]{9B9B9B} & \cellcolor[HTML]{9B9B9B} & \cellcolor[HTML]{9B9B9B} & \cellcolor[HTML]{9B9B9B} & \cellcolor[HTML]{9B9B9B} & & \cellcolor[HTML]{9B9B9B} & & & & & & \cellcolor[HTML]{9B9B9B} \\ \hline
		Twitter & & & & & & & & & & & & & & & & & & & & & & & & & & & & & & & \cellcolor[HTML]{9B9B9B} & & \cellcolor[HTML]{9B9B9B} & & & & & & \\ \hline
		Viber & & & & & & & & & & & & & & & & & & & & & \cellcolor[HTML]{9B9B9B} & \cellcolor[HTML]{9B9B9B} & \cellcolor[HTML]{9B9B9B} & \cellcolor[HTML]{9B9B9B} & & & & \cellcolor[HTML]{9B9B9B} & \cellcolor[HTML]{9B9B9B} & \cellcolor[HTML]{9B9B9B} & & & \cellcolor[HTML]{9B9B9B} & & & & & & \\ \hline
		VK & & & & & & & & & & & & & & & & & & & \cellcolor[HTML]{9B9B9B}{\color[HTML]{333333} } & & \cellcolor[HTML]{9B9B9B} & \cellcolor[HTML]{9B9B9B} & & \cellcolor[HTML]{9B9B9B} & & & & & & & \cellcolor[HTML]{9B9B9B} & & \cellcolor[HTML]{9B9B9B} & & & & \cellcolor[HTML]{9B9B9B} & & \\ \hline
		WeChat & & & & & & & & & & & & & & & & & & & & & \cellcolor[HTML]{9B9B9B} & \cellcolor[HTML]{9B9B9B} & \cellcolor[HTML]{9B9B9B} & \cellcolor[HTML]{9B9B9B} & & & & \cellcolor[HTML]{9B9B9B} & \cellcolor[HTML]{9B9B9B} & \cellcolor[HTML]{9B9B9B} & & & \cellcolor[HTML]{9B9B9B} & & & & & & \\ \hline
		WhatsApp & & & & & & & & & & & & & & & & & & & \cellcolor[HTML]{9B9B9B} & & \cellcolor[HTML]{9B9B9B} & & & & & & & \cellcolor[HTML]{9B9B9B} & \cellcolor[HTML]{9B9B9B} & \cellcolor[HTML]{9B9B9B} & \cellcolor[HTML]{9B9B9B} & & \cellcolor[HTML]{9B9B9B} & & & & \cellcolor[HTML]{9B9B9B} & & \cellcolor[HTML]{9B9B9B} \\ \hline
	\end{tabu}
}
\end{table}

\begin{figure}[!t]
 \begin{center}
   \includegraphics[width=3.1in,clip,keepaspectratio]{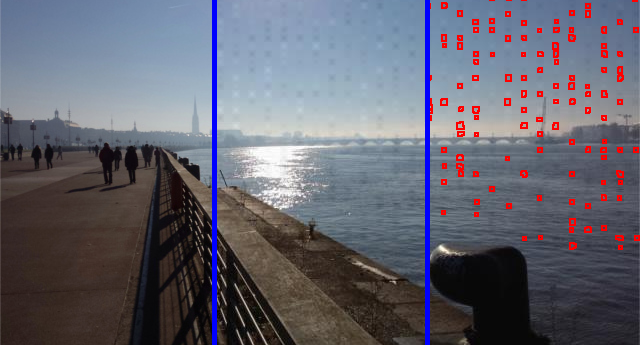}
  \caption{The original image portion \emph{on the left}, the portion of the SilentEye outcome \emph{in the middle}, and the same SilentEye outcome with red circles highlighting artefacts \emph{on the right}.}\label{fig:mod}
 \end{center}
\end{figure}

Table \ref{tab:col} shows the results of the watermark extraction process for each resolution class respectively. In particular, we denote with a grey cell if the extraction process successfully retrieved the original watermark from the downloaded images. SilentEye and Steghide are not able to embed the watermark with large images and SteganPEG returns ``image capacity exceeded'' error with small images. For these reasons, we consider these 3 algorithms as a failure. Since Flickr and Google+ do not apply any compression, the watermark is preserved for each watermarking algorithm and resolution class. None of the selected methods is able to create a robust watermark that can successfully pass through Instagram and LinkedIn. Experiments also revealed that spatial domain based watermarking methods, like BlindHide, HideSeek, FilterFirst, BattleSteg, Dynamic FilterFirst, Dynamic BattleSteg and SteganPEG are too weak to be applied to shared images on SNs. The other methods, like F5, OpenPuff, SecretBook and SilentEye, obtain better results with standard and small images. However, SilentEye heavily transforms the image that appears visually modified and degraded to the user. Figure \ref{fig:mod} shows the effects of SilentEye algorithm. The watermarked image is affected by a diffuse noise if compared with the original one. We highlighted the vertical artefacts on the right-hand side with red circles. The same artefacts can be found in the middle portion of the image.

Based on findings in \emph{user explicit watermarking} test, \emph{transform domain watermarking} methods are a good candidate to apply a personal watermark to the images to be shared on SNs. However, as shown in Table \ref{tab:col}, there is not a single watermarking technique that can be successfully used across all SNs. Moreover, the user has to act consciously to embed a watermark into his/her personal photos. However, this is less likely to happen in scenarios of instant picture sharing through mobiles on social platforms like Instagram and WhatsApp.

\section{User Unaware Watermarking}\label{sec:implicit}
Smartphones have a big contribution to a large number of images being shared on SNs \cite{salehan2013social}. For this reason, in this section, we explore whether the noise left by the sensor of the smartphone's camera can be successfully used to watermark the images being uploaded on SNs.

To perform our investigation, we selected six smartphones from three different brands with three pairs of identical models (Table \ref{tab:smart}). The reason of picking the identical models was to verify if the test can work across two different phones of the same models, as identical models mount identical components. For each device, we considered both the front and rear camera that usually present different characteristics in term of sensor, quality, and resolution. The \emph{iPhone 6} and \emph{iPhone 6 Plus} both have an identical front and rear camera. For each camera of each device, we took 700 photos. In total the dataset consists of 8,400 images \footnote{The dataset is available from \url{http://smartdata.cs.unibo.it/datasets\#images}}. Then for each camera, we kept a subset of 50 original images and we uploaded and downloaded 50 images on each of the 13 SNs in Table \ref{tab:sn}.

\begin{table}[!h]
	\renewcommand{\arraystretch}{0.85}
	\centering
	\caption{The six smartphones used for the user unaware watermarking tests and the pixel resolution of the front and rear cameras.}\label{tab:smart}
	\begin{tabular}{|c|l|l|c|c|}
		\hline
		\textbf{ID} & \textbf{Brand} & \textbf{Model} & \textbf{Front Camera} & \textbf{Rear Camera} \\ \hline \hline
		1 & Apple & iPhone 6 & 1280$\times$960 & 3264$\times$2448 \\ \hline 
		2 & Apple & iPhone 6 Plus & 1280$\times$960 & 3264$\times$2448 \\ \hline 
		3 & LG & Nexus 5 & 1280$\times$960 & 3264$\times$2448 \\ \hline 
		4 & LG & Nexus 5 & 1280$\times$960 & 3264$\times$2448 \\ \hline 
		5 & Samsung & Galaxy S2 & 1600$\times$1200 & 3264$\times$2448 \\ \hline 
		6 & Samsung & Galaxy S2 & 1600$\times$1200 & 3264$\times$2448 \\ \hline 
	\end{tabular}
\end{table}

Firstly, we provide a small background in image processing, and then we describe our methodology on original images. Next, we discuss the results of three different tests that is \emph{profile attribution}, \emph{intra-layer user profiles linking}, and \emph{inter-layer user profiles linking} (see Section \ref{sec:intro_contr}) on the downloaded images.

\subsection{User Unaware Watermarking through PRNU}
The unavoidable noise that affects any image can be categorised into \emph{shot noise} and \emph{pattern noise}. The first one is introduced due to external factors such as brightness, temperature and humidity; the second component is regular and systematic. We exploit a PRNU-based method \cite{lukas2006digital} to extract the dominant part of the \emph{pattern noise} and define this noise as the \emph{user unaware watermark}.

In order to extract the noise, we tested several different algorithms, like Diffusion Anisotropic \cite{weickert1998anisotropic}, Diffusion Isotropic \cite{perona1990scale}, Block Matching 3D (BM3D) \cite{dabov2006image}, Wavelet Soft Threshold \cite{donoho1995noising}, Wavelet Hard Threshold \cite{donoho1995adapting}, and Wavelet Multi Frame \cite{mayer2012wavelet}. The BM3D approach provided the best results, especially combined with the Y channel of the images, that is the luminance component in YCbCr colour space format. The fingerprint $FP_i$ of the smartphone camera $i$ was approximated as the average of residual PRNUs of $n$ pictures captured by that device:
\begin{equation}\label{equ:fingerprint}
	FP_i = \frac{1}{n}\sum_{j=1}^n BM3D(Y(I_j))
\end{equation}
where the function $Y()$ extracts the Y channel and $BM3D()$ extracts the PRNU, and the resultant $FP_i$ is a matrix as big as the original image. In order to evaluate if a generic image $I_k$ had been taken by the same camera, the standard normalized correlation was applied between the PRNU ($N_k$) extracted from $I_k$ and the fingerprint $FP_i$:
\begin{equation}\label{equ:correlation}
	corr(N_k, FP_i) = \frac{(N_k - \overline{N_k})(FP_i - \overline{FP_i})}{\|(N_k - \overline{N_k})\|\|(FP_i - \overline{FP_i})\|}
\end{equation}
where $N_k$ is equal to $BM3D(Y(I_k))$, and $\overline{N_k}$ and $\overline{FP_i}$ are scalars that represent the mean value of $N_k$ and $FP_i$ matrices. The correlation value can vary from 0 (different source) to 1 (same source). In order to correlate images with different sizes, large images were scaled to the small one.

\begin{figure}[!t]
    \centering
    \subfloat[][]{\includegraphics[height=1in, width=1.5in]{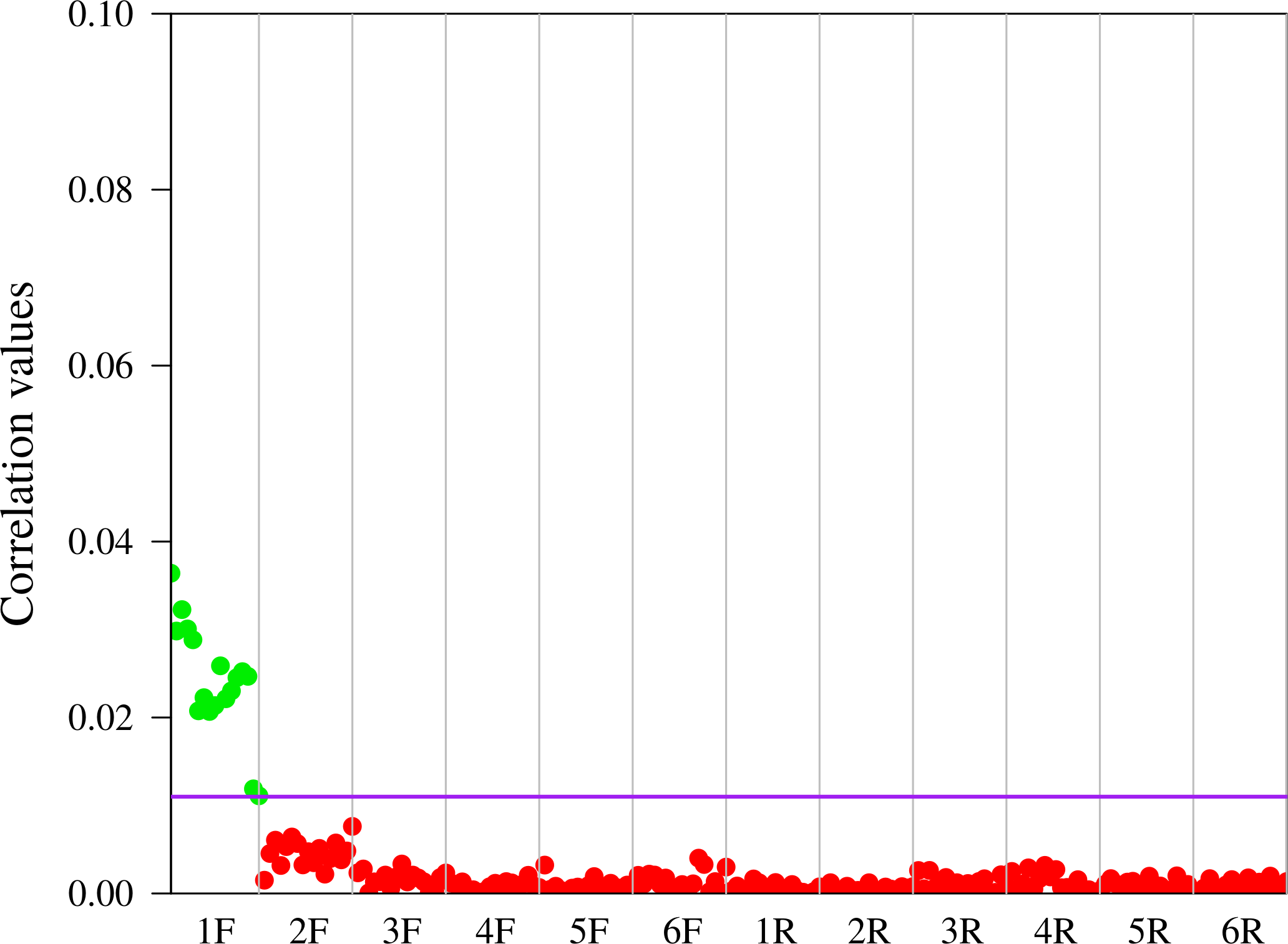}\label{fig:org_vs_org_1f}}\hspace{0.2mm}
    \subfloat[][]{\includegraphics[height=1in, width=1.5in]{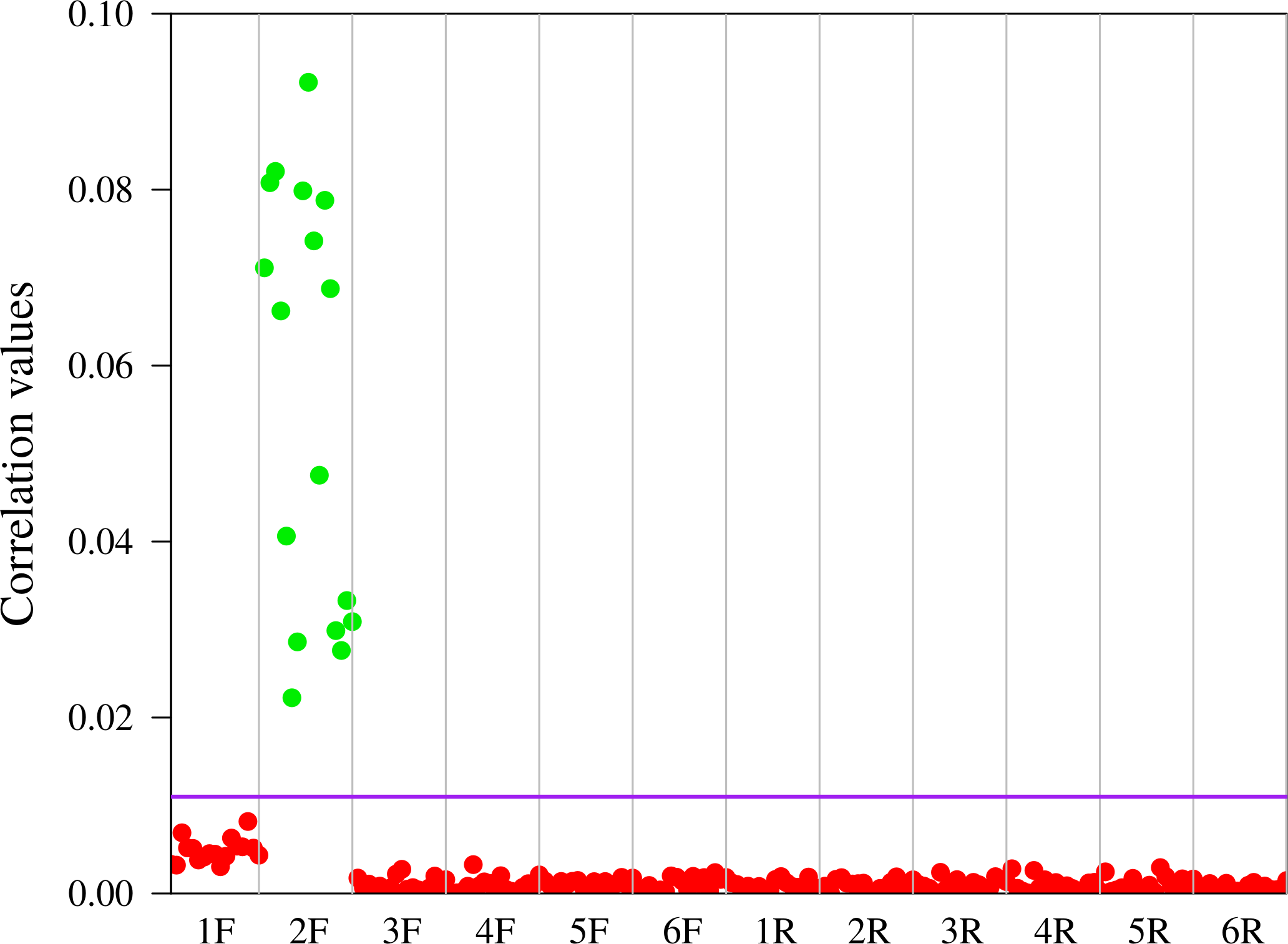}\label{fig:org_vs_org_2f}}\hspace{0.2mm}
    \subfloat[][]{\includegraphics[height=1in, width=1.5in]{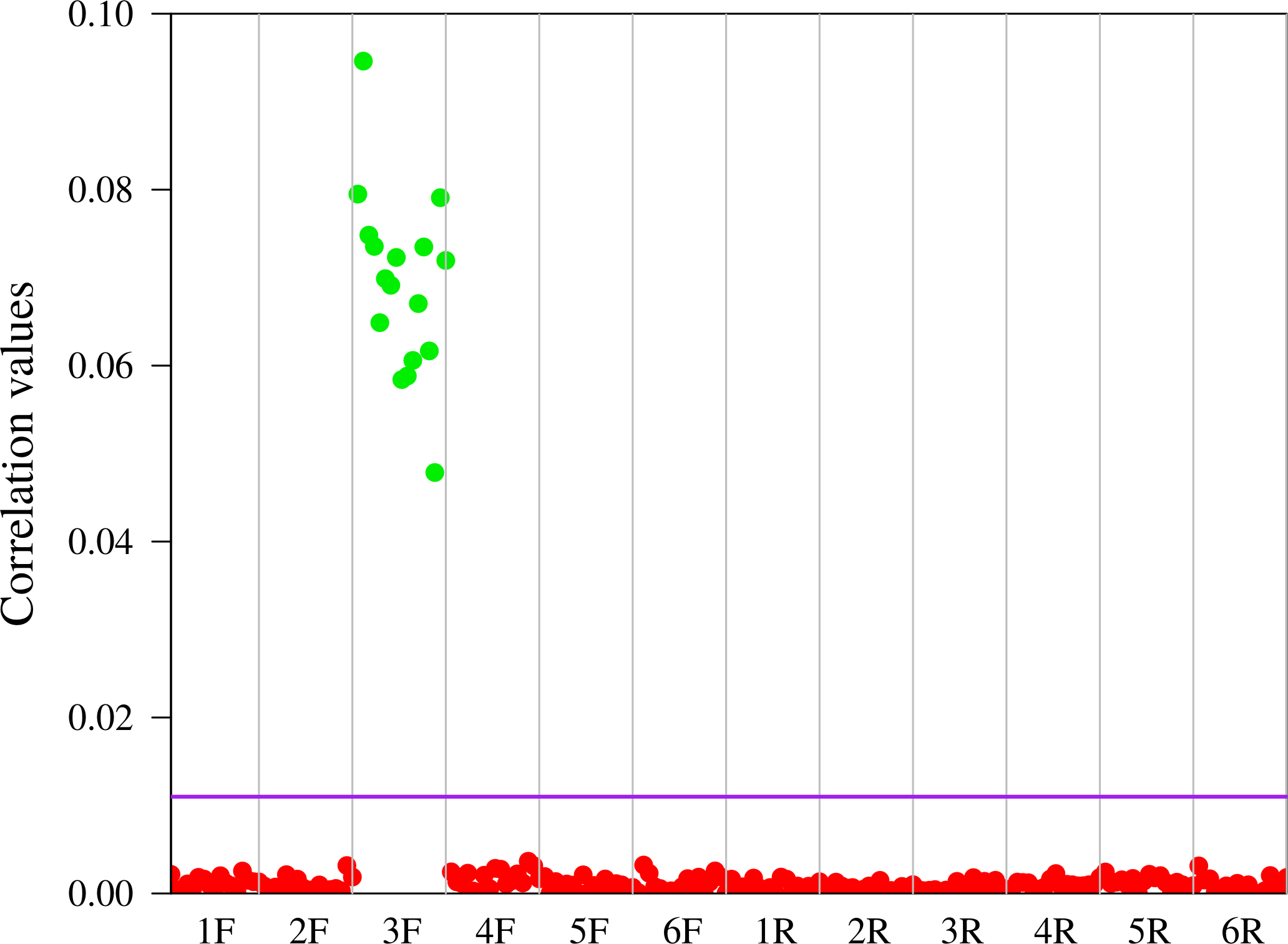}\label{fig:org_vs_org_3f}}\hspace{0.2mm}\\
    \subfloat[][]{\includegraphics[height=1in, width=1.5in]{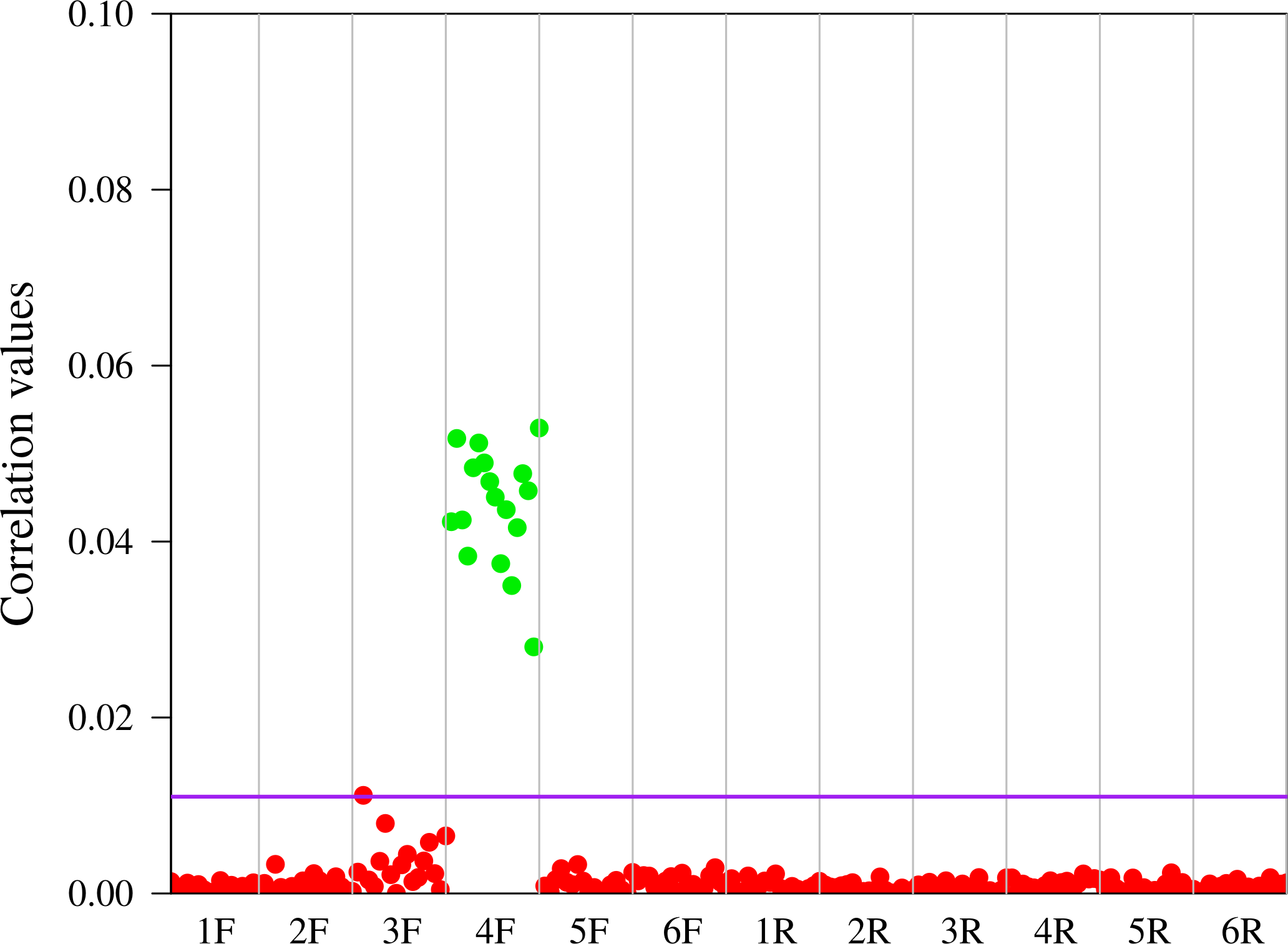}\label{fig:org_vs_org_4f}}\hspace{0.2mm}
    \subfloat[][]{\includegraphics[height=1in, width=1.5in]{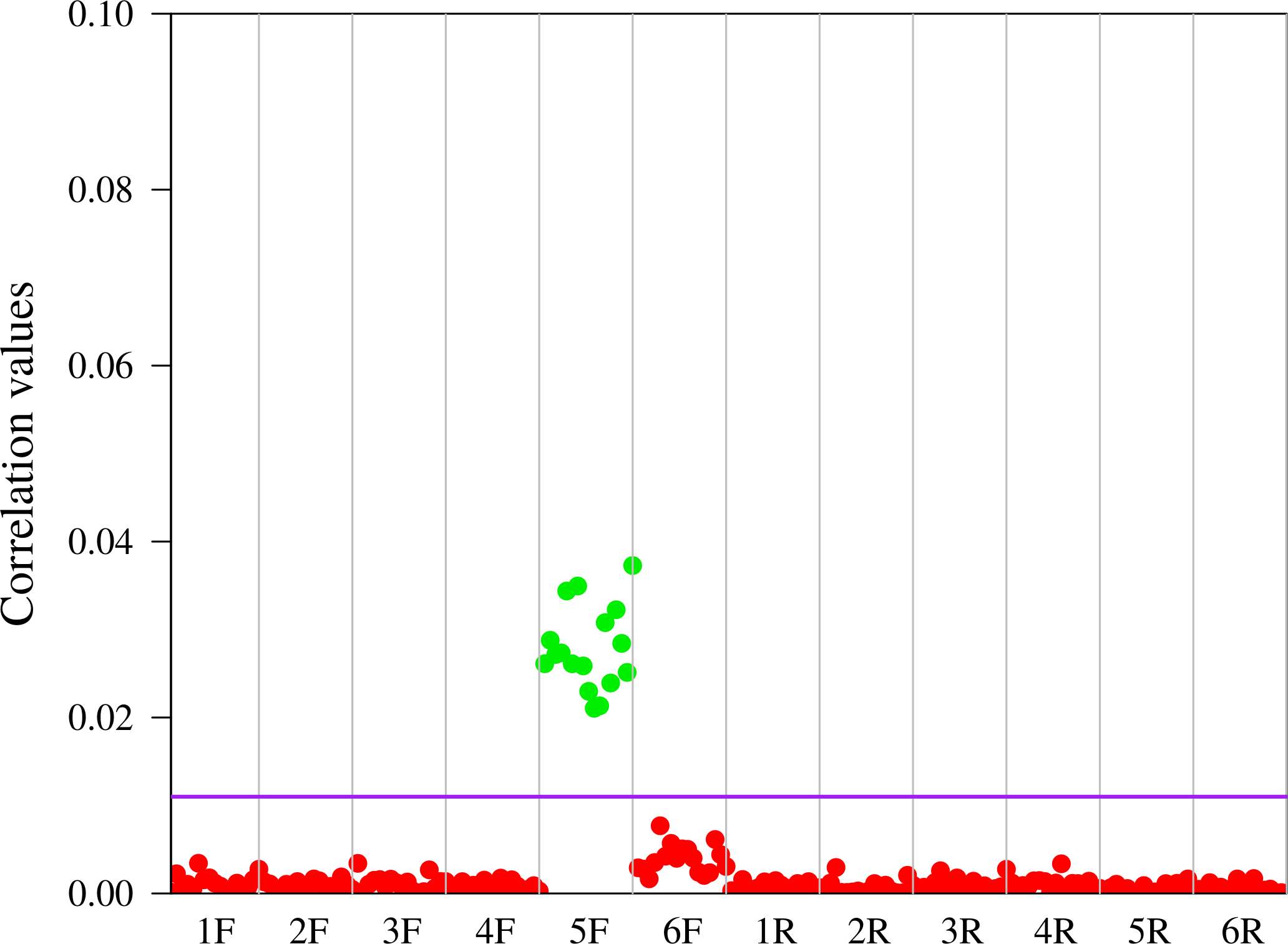}\label{fig:org_vs_org_5f}}\hspace{0.2mm}
    \subfloat[][]{\includegraphics[height=1in, width=1.5in]{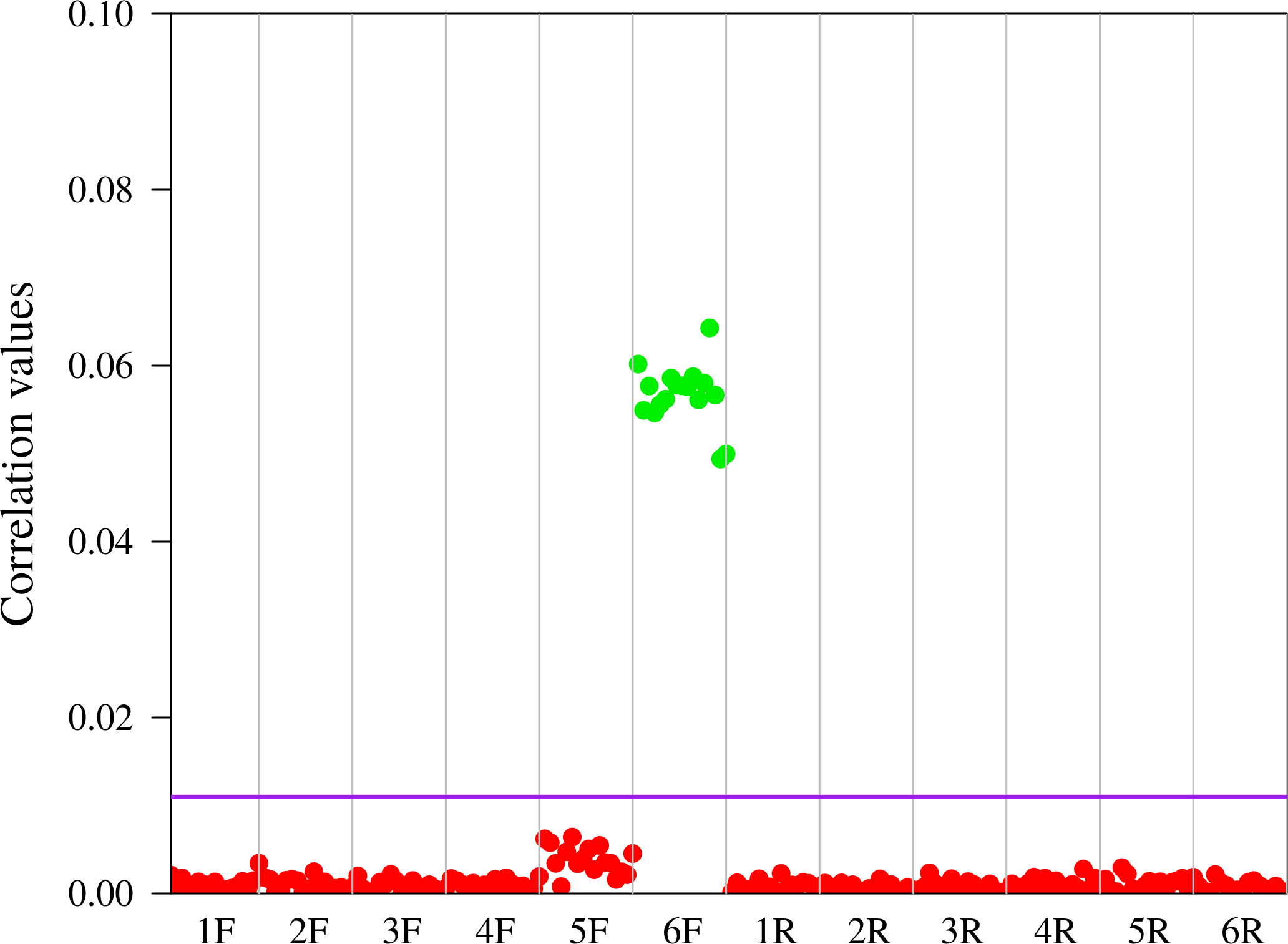}\label{fig:org_vs_org_6f}}\\
    \subfloat[][]{\includegraphics[height=1in, width=1.5in]{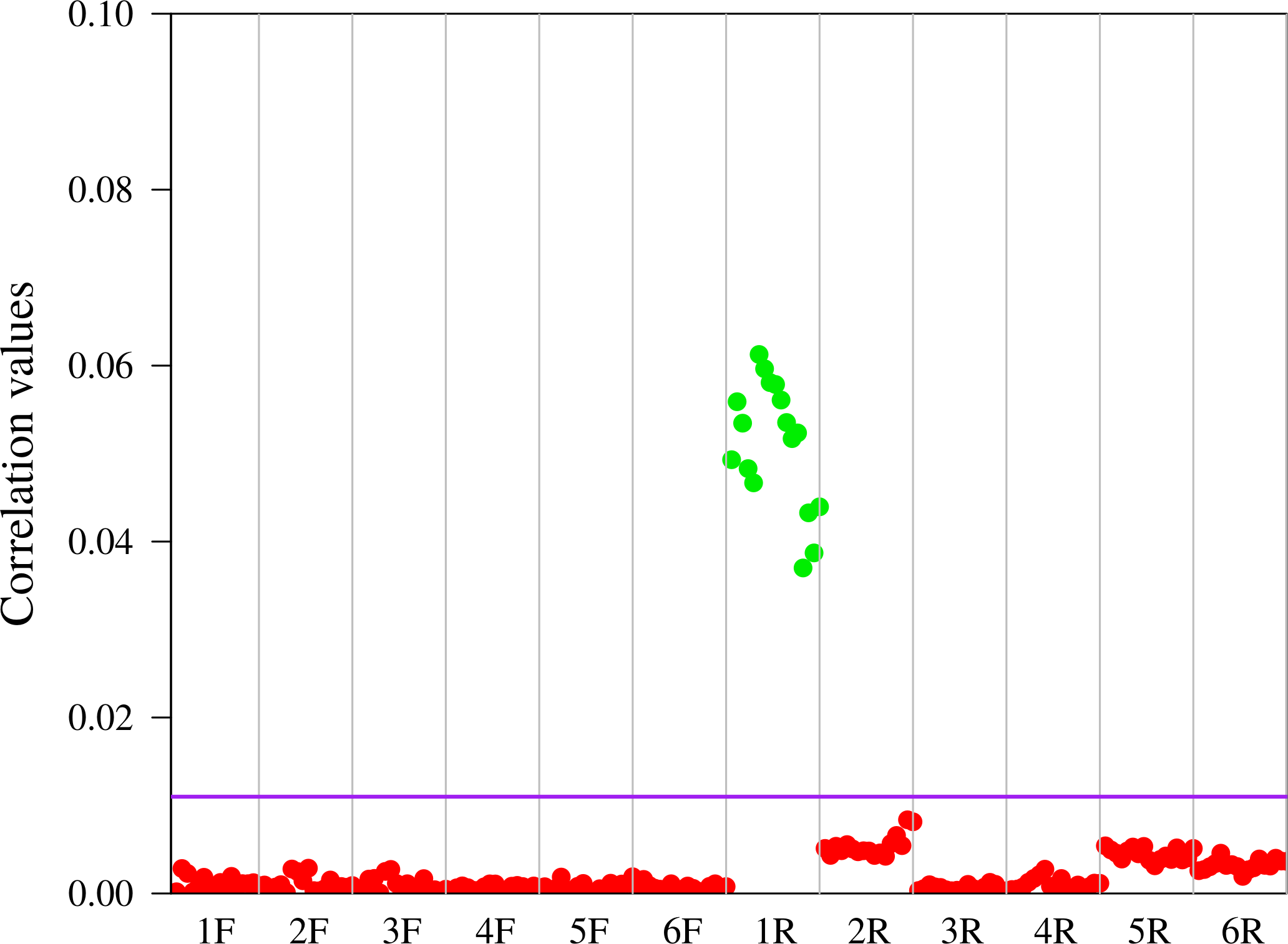}\label{fig:org_vs_org_1r}}\hspace{0.2mm}
    \subfloat[][]{\includegraphics[height=1in, width=1.5in]{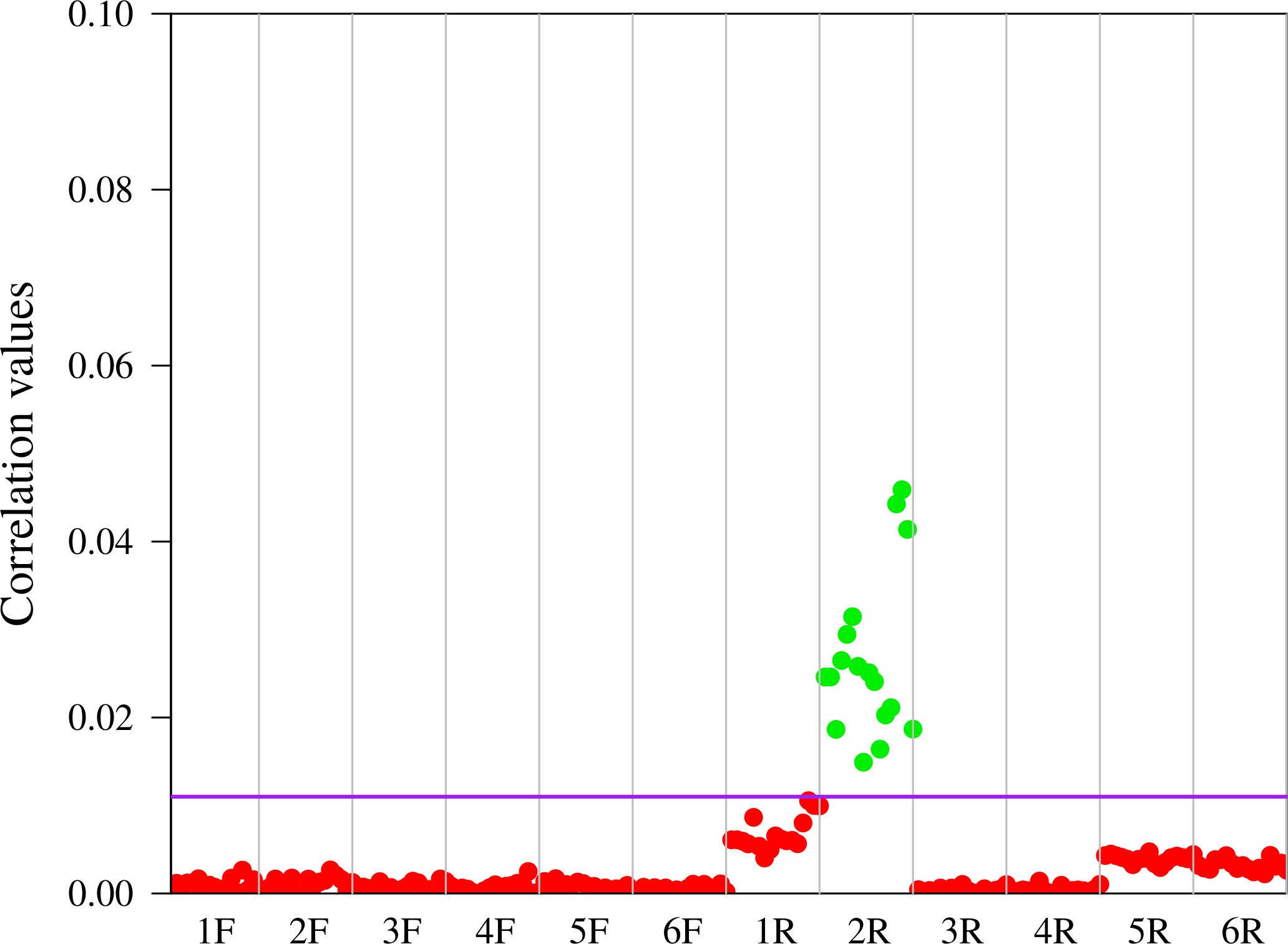}\label{fig:org_vs_org_2r}}\hspace{0.2mm}
    \subfloat[][]{\includegraphics[height=1in, width=1.5in]{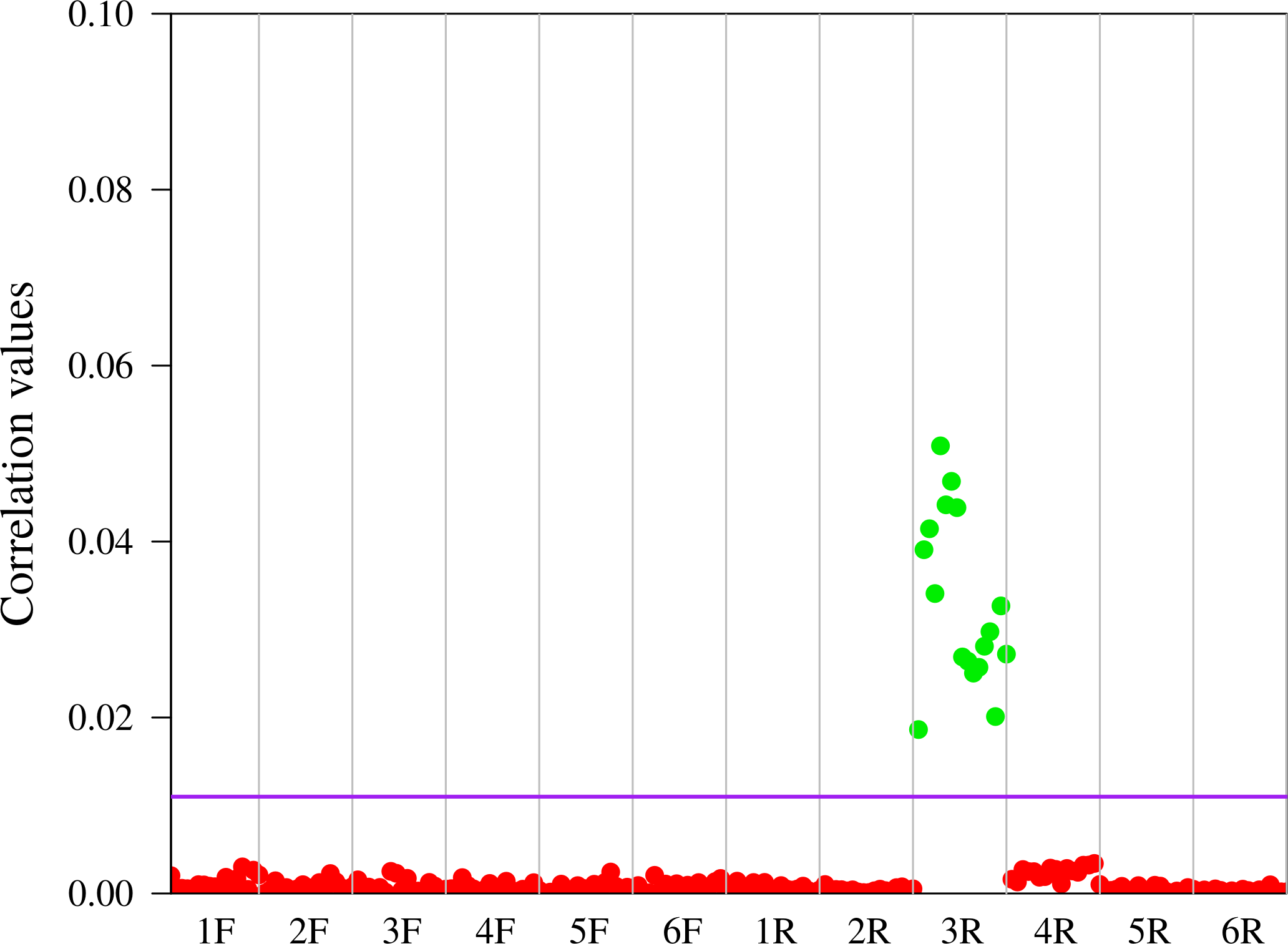}\label{fig:org_vs_org_3r}}\hspace{0.2mm}\\
    \subfloat[][]{\includegraphics[height=1in, width=1.5in]{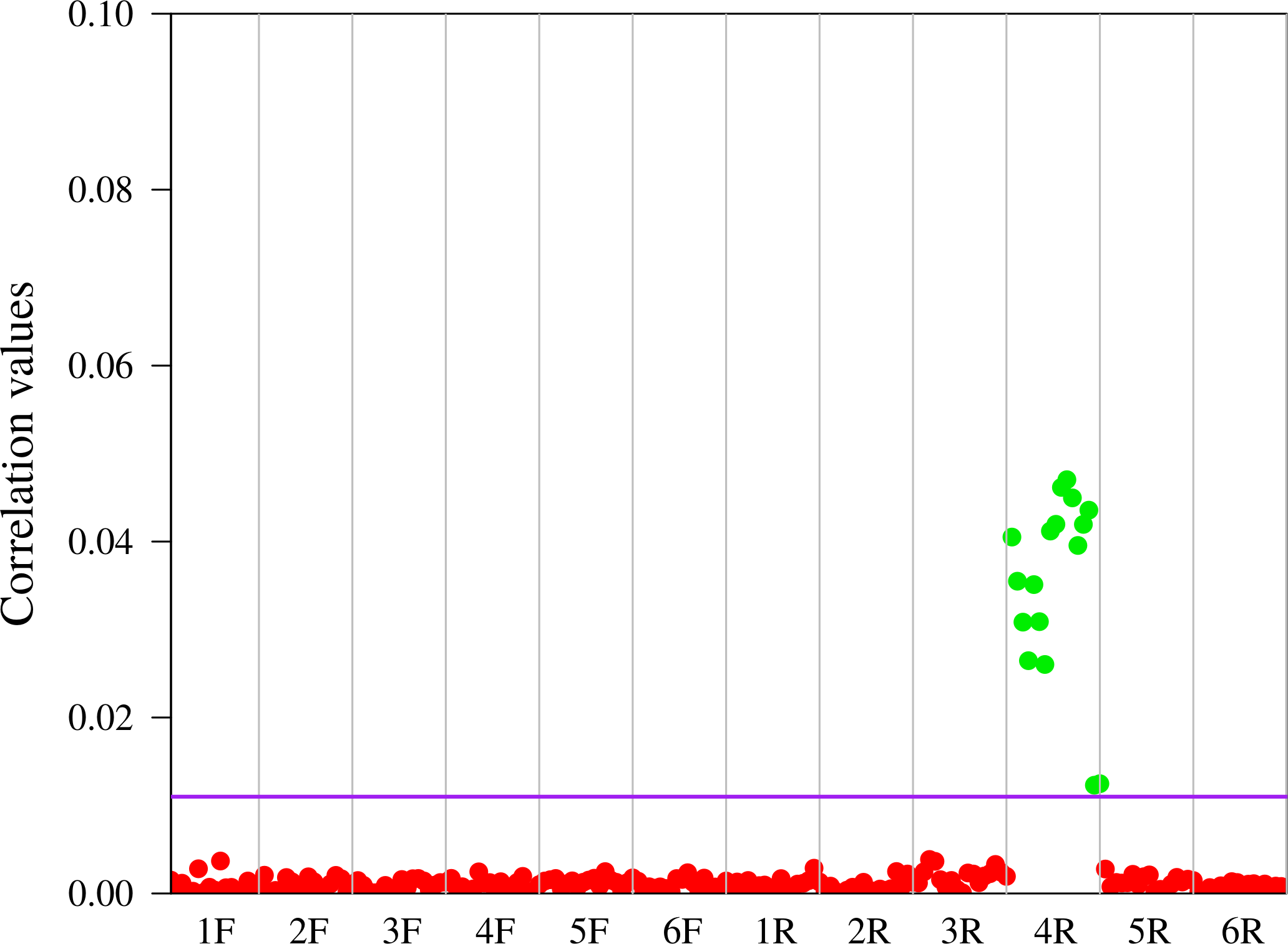}\label{fig:org_vs_org_4r}}\hspace{0.2mm}
    \subfloat[][]{\includegraphics[height=1in, width=1.5in]{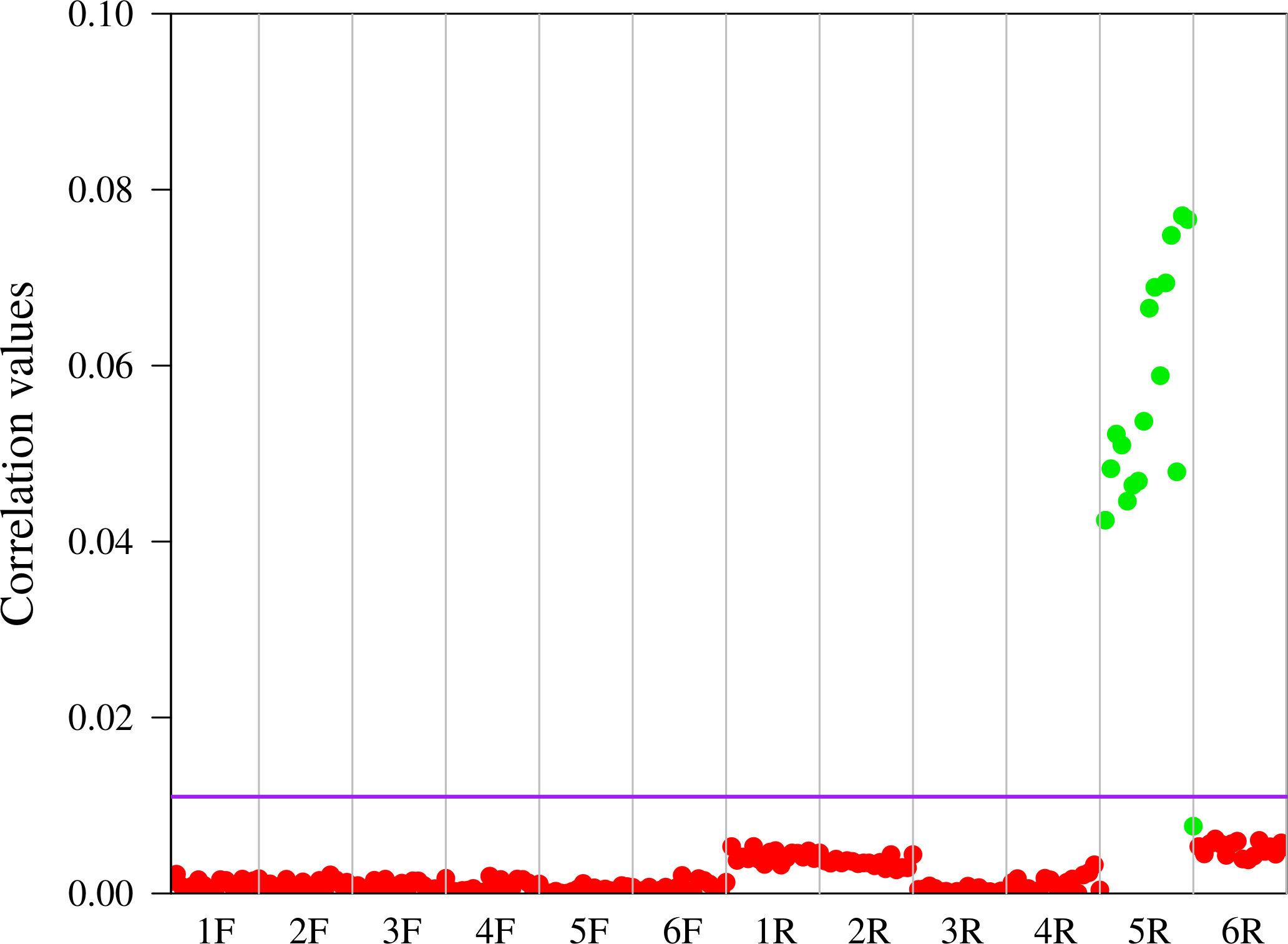}\label{fig:org_vs_org_5r}}\hspace{0.2mm}
    \subfloat[][]{\includegraphics[height=1in, width=1.5in]{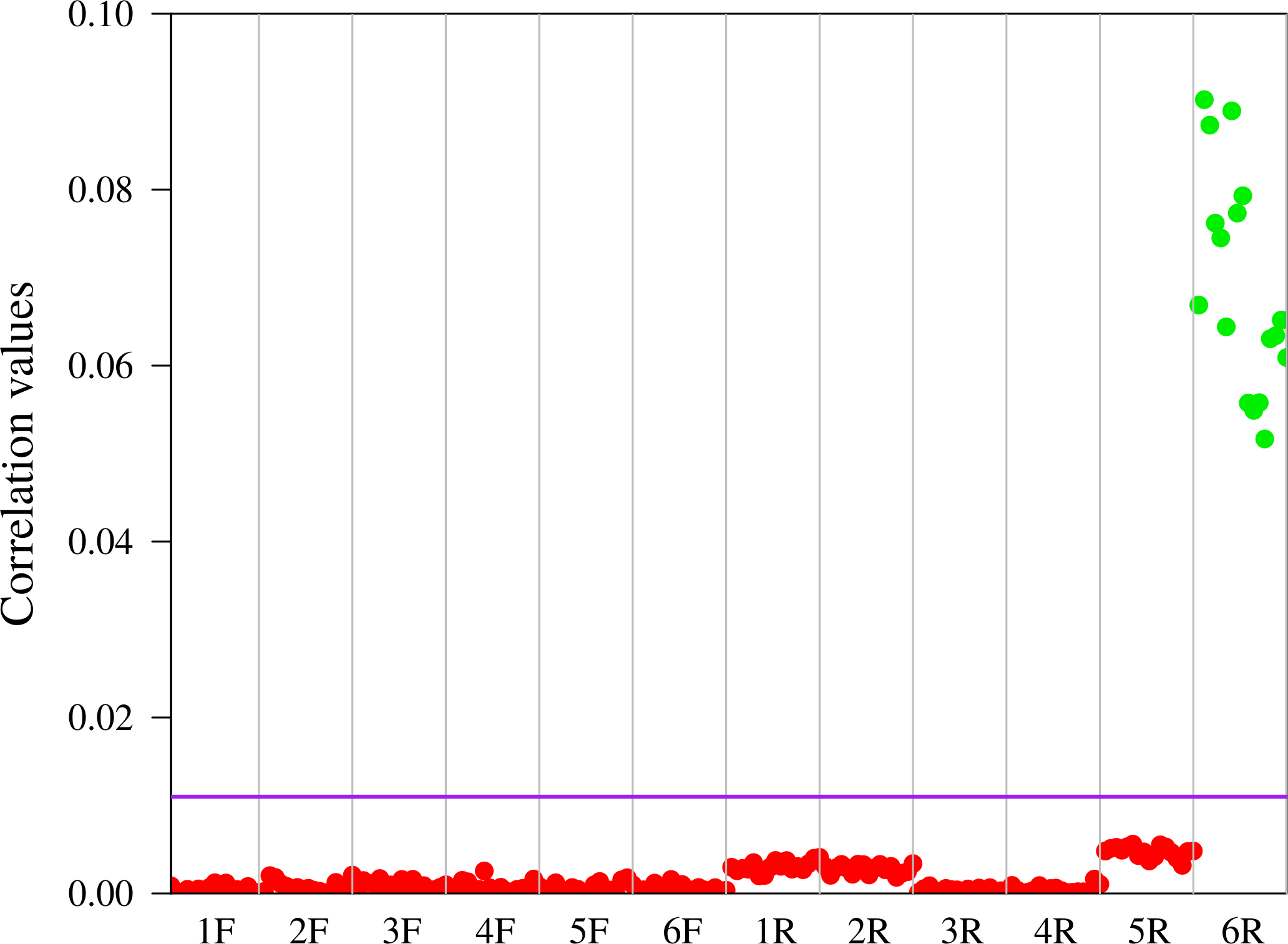}\label{fig:org_vs_org_6r}}\\
    \caption{Correlation results for each of the twelve fingerprints, six for the front cameras (first and second rows) and six for the rear cameras (third and fourth rows). $X$F and $X$R identify the front and the rear camera of the smartphone $X$, respectively.}\label{fig:org_vs_org}
    \label{fig:globfig}
\end{figure}

Figure \ref{fig:org_vs_org} shows the results using the methodology described above on original images. The graphs in the first two rows represent the front cameras of each device, while those in the second two rows represent the rear cameras. For each smartphone's camera in Table \ref{tab:smart}, we used a \textbf{training set} of 33 images (two-thirds) to define $FP$, and a \textbf{test set} of 17 images (one-thirds) to compose the set of images to be correlated. In accordance with \cite{bertini2015profile}, using a training set of 20 is enough to obtain good results. Each graph represents the values obtained by correlating a specific $FP_i$ with all the images in the 12 test sets. High values were achieved when the $FP_i$ and the test set come from the same source (green dots). All obtained correlation values were in the range from 0 to 0.1. However, a heuristic threshold of 0.011 (purple horizontal line) that maximises both the whole positive and negative predictive values allowed to correctly classify all the images, in line with previous evidence \cite{castiglione2013experimentations}. In the subsequent subsections, we discuss the results of our three experiments (e.g., \emph{profile attribution}, \emph{intra-layer} and \emph{inter-layer user profiles linking}) on the images downloaded from all the 13 SNs.

\begin{figure}[!t]
    \centering
    \subfloat[][]{\includegraphics[height=1in, width=1.5in]{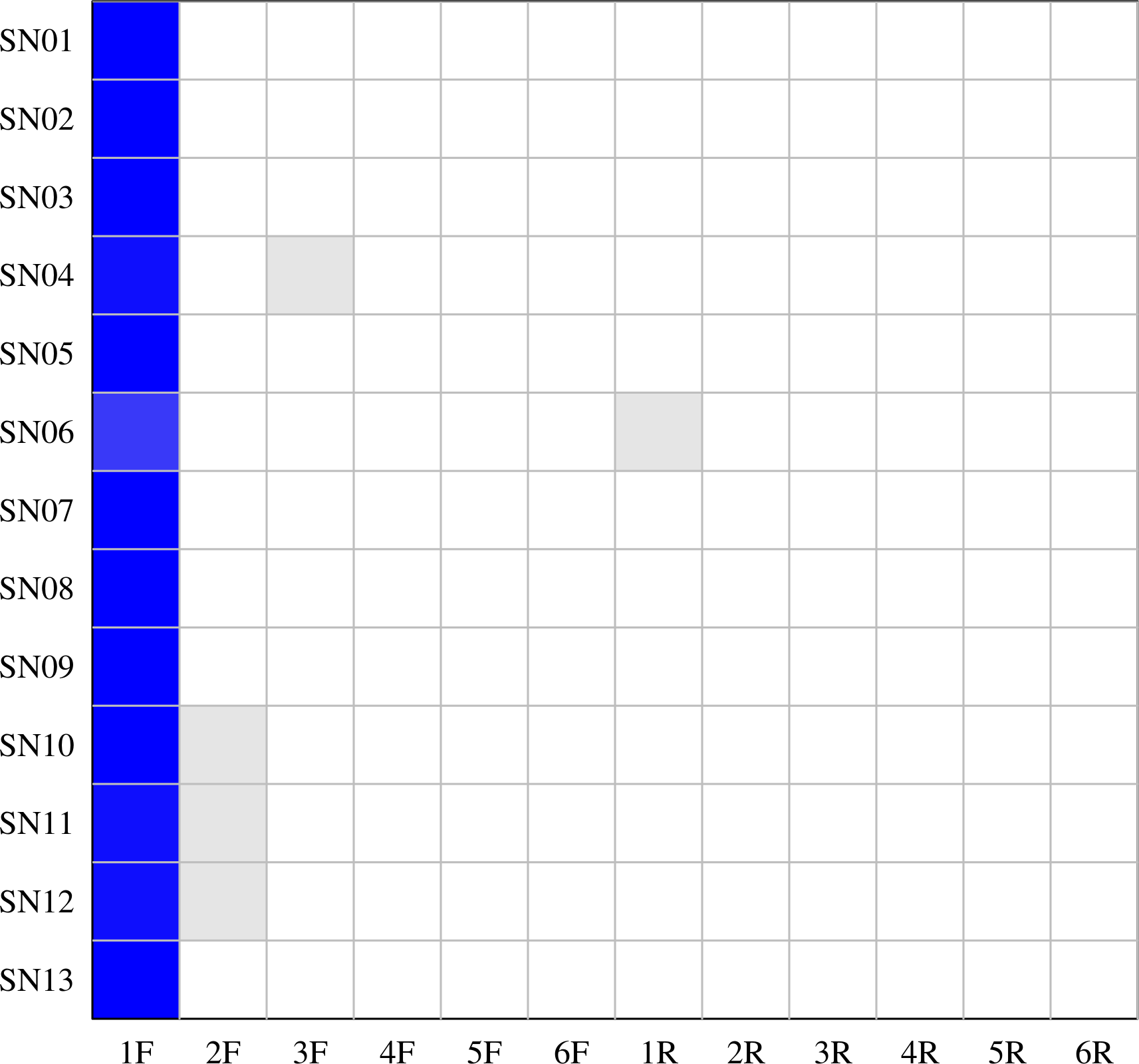}\label{fig:org_vs_soc_1f}}\hspace{0.2mm}
    \subfloat[][]{\includegraphics[height=1in, width=1.5in]{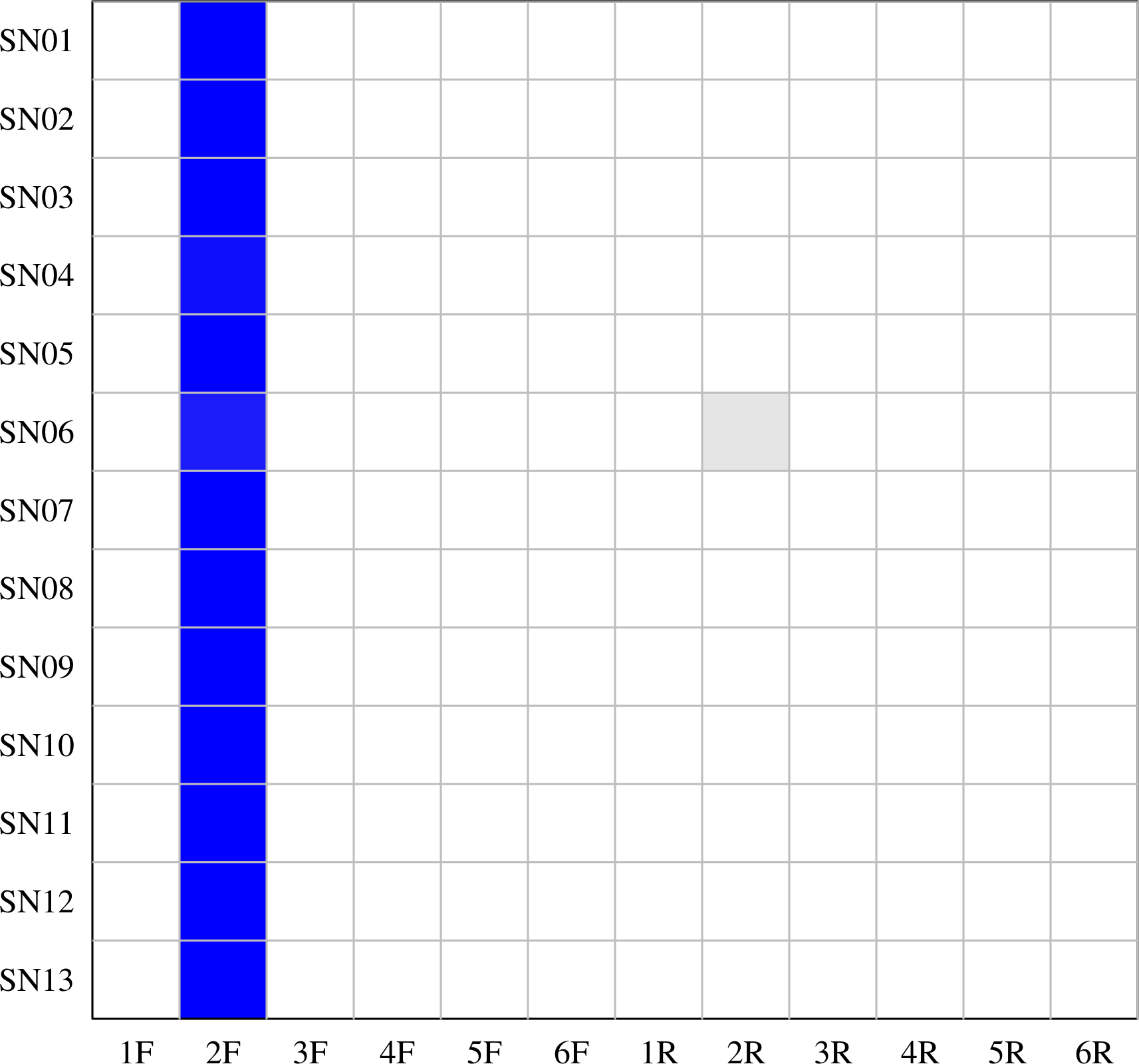}\label{fig:org_vs_soc_2f}}\hspace{0.2mm}
    \subfloat[][]{\includegraphics[height=1in, width=1.5in]{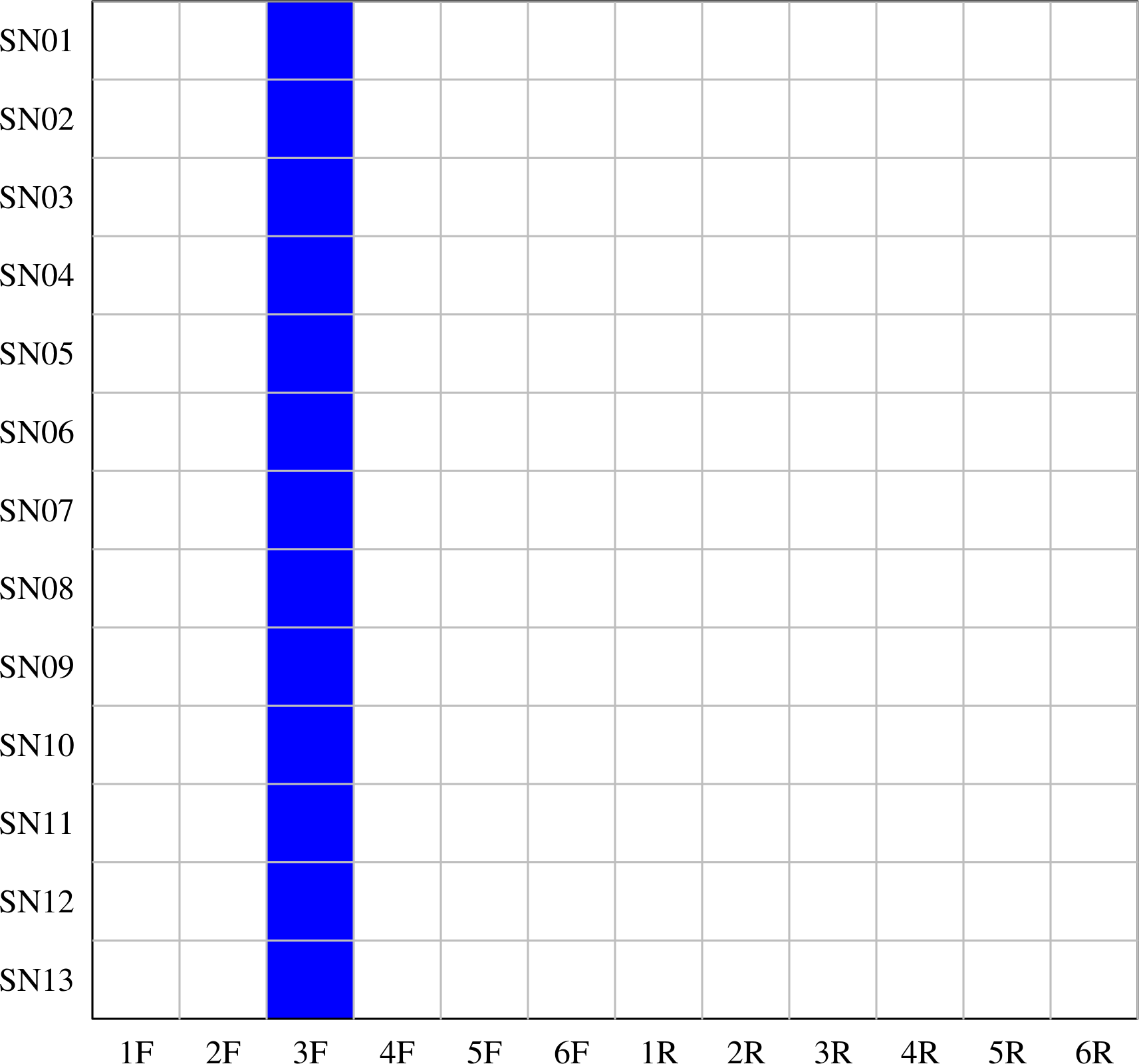}\label{fig:org_vs_soc_3f}}\hspace{0.2mm}
    \subfloat[][]{\includegraphics[height=1in, width=1.5in]{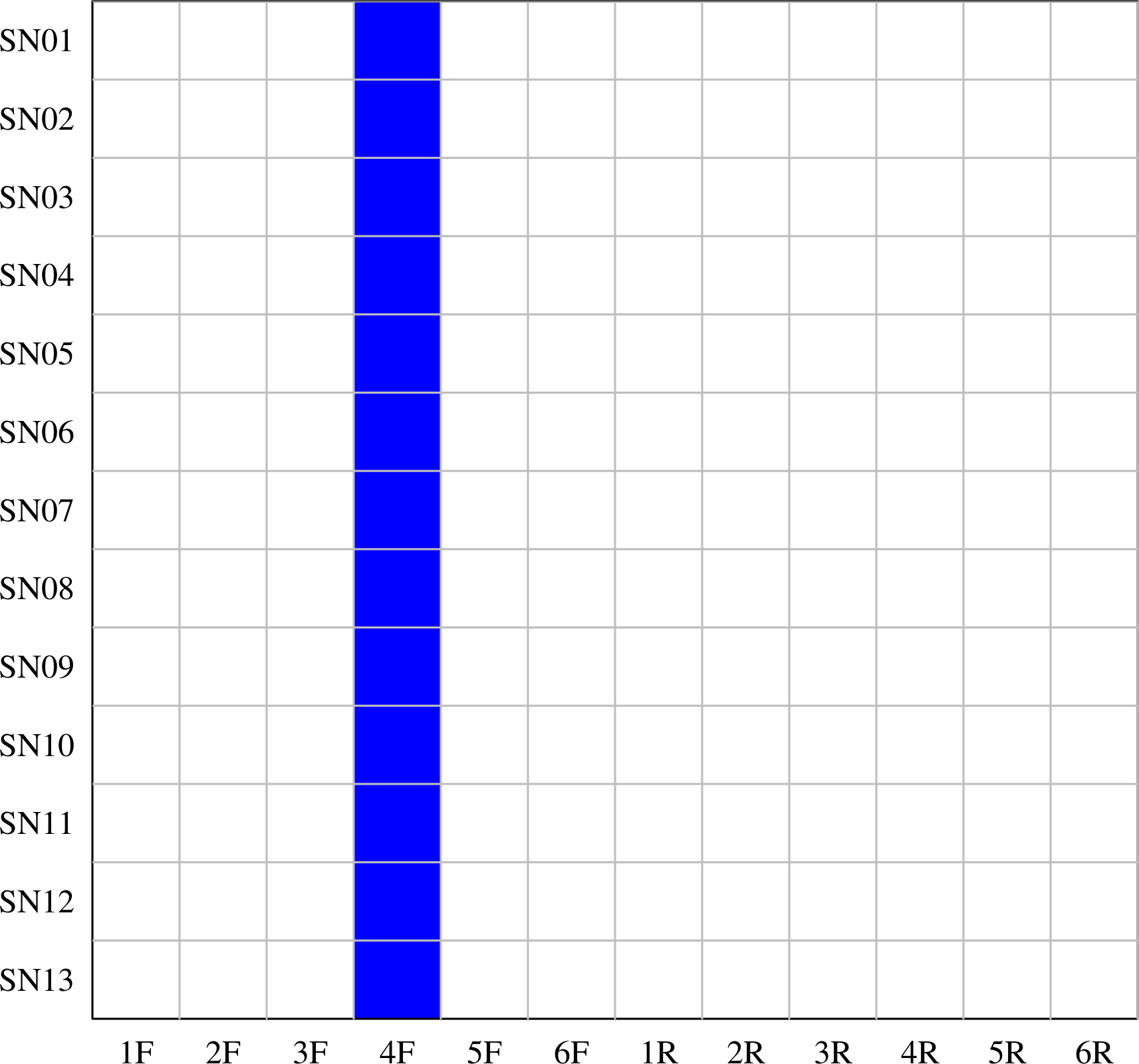}\label{fig:org_vs_soc_4f}}\hspace{0.2mm}
    \subfloat[][]{\includegraphics[height=1in, width=1.5in]{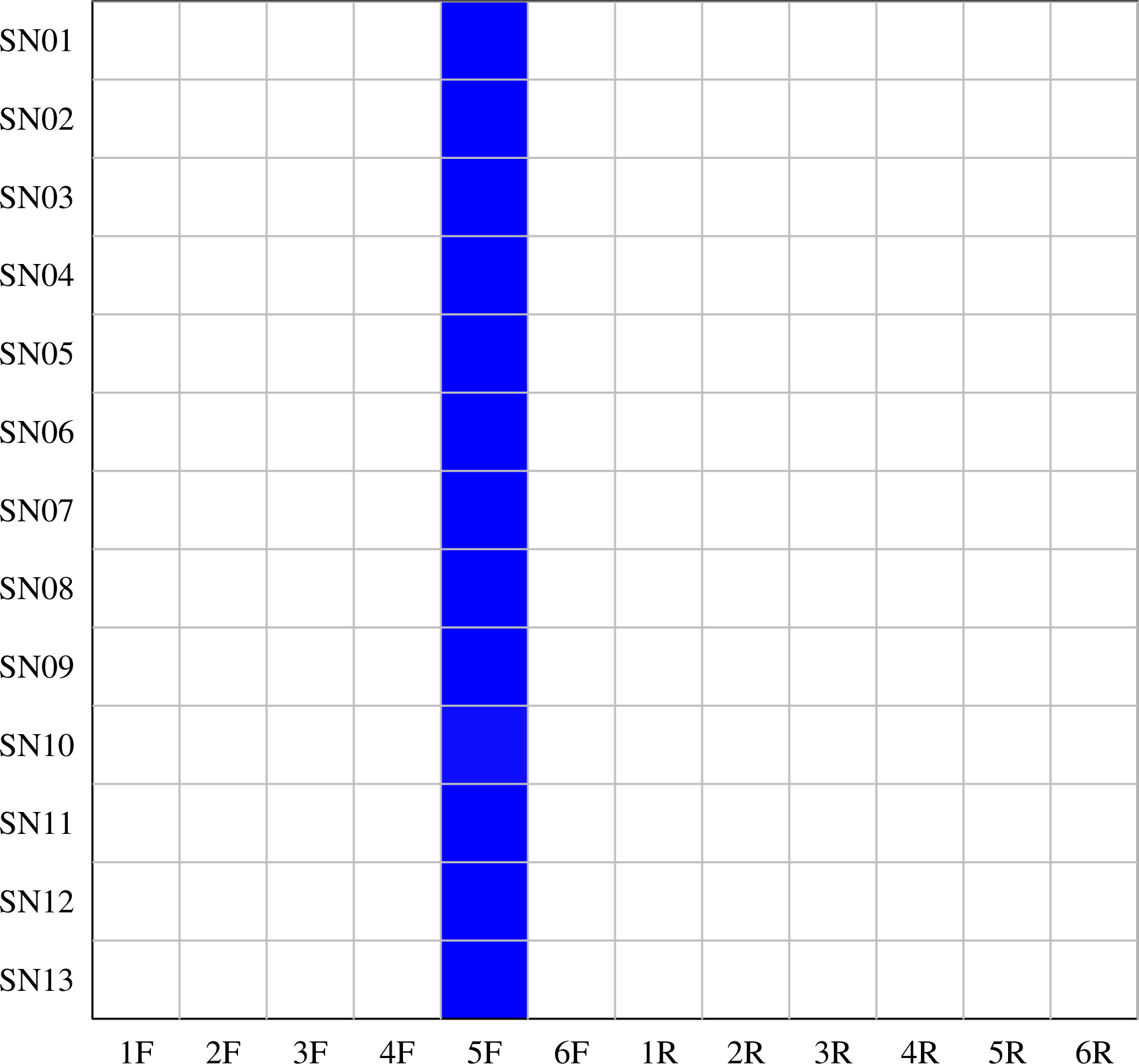}\label{fig:org_vs_soc_5f}}\hspace{0.2mm}
    \subfloat[][]{\includegraphics[height=1in, width=1.5in]{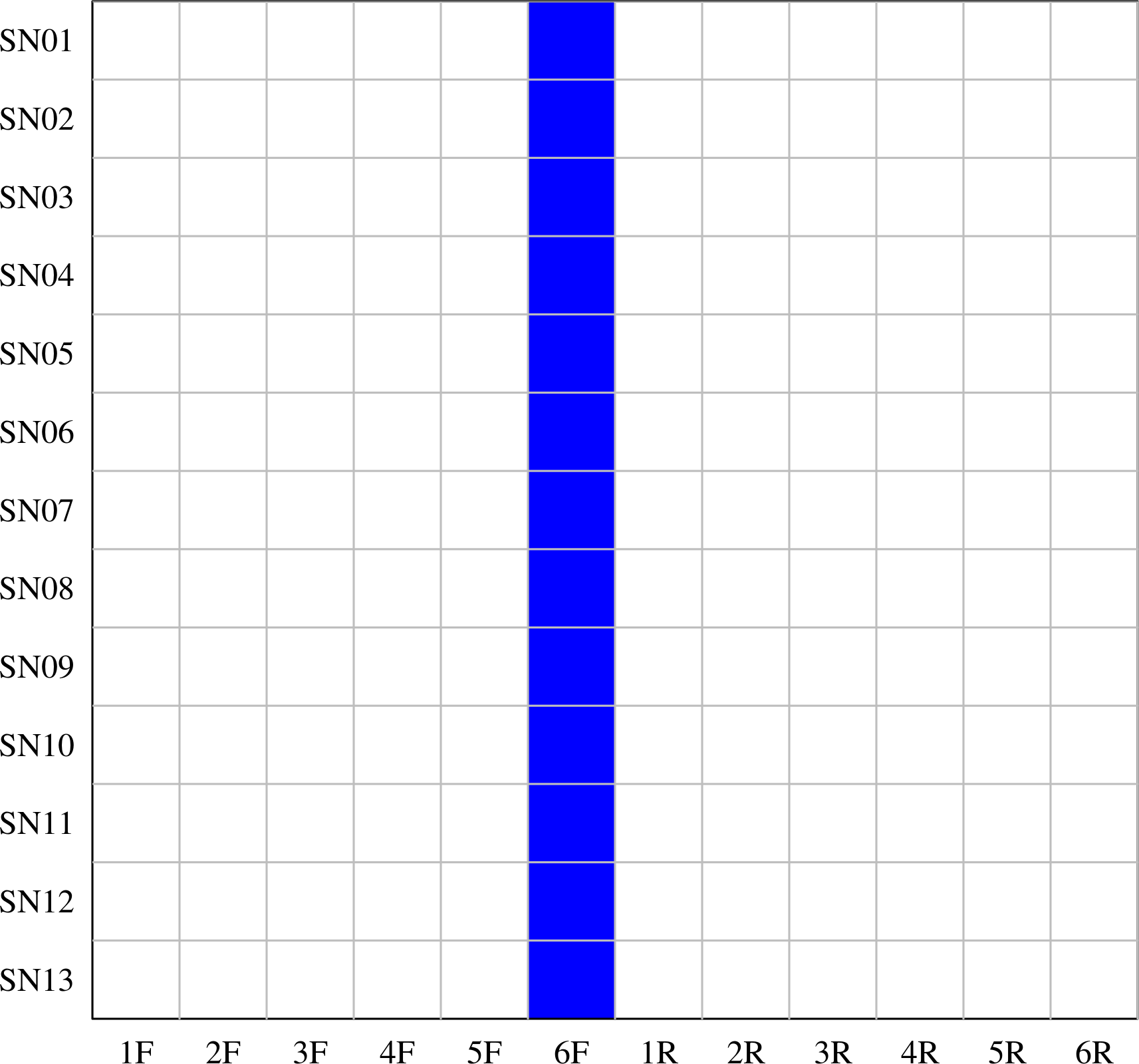}\label{fig:org_vs_soc_6f}}\\
    \subfloat[][]{\includegraphics[height=1in, width=1.5in]{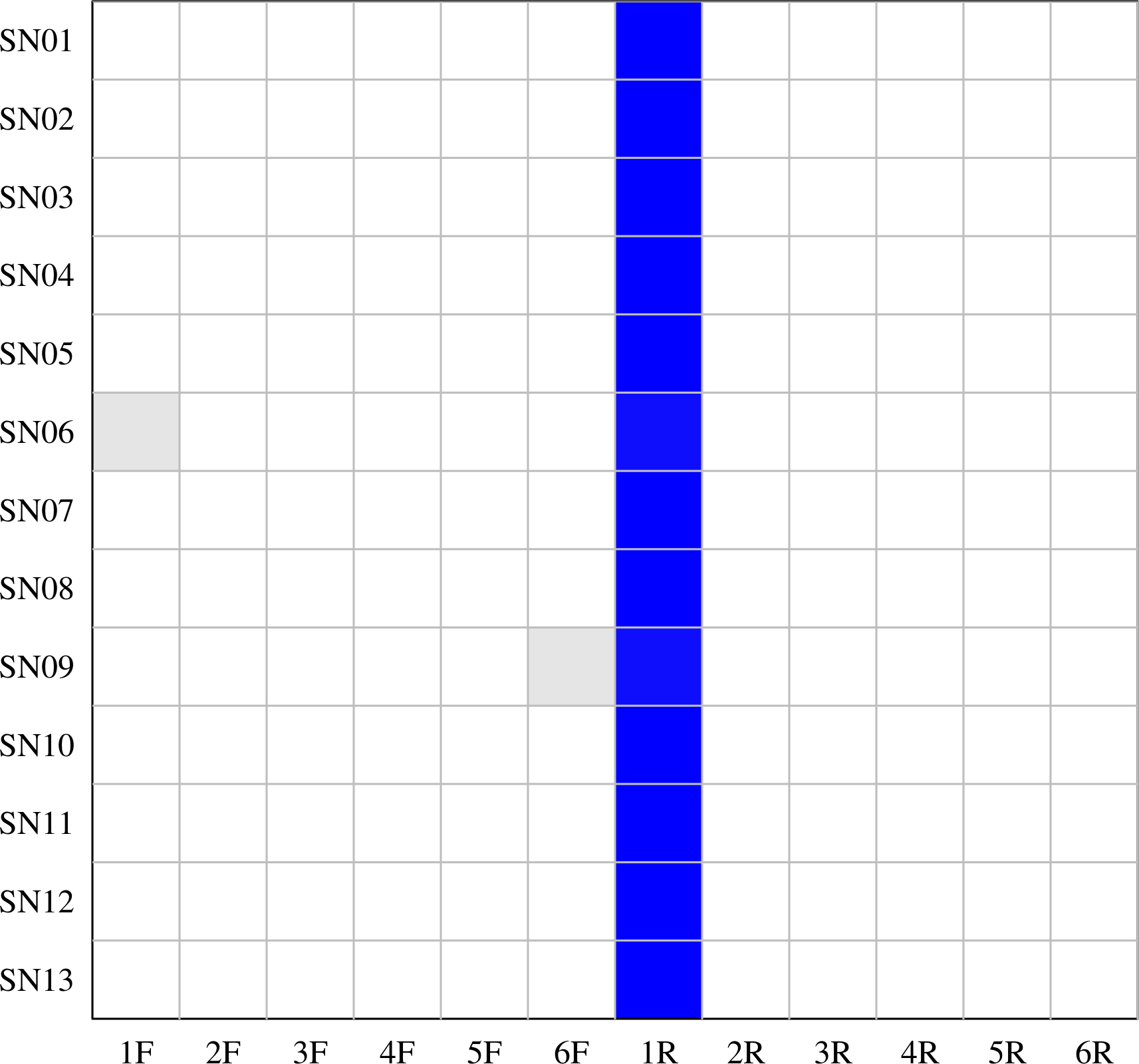}\label{fig:org_vs_soc_1r}}\hspace{0.2mm}
    \subfloat[][]{\includegraphics[height=1in, width=1.5in]{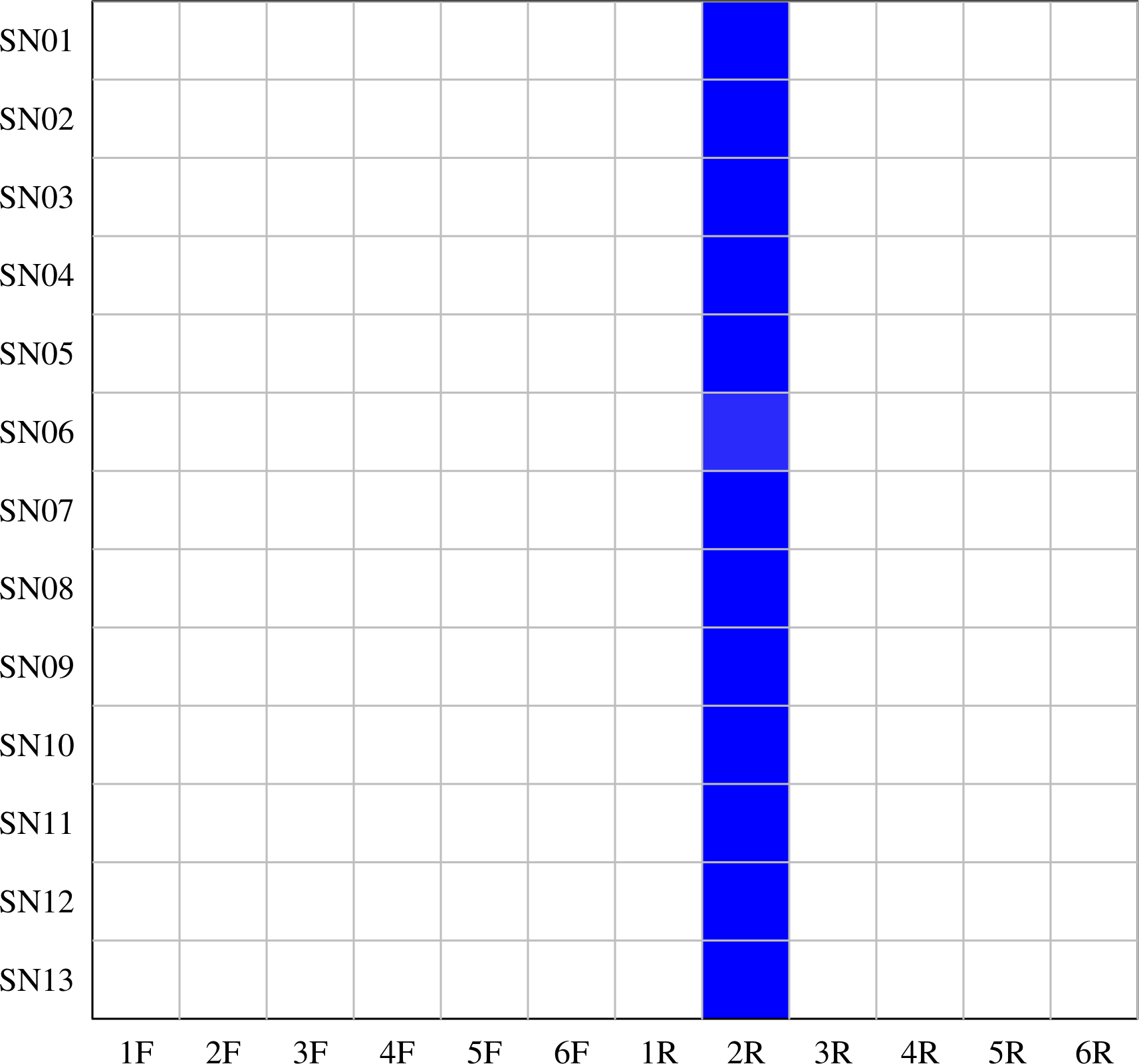}\label{fig:org_vs_soc_2r}}\hspace{0.2mm}
    \subfloat[][]{\includegraphics[height=1in, width=1.5in]{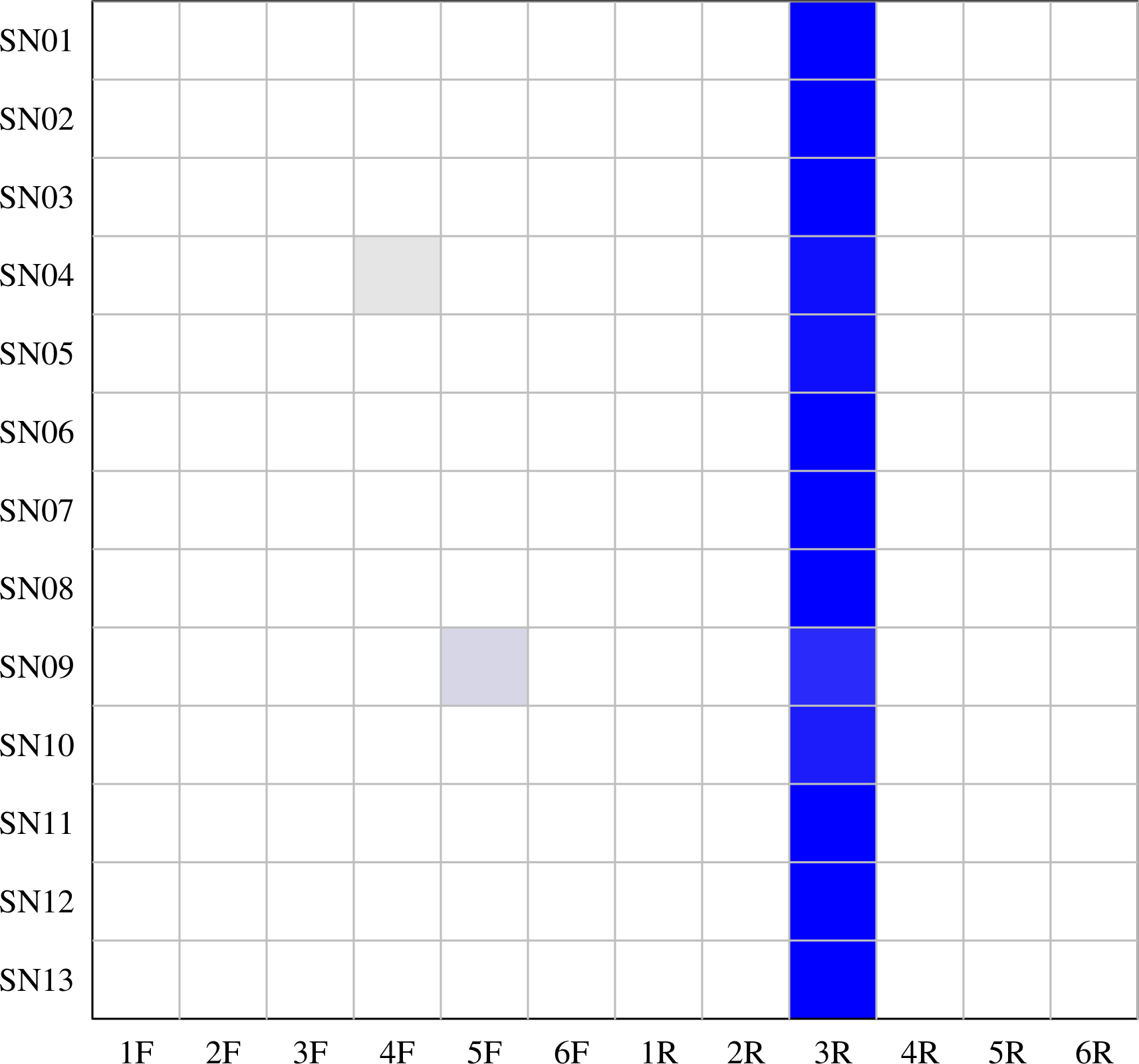}\label{fig:org_vs_soc_3r}}\hspace{0.2mm}
    \subfloat[][]{\includegraphics[height=1in, width=1.5in]{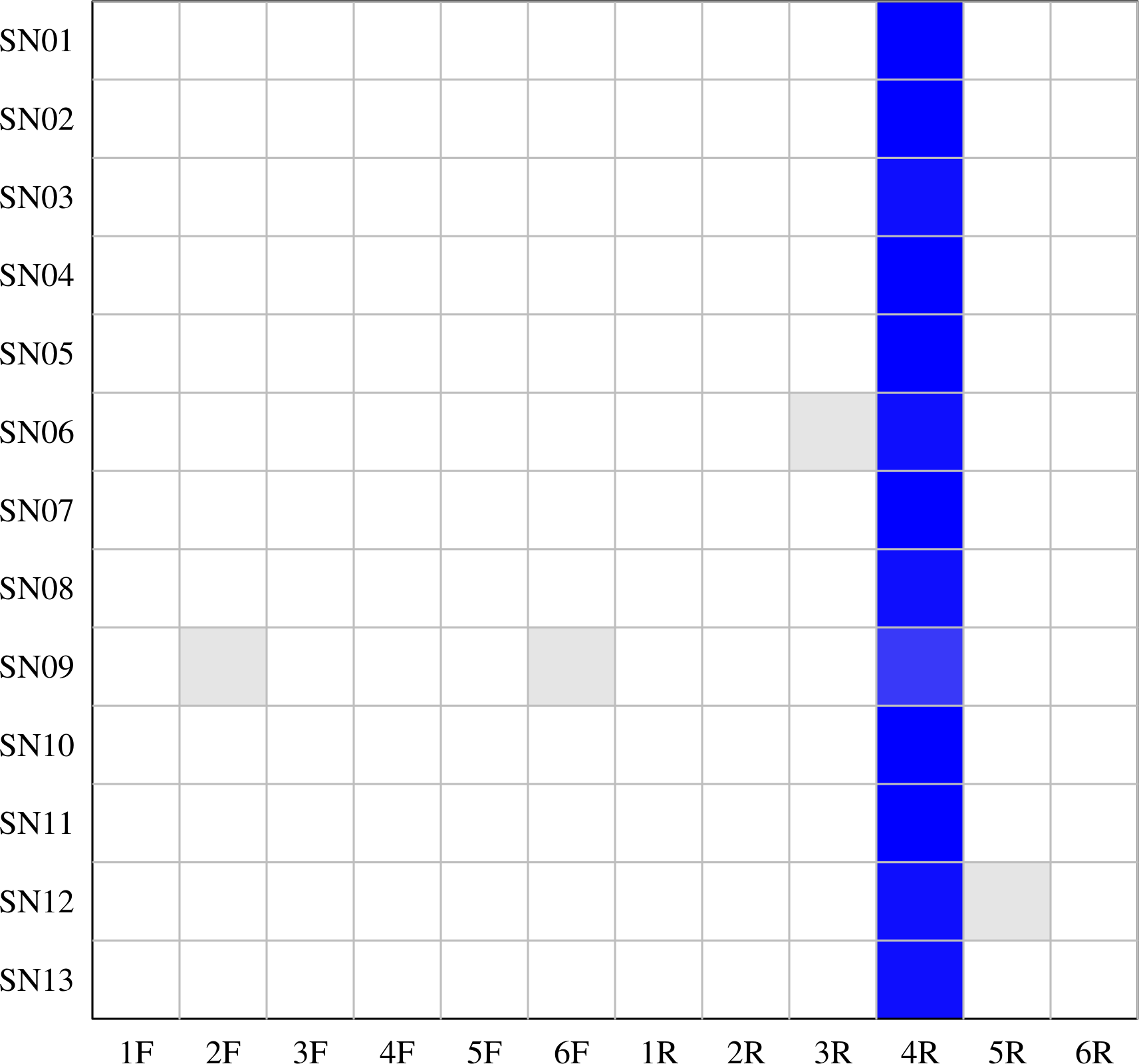}\label{fig:org_vs_soc_4r}}\hspace{0.2mm}
    \subfloat[][]{\includegraphics[height=1in, width=1.5in]{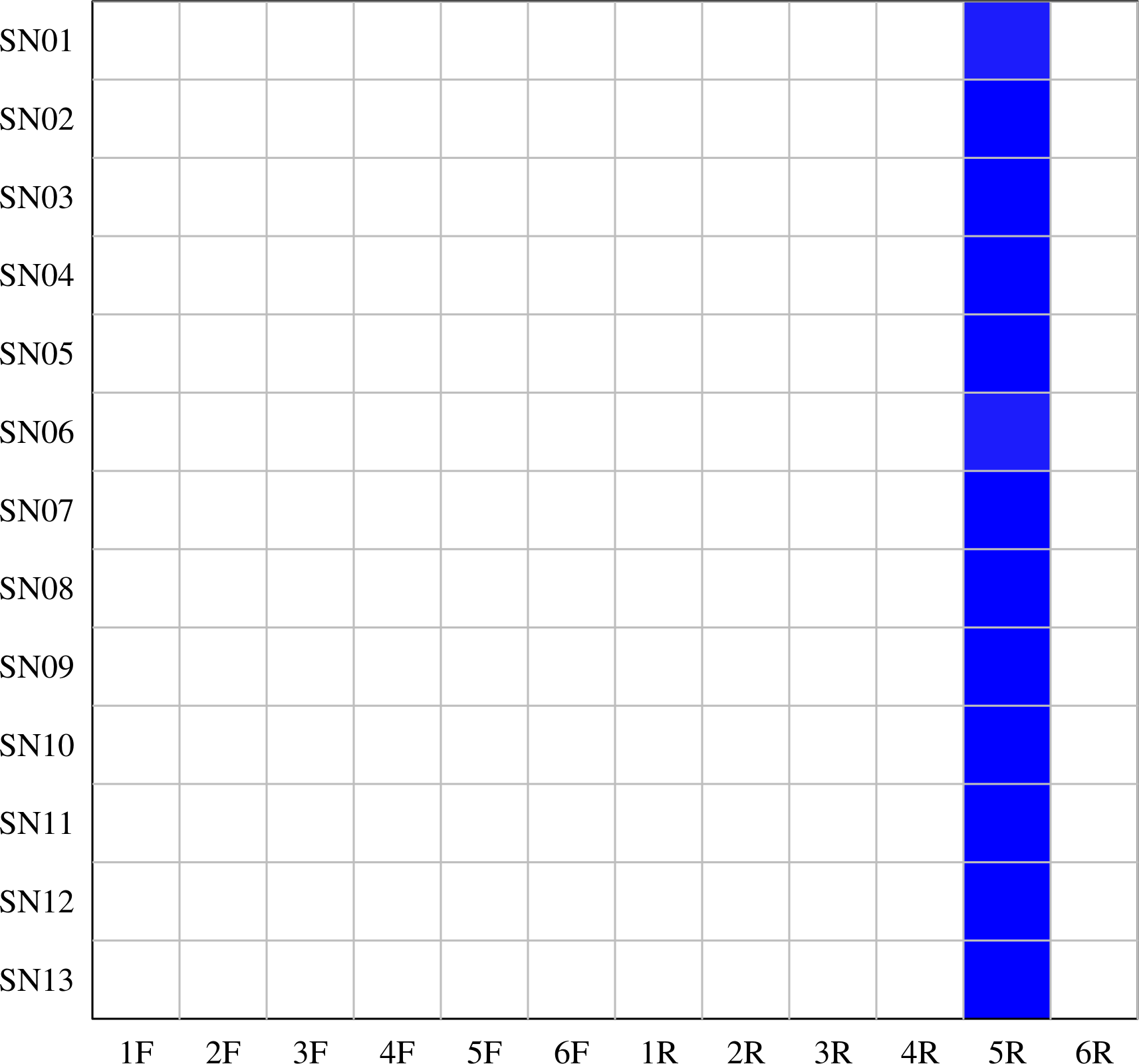}\label{fig:org_vs_soc_5r}}\hspace{0.2mm}
    \subfloat[][]{\includegraphics[height=1in, width=1.5in]{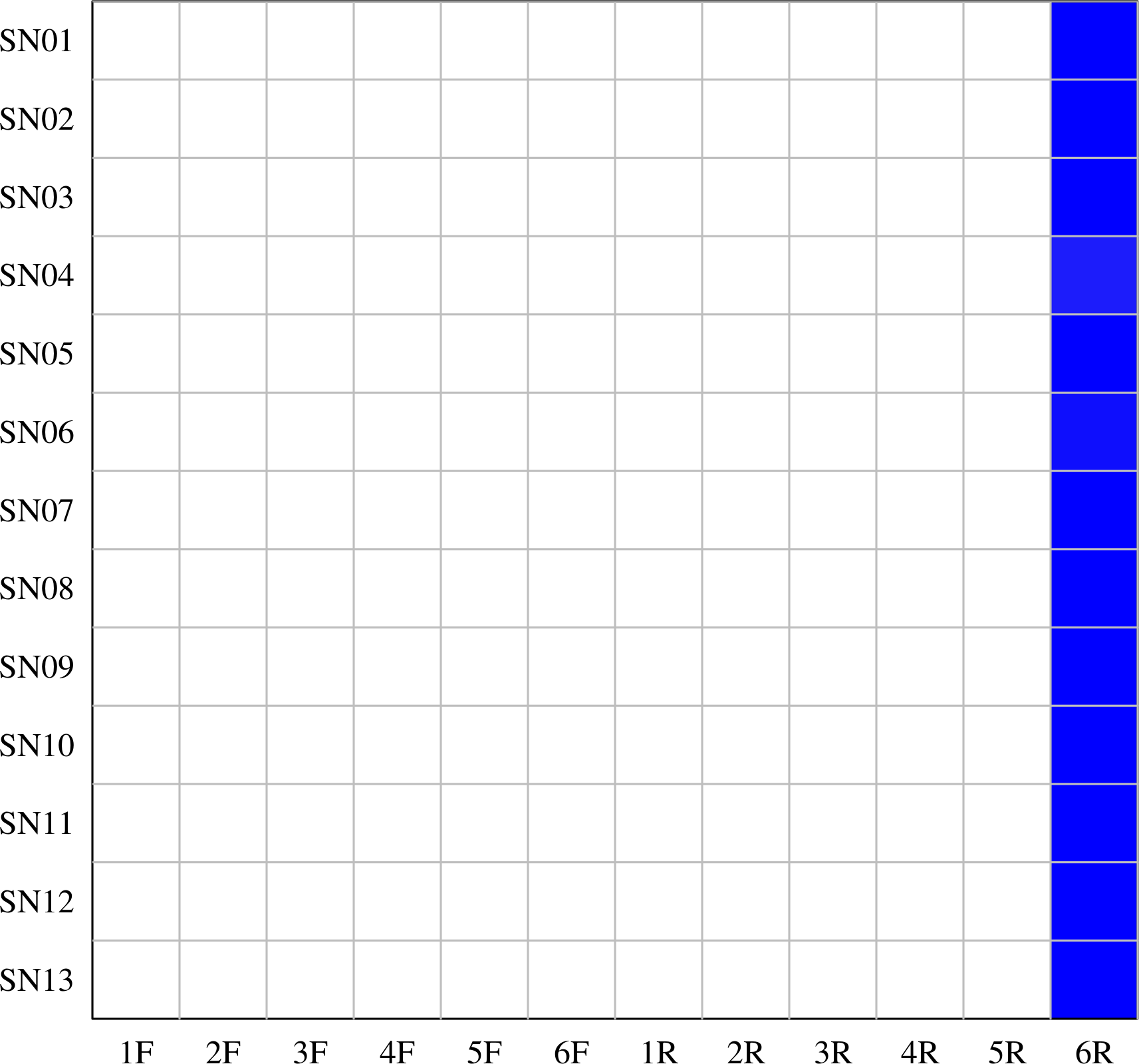}\label{fig:org_vs_soc_6r}}\\
    \caption{\emph{Profile attribution} results. Each graph groups the results of a single source on all thirteen SNs, six for the front cameras (first and second rows) and six for the rear cameras (third and fourth rows). The number of images in each cell is identified through a white-to-blue scale, from 0 (white) to 17 (blue).}\label{fig:org_vs_soc}
    \label{fig:globfig}
\end{figure}

\subsection{Profile Attribution}
\emph{Profile attribution} task allows us to verify through which smartphone the user has taken and uploaded the images on a SN (see case (a) in Figure \ref{fig:multilayer}). For this test, for each SN in Table \ref{tab:sn} and for each smartphone's camera in Table \ref{tab:smart}, we also used a \textbf{training set} of 33 original images and a \textbf{test set} of 17 downloaded images. This means that we obtained thirteen graphs for each smartphone's camera, one for each SN, that look like as a single graph in Figure \ref{fig:org_vs_org}. In this case, in order to define the threshold that allows to classify the images we used a simple generalized linear model.

Figure \ref{fig:org_vs_soc} shows the \emph{profile attribution} results for each smartphone's camera, results for the front cameras are in the first two rows and for the rear cameras are in the second two rows. In particular, each row in each graph represents the classification results for a specific SN and each white-to-blue scale cell identifies the 17 images of test set assigned to the source camera in that SN, from 0 (white) to 17 (blue). This aggregation allows to highlight the \emph{user unaware watermarking} capability of each smartphone's camera in each SN. If all the images are correctly assigned, in each graph we see an intense blue column corresponding to the right source and no other ``switched on'' cells out of that column. The results show that for all source cameras in each SN, the fingerprint $FP$ allows to correctly classify almost all the 17 downloaded images. Moreover, the erroneously assigned images do not compromise the identification of the source since only in 5 graphs out of 12, that is (a), (b), (g), (i) and (l), we can see very few images wrongly assigned to the right source.

\subsection{Intra-layer User Profiles Linking}
In this section, we discuss results in the context of \emph{intra-layer user profiles linking} that is the task to identify if a set of images from two different user profiles within the same SN belong to the same user (see case (b) in Figure \ref{fig:multilayer}). For this reason, both the \textbf{training set} (33 downloaded images) and the \textbf{test set} (17 downloaded images) were taken from the same SN. Thus, we obtained thirteen graphs (one for each SN) for each smartphone's camera, and we used again a generalized linear model for the classification process.

\begin{figure}[!t]
    \centering
    \subfloat[][]{\includegraphics[height=1in, width=1.5in]{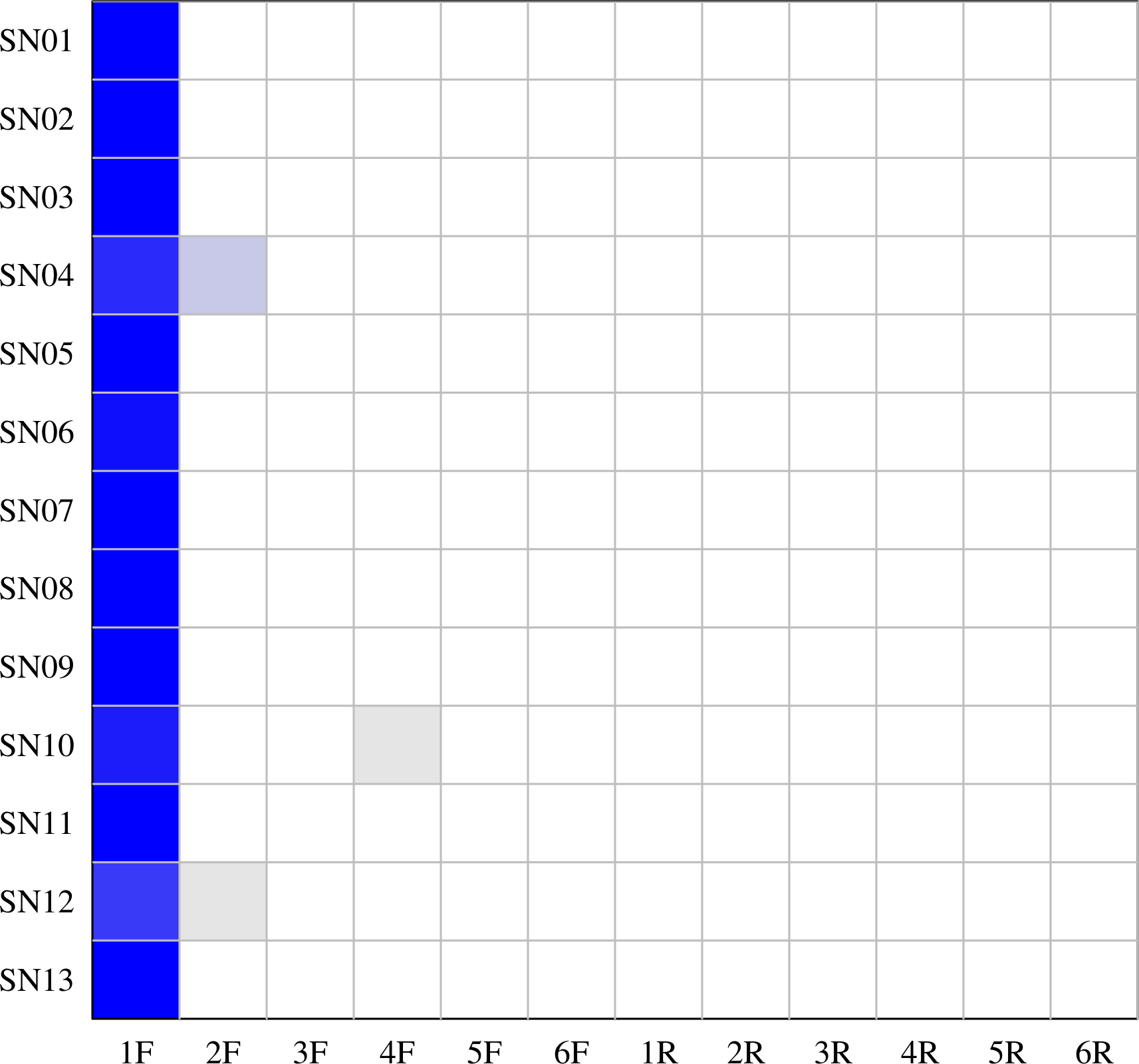}\label{fig:soc_vs_soc_1f}}\hspace{0.2mm}
    \subfloat[][]{\includegraphics[height=1in, width=1.5in]{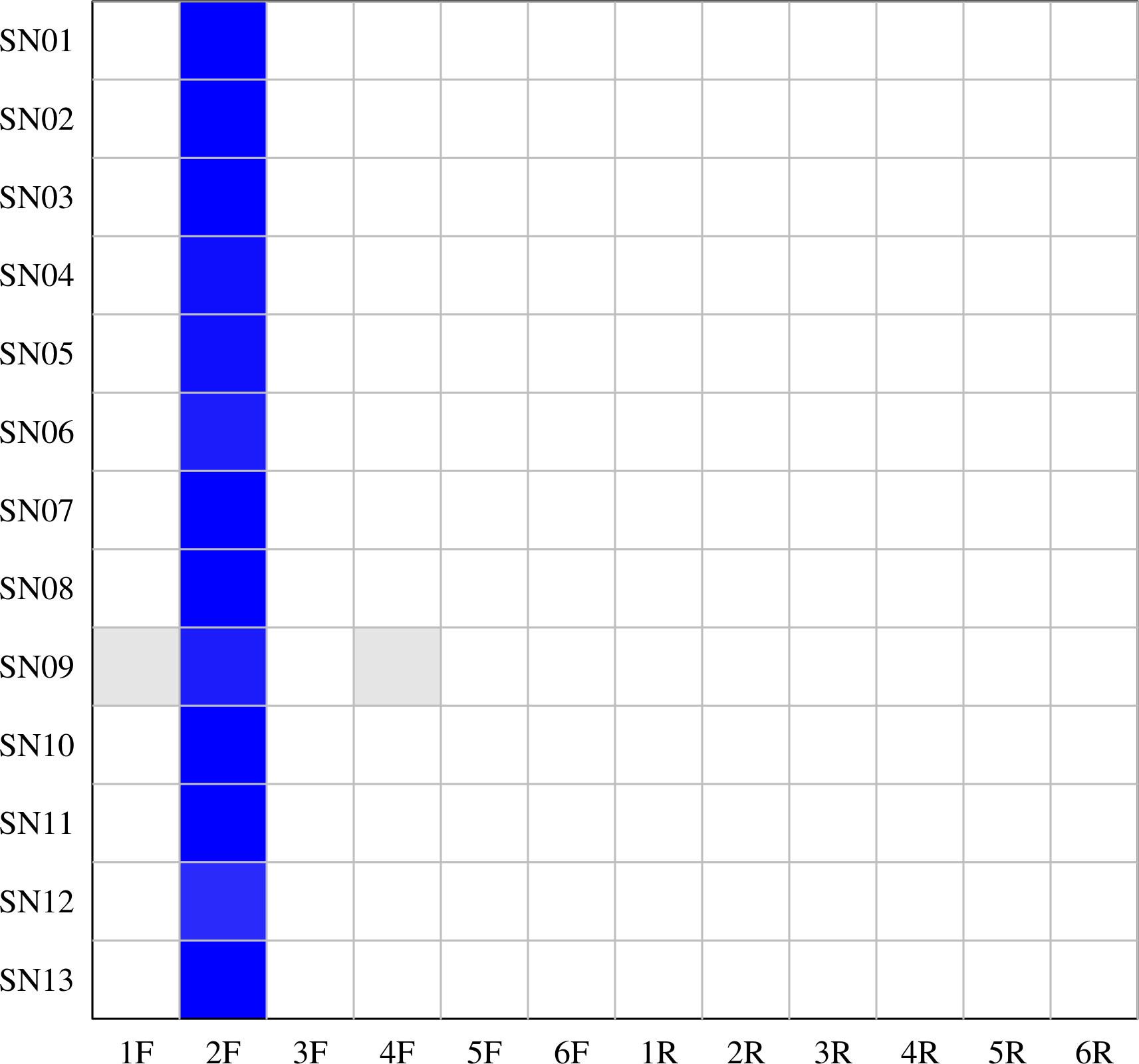}\label{fig:soc_vs_soc_2f}}\hspace{0.2mm}
    \subfloat[][]{\includegraphics[height=1in, width=1.5in]{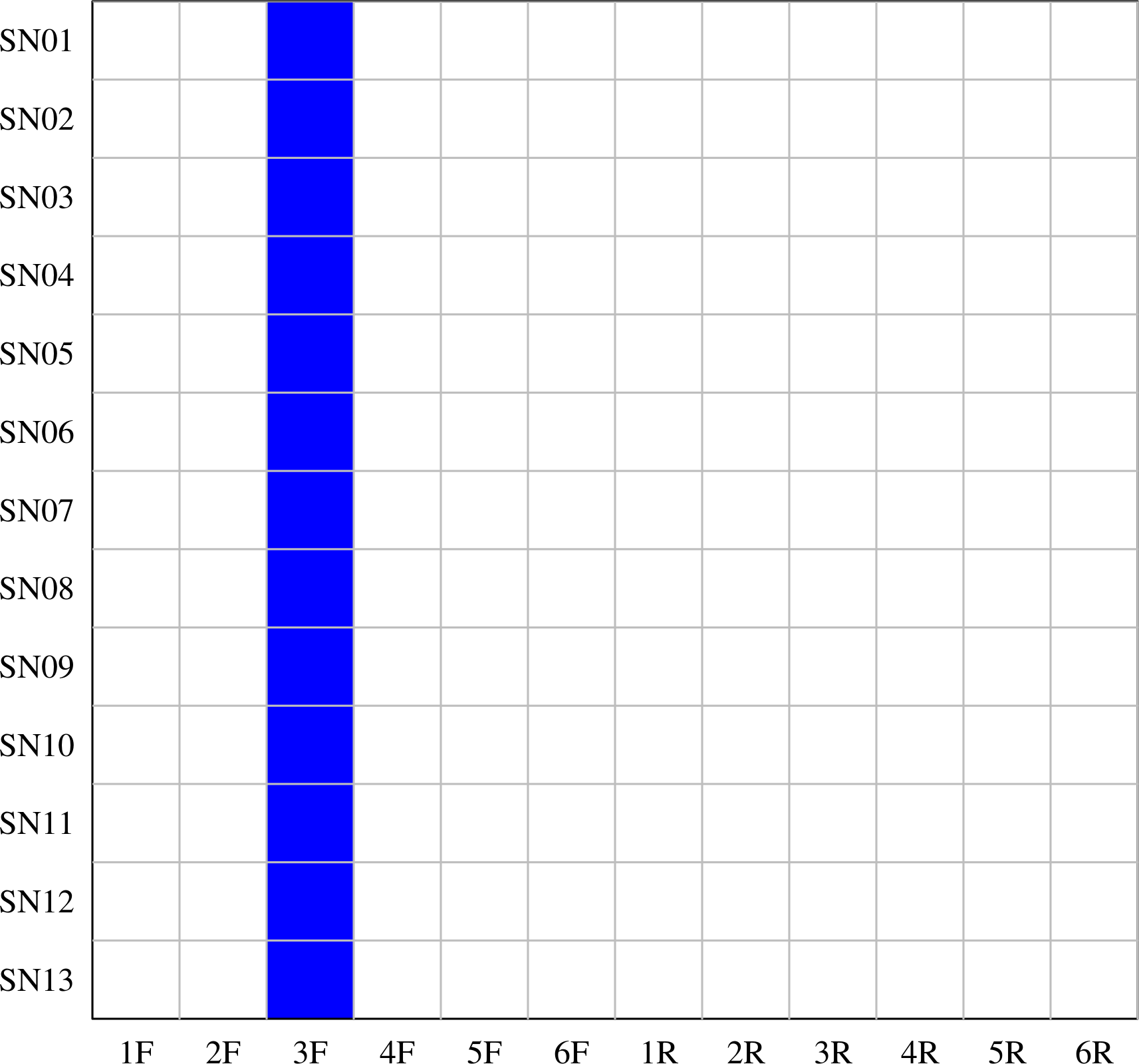}\label{fig:soc_vs_soc_3f}}\hspace{0.2mm}
    \subfloat[][]{\includegraphics[height=1in, width=1.5in]{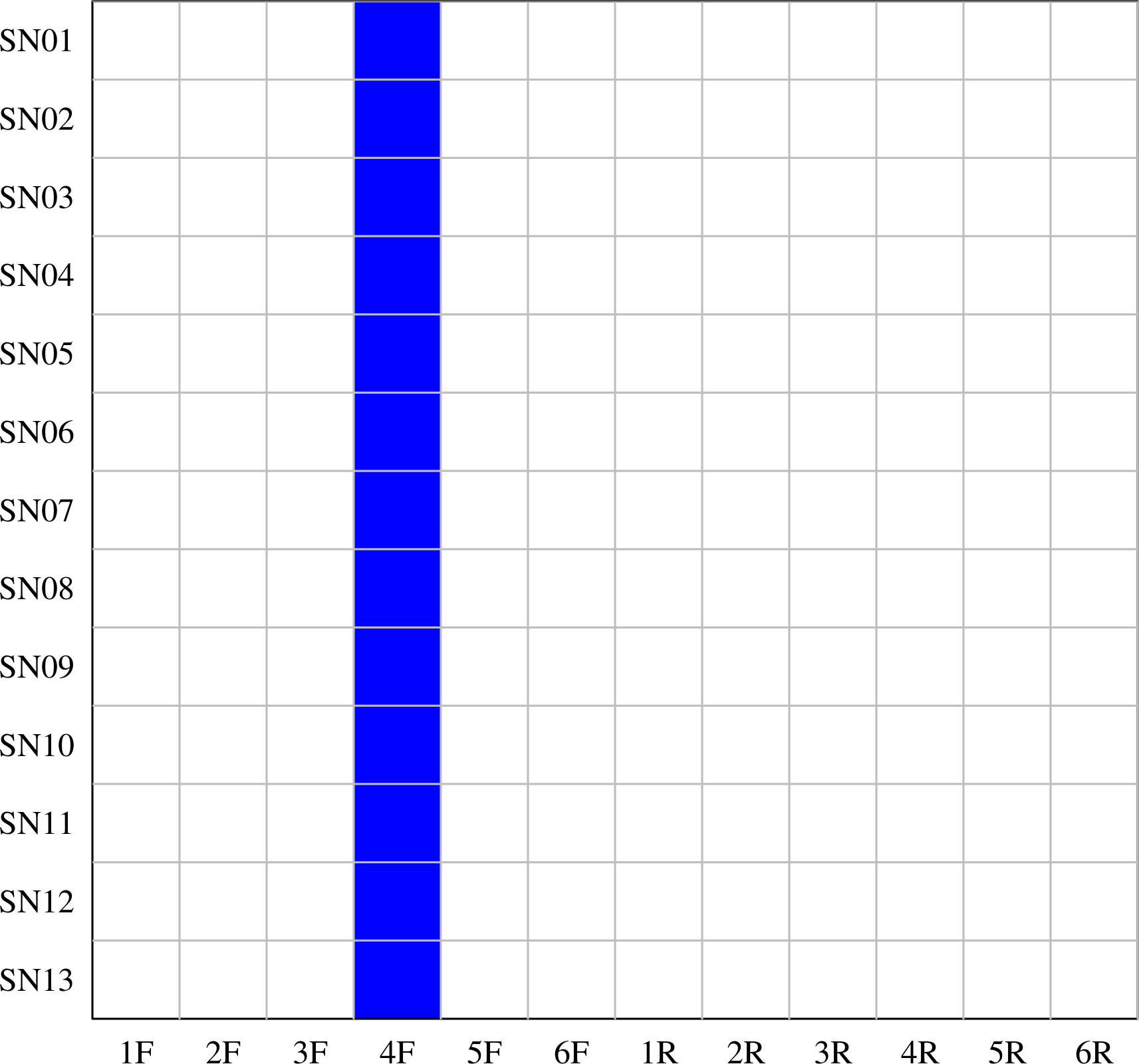}\label{fig:soc_vs_soc_4f}}\hspace{0.2mm}
    \subfloat[][]{\includegraphics[height=1in, width=1.5in]{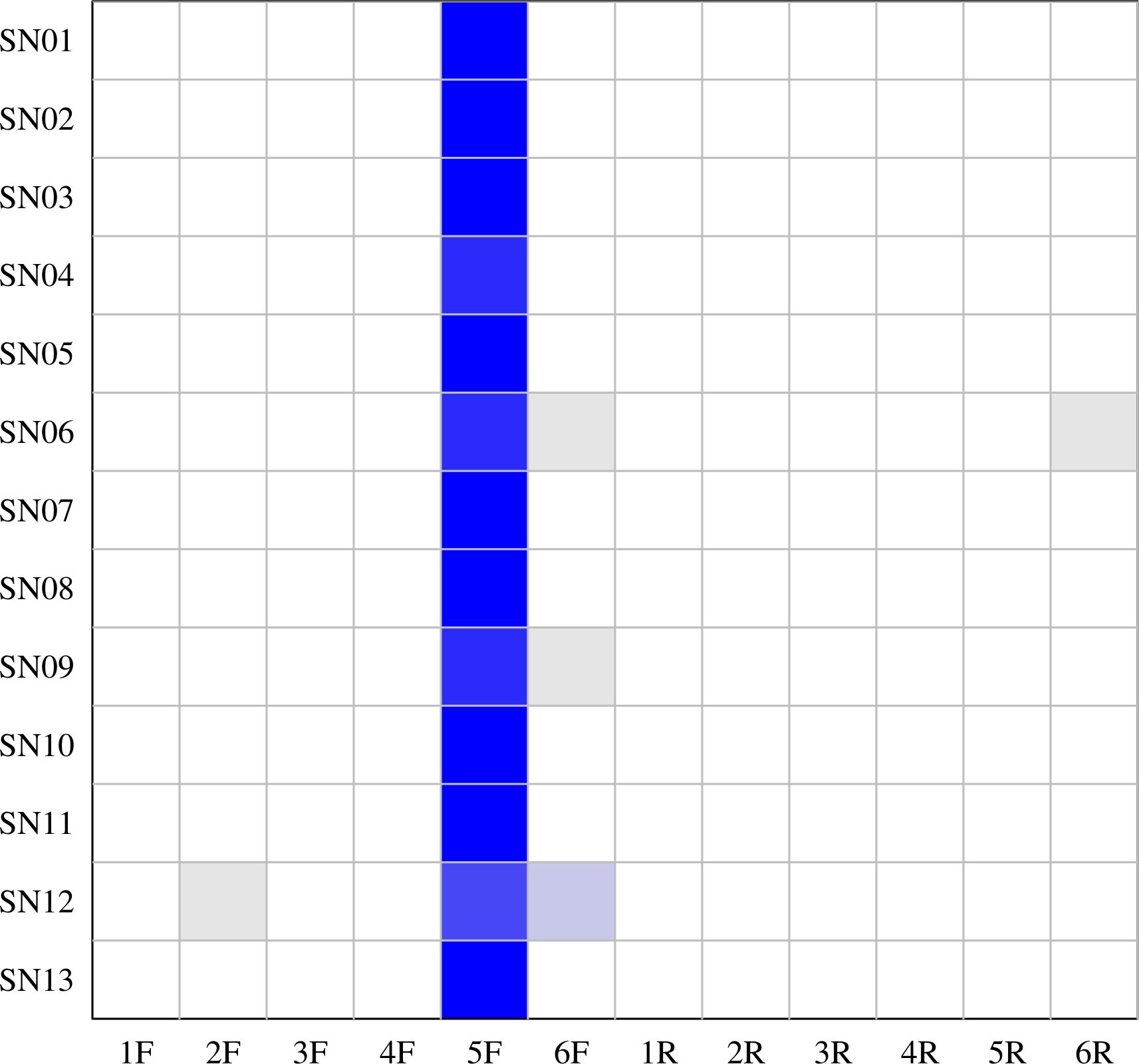}\label{fig:soc_vs_soc_5f}}\hspace{0.2mm}
    \subfloat[][]{\includegraphics[height=1in, width=1.5in]{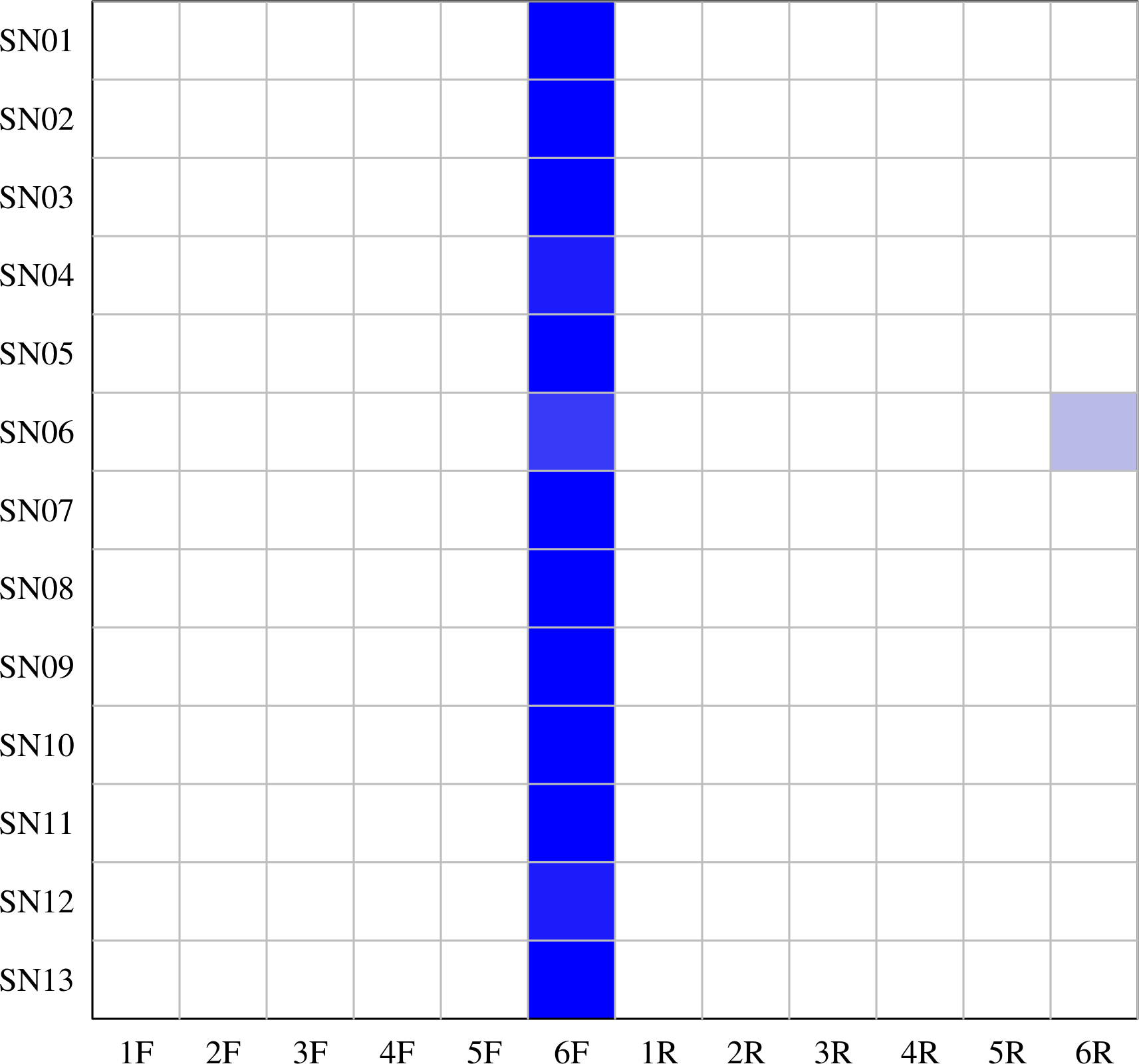}\label{fig:soc_vs_soc_6f}}\\
    \subfloat[][]{\includegraphics[height=1in, width=1.5in]{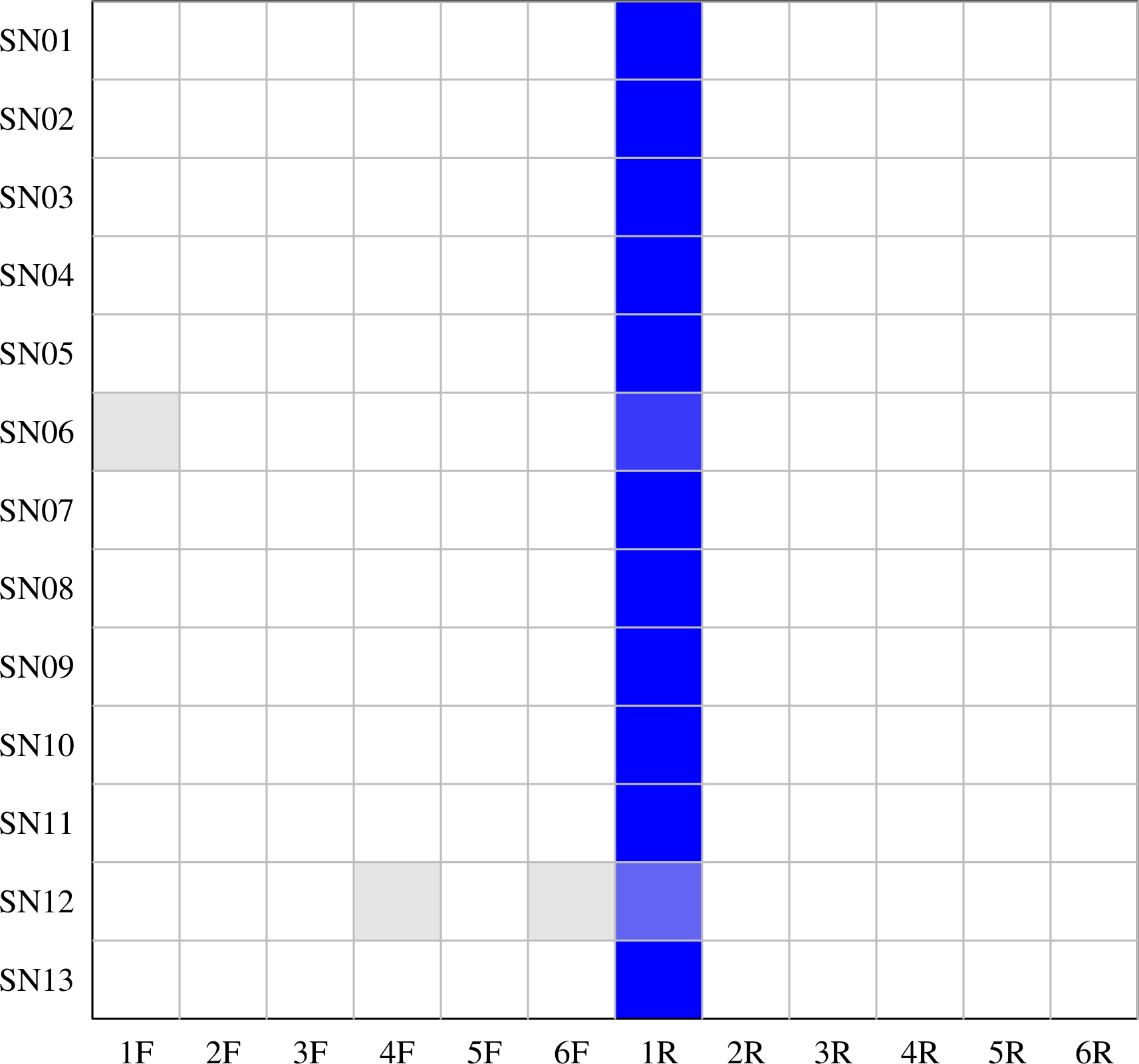}\label{fig:soc_vs_soc_1r}}\hspace{0.2mm}
    \subfloat[][]{\includegraphics[height=1in, width=1.5in]{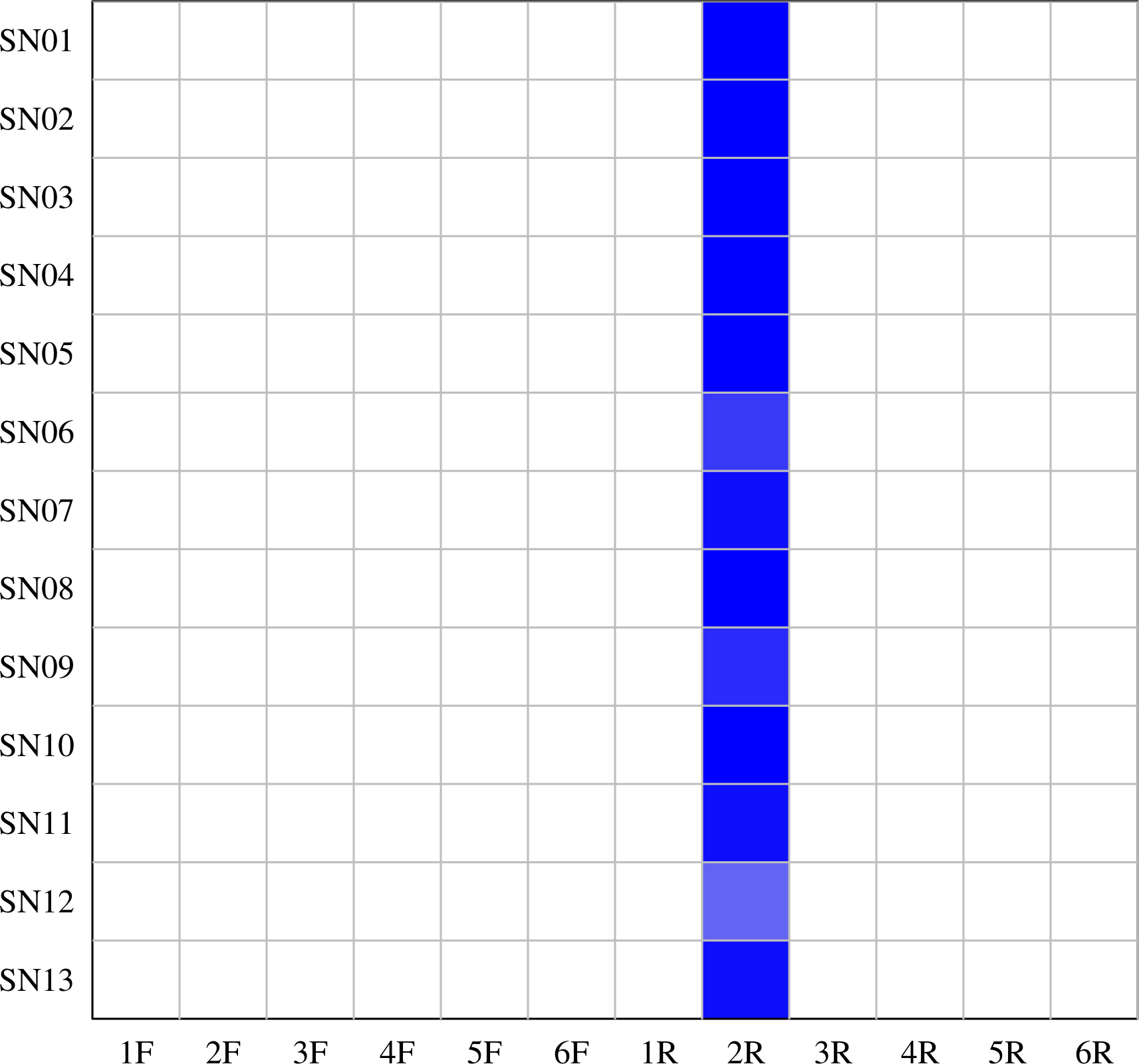}\label{fig:soc_vs_soc_2r}}\hspace{0.2mm}
    \subfloat[][]{\includegraphics[height=1in, width=1.5in]{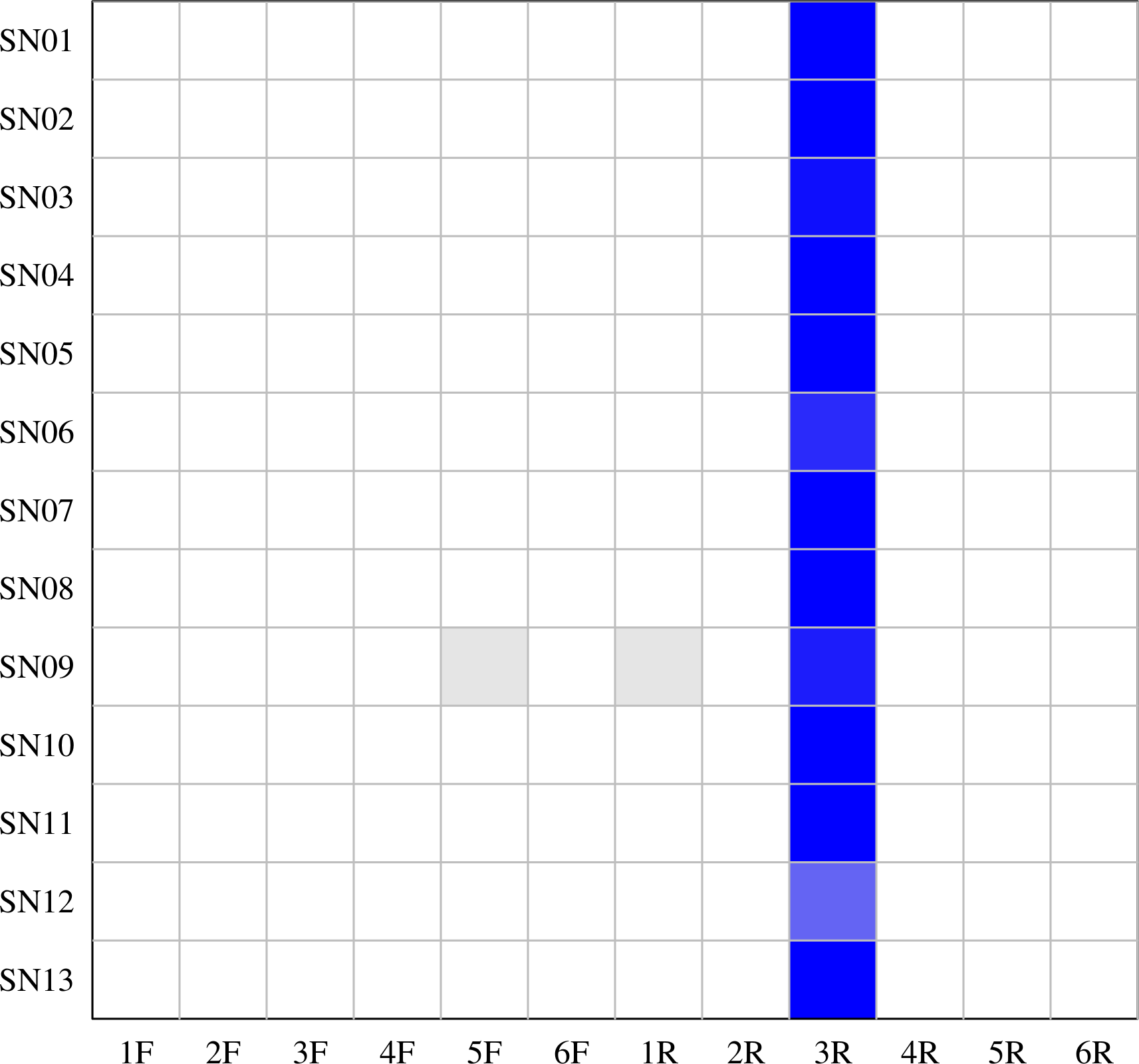}\label{fig:soc_vs_soc_3r}}\hspace{0.2mm}
    \subfloat[][]{\includegraphics[height=1in, width=1.5in]{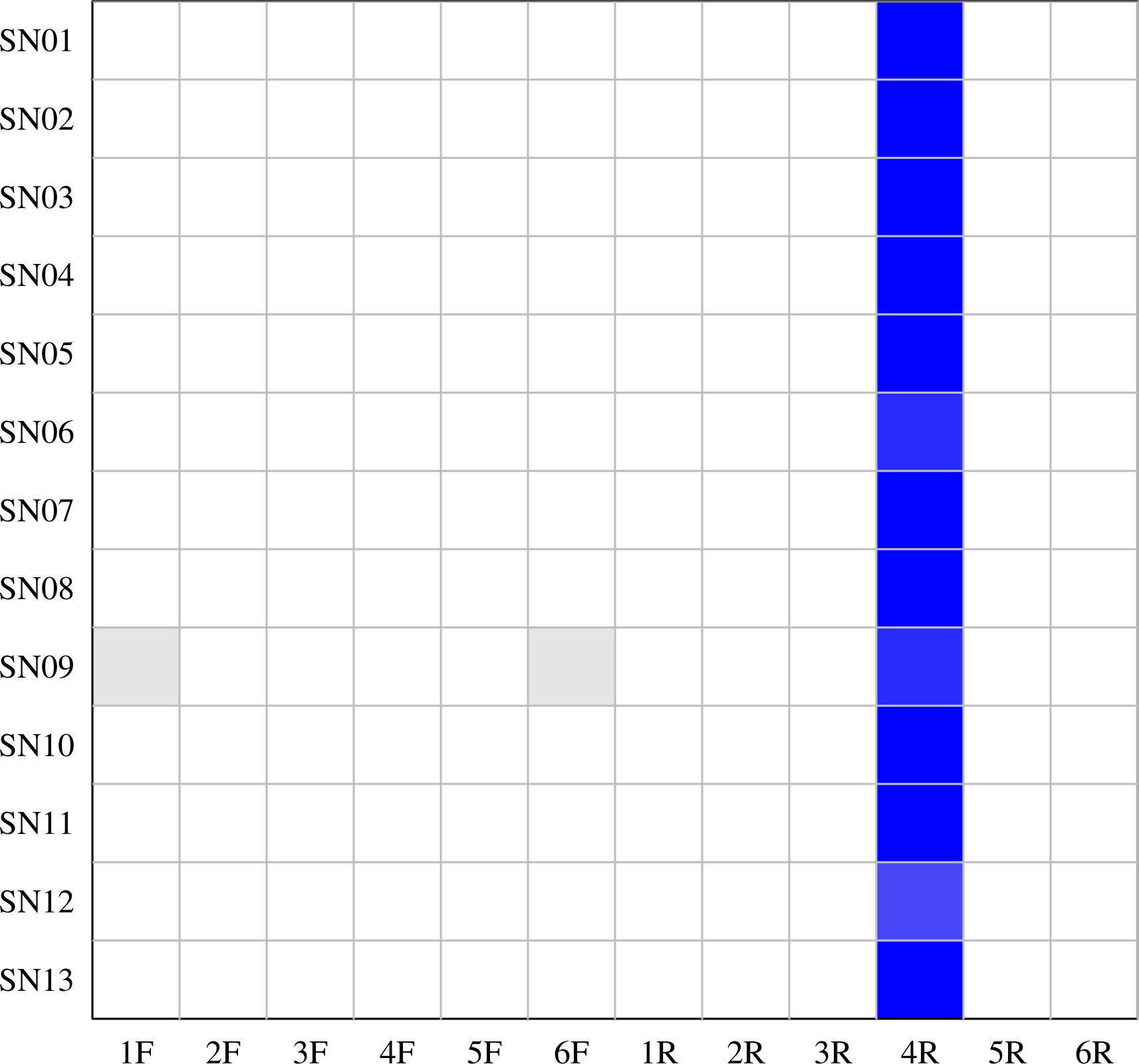}\label{fig:soc_vs_soc_4r}}\hspace{0.2mm}
    \subfloat[][]{\includegraphics[height=1in, width=1.5in]{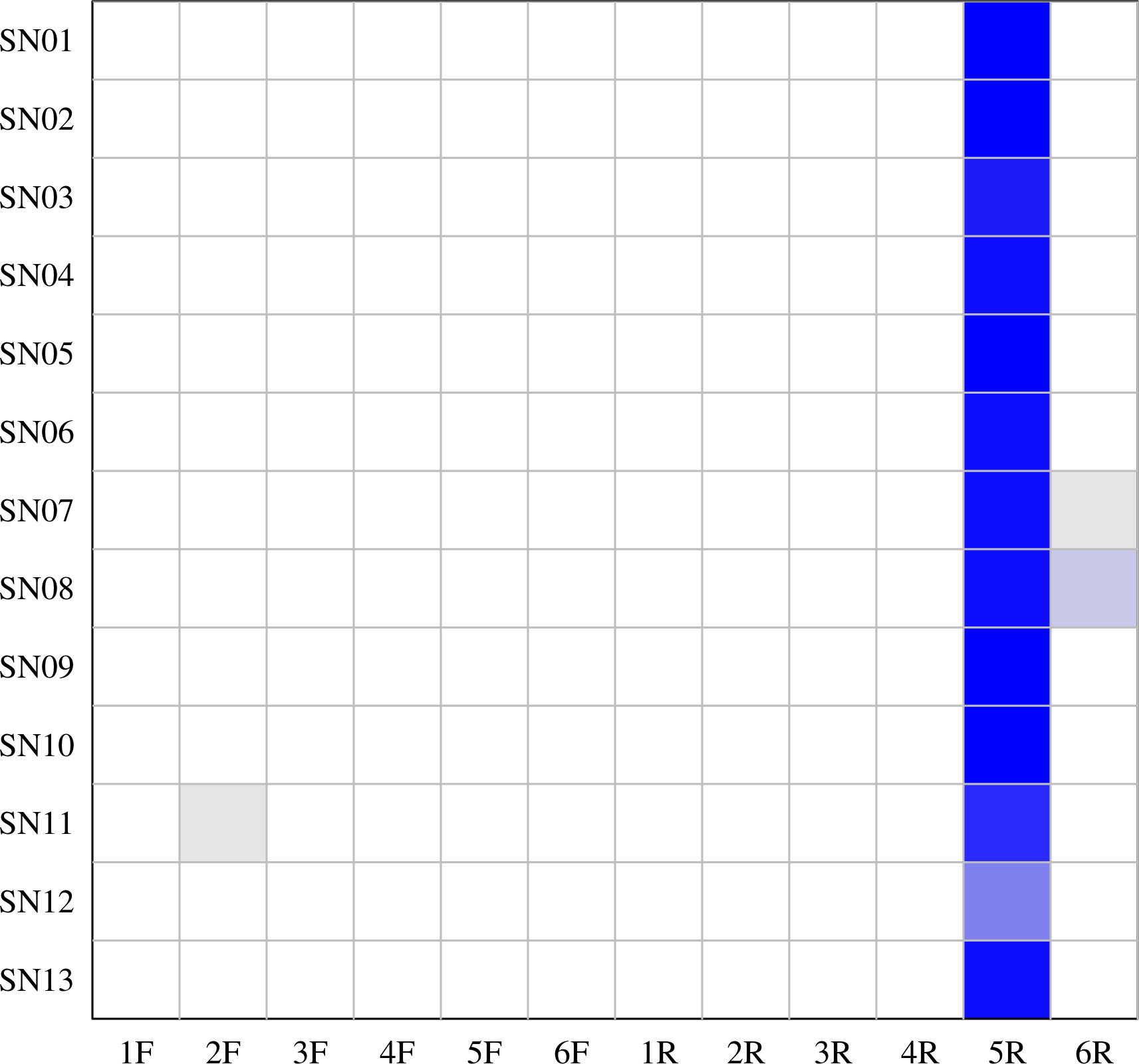}\label{fig:soc_vs_soc_5r}}\hspace{0.2mm}
    \subfloat[][]{\includegraphics[height=1in, width=1.5in]{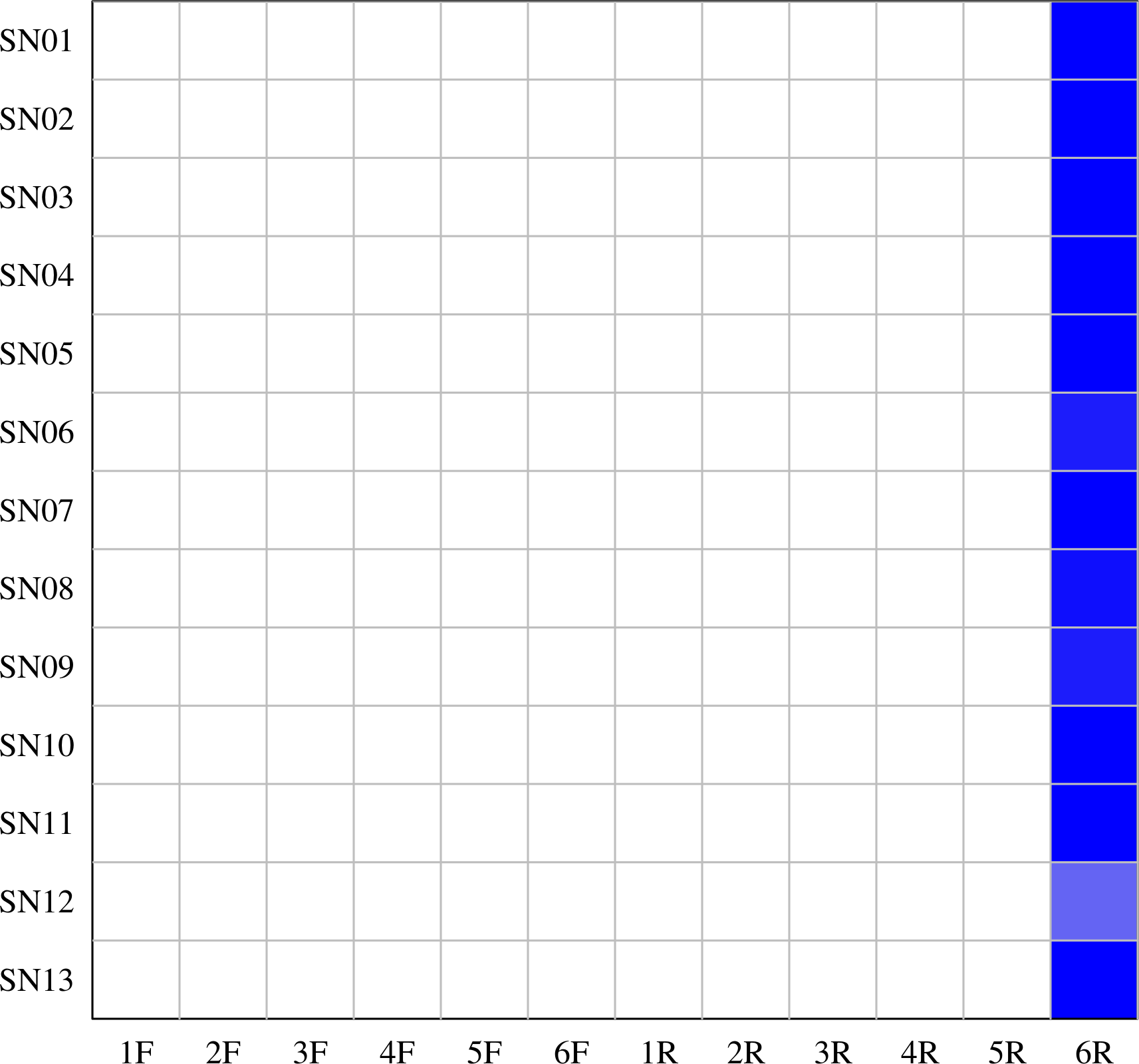}\label{fig:soc_vs_soc_6r}}\\
    \caption{\emph{Intra-layer user profiles linking} results. Each graph groups the results of a single source on all thirteen SNs, six for the front cameras (first and second rows) and six for the rear cameras (second and third rows). The number of images in each cell is identified through a white-to-blue scale, from 0 (white) to 17 (blue).}\label{fig:soc_vs_soc}
\end{figure}

Figure \ref{fig:soc_vs_soc} shows the \emph{intra-layer user profiles linking} performances for each smartphone's camera, front cameras are in the first two rows and rear cameras are in the second two rows. In particular, each row in each graph represents the classification results for a specific SN and each white-to-blue scale cell identifies the images assigned to the source camera in that SN, from 0 (white) to 17 (blue). The results show that for each front source camera, the fingerprint $FP$ allows  to correctly classify almost all the 17 downloaded images in each SN. WeChat (penultimate row from the top in each graph) returns worst results for the rear cameras. This is probably because WeChat provides a high compression level with large images, as shown in Table \ref{tab:compr}. However, in all the graphs we see a well-defined blue column corresponding to the right source.

\subsection{Inter-layer User Profiles Linking}
\emph{Inter-layer user profiles linking} is the most challenging task and allows to verify if two sets of images from different user profiles on different SNs belong to the same user (see case (c) in Figure \ref{fig:multilayer}). Since the combination of all SNs produces a very large set of results (i.e., 12 graphs for each SN for each camera), we selected a representative subset of SNs composed of Facebook (SN01), Instagram (SN04), Telegram (SN07), and WhatsApp (SN13). In this case, the \textbf{training set} of 33 downloaded images and the \textbf{test set} of 17 downloaded images were selected from different SNs among the selected ones. For instance, while we used 33 images from Facebook to define the fingerprint $FP$, all the 17-images test sets were selected from Instagram. In this test, we also used a generalized linear model for the classification process.

\begin{figure}[!t]
    \centering
    \subfloat[][]{\includegraphics[height=1in, width=1.5in]{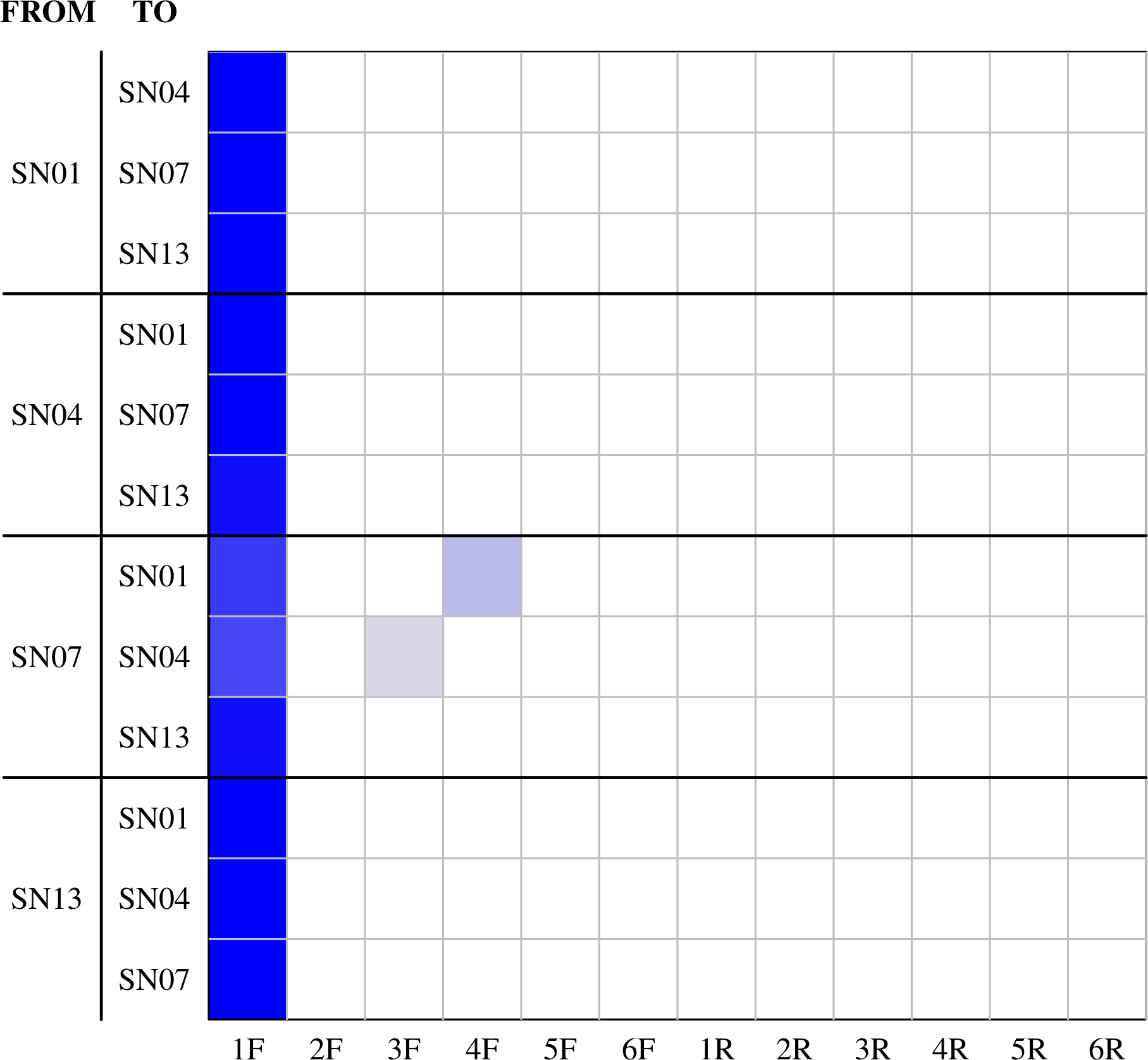}\label{fig:xsoc_vs_xsoc_1f}}\hspace{0.2mm}
    \subfloat[][]{\includegraphics[height=1in, width=1.5in]{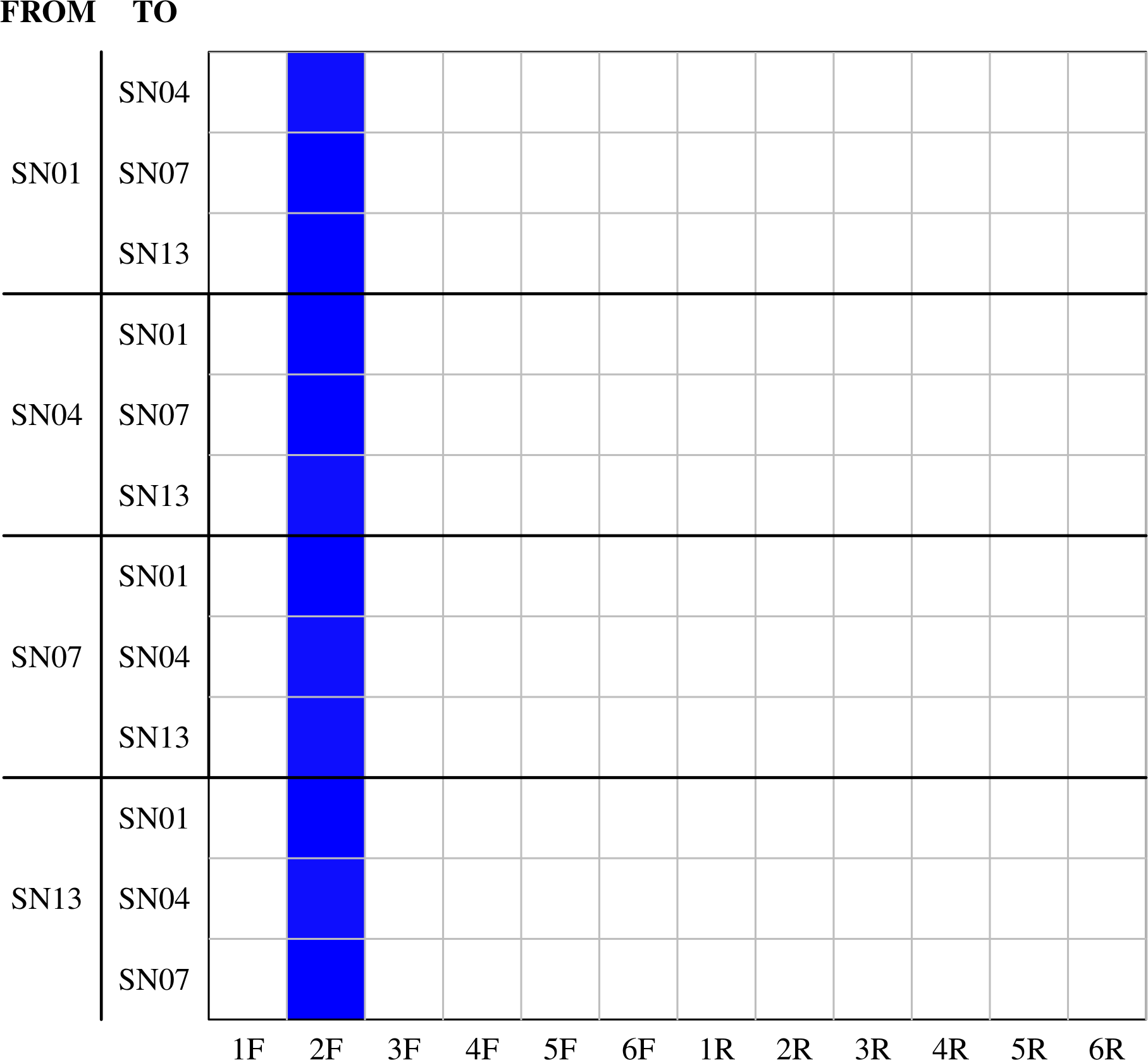}\label{fig:xsoc_vs_xsoc_2f}}\hspace{0.2mm}
    \subfloat[][]{\includegraphics[height=1in, width=1.5in]{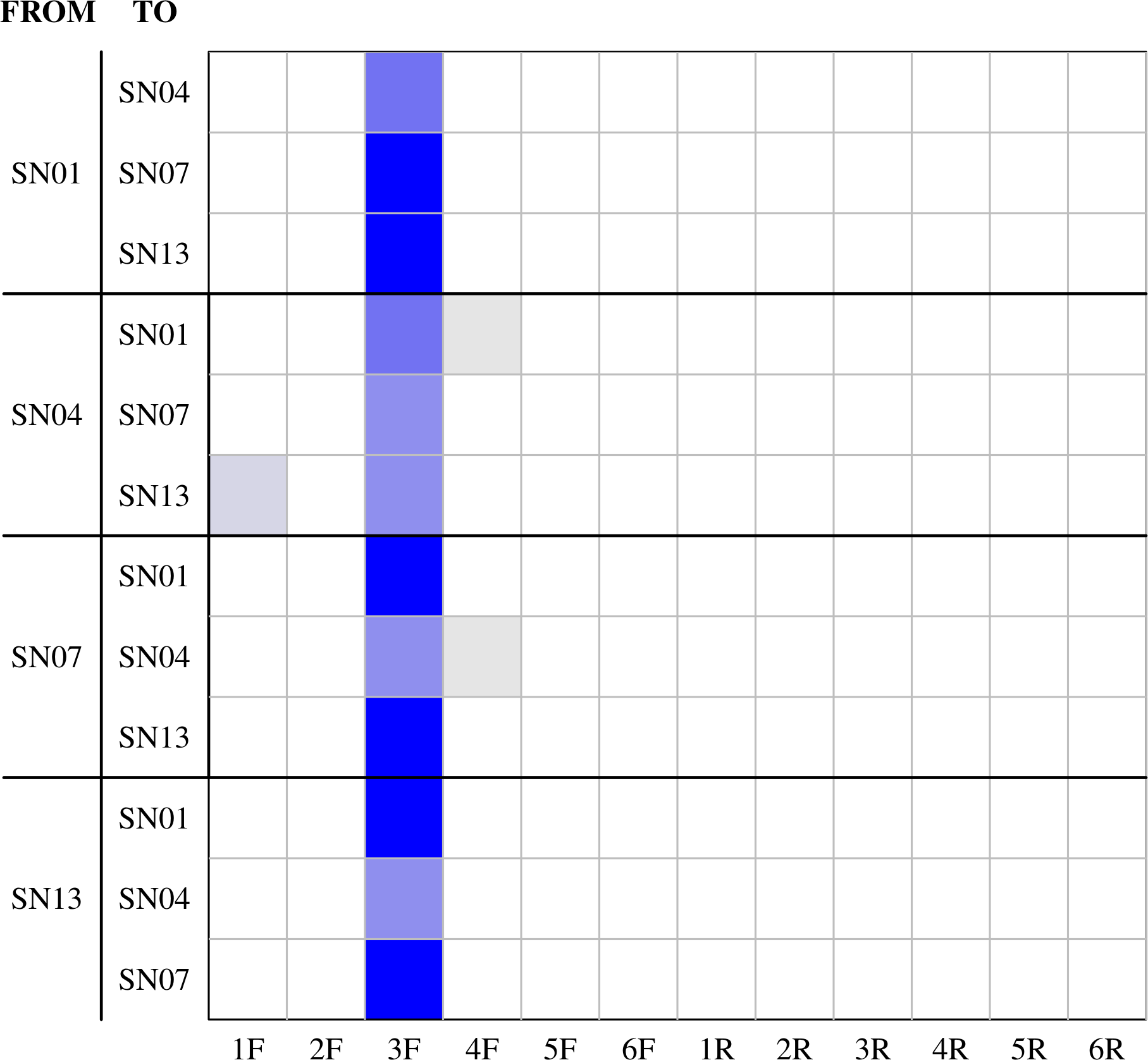}\label{fig:xsoc_vs_xsoc_3f}}\hspace{0.2mm}
    \subfloat[][]{\includegraphics[height=1in, width=1.5in]{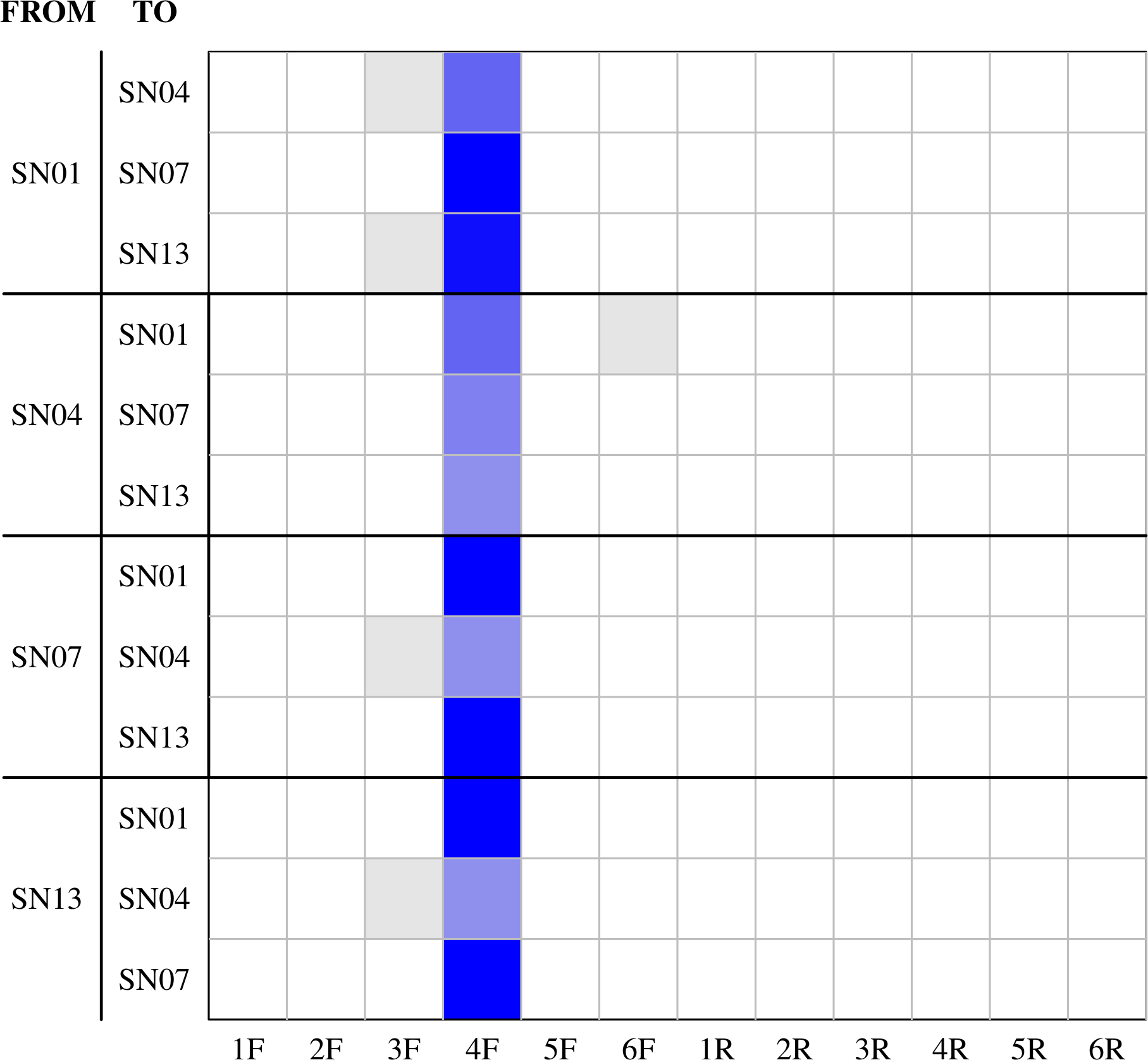}\label{fig:xsoc_vs_xsoc_4f}}\hspace{0.2mm}
    \subfloat[][]{\includegraphics[height=1in, width=1.5in]{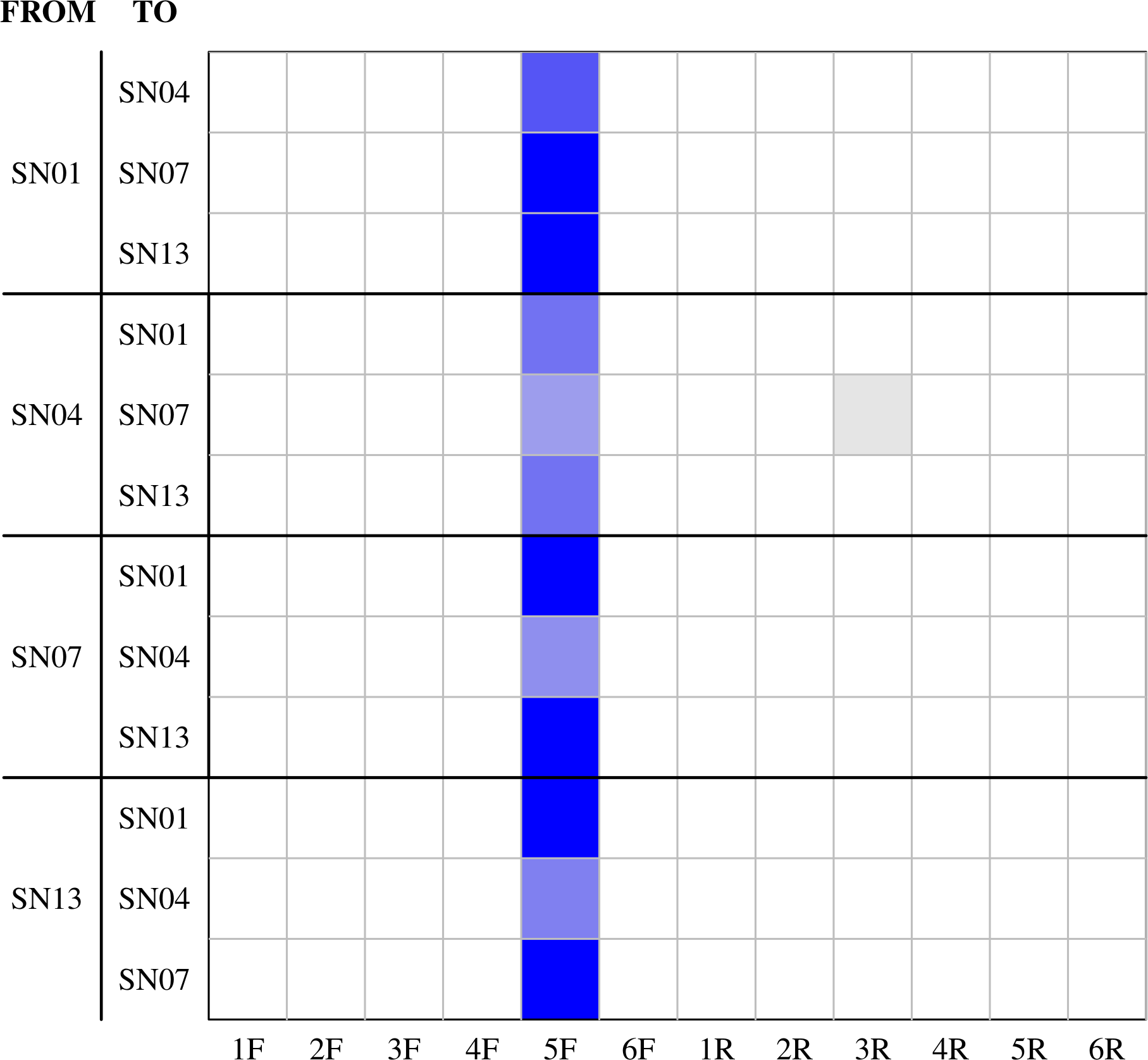}\label{fig:xsoc_vs_xsoc_5f}}\hspace{0.2mm}
    \subfloat[][]{\includegraphics[height=1in, width=1.5in]{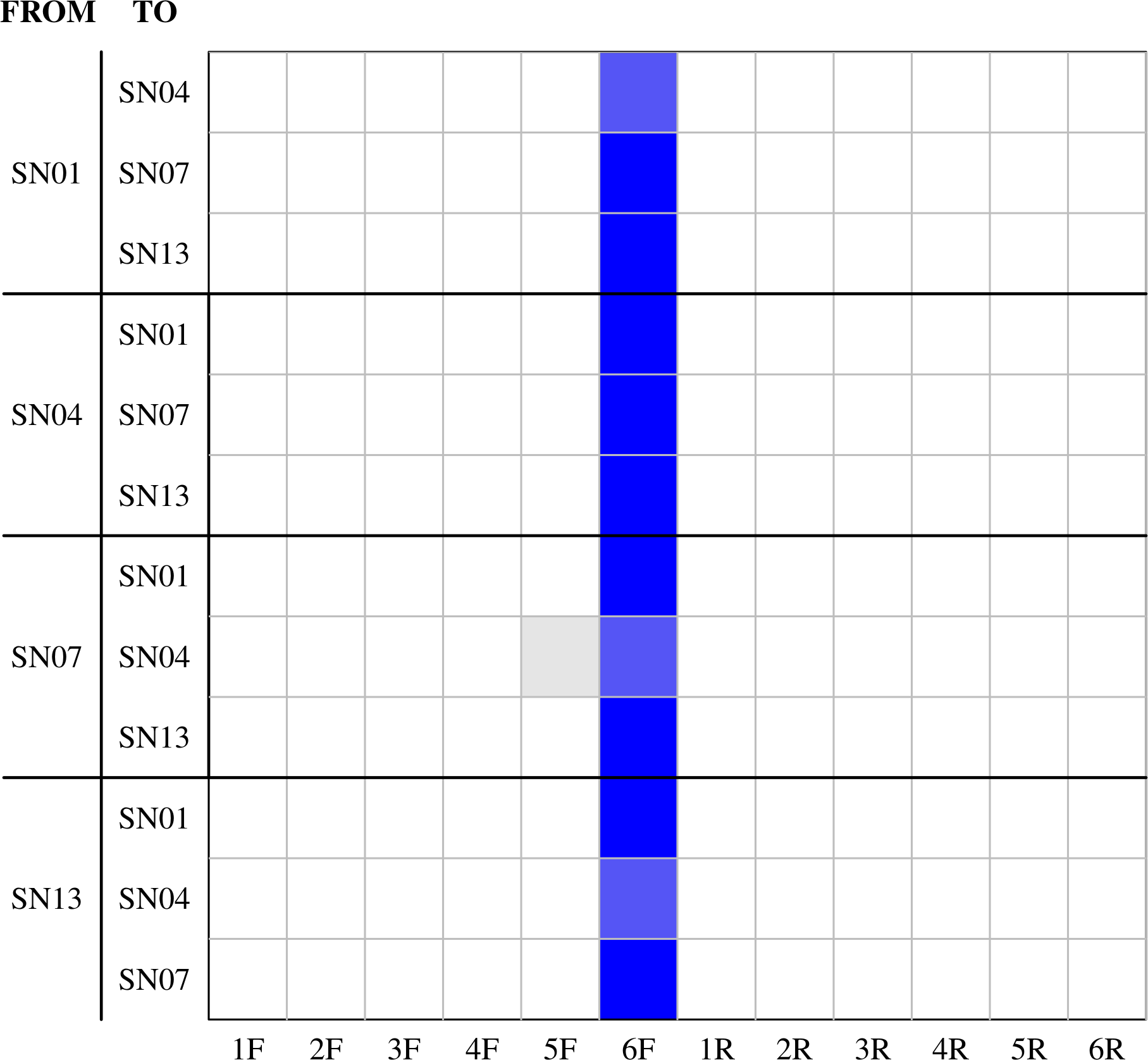}\label{fig:xsoc_vs_xsoc_6f}}\\
    \subfloat[][]{\includegraphics[height=1in, width=1.5in]{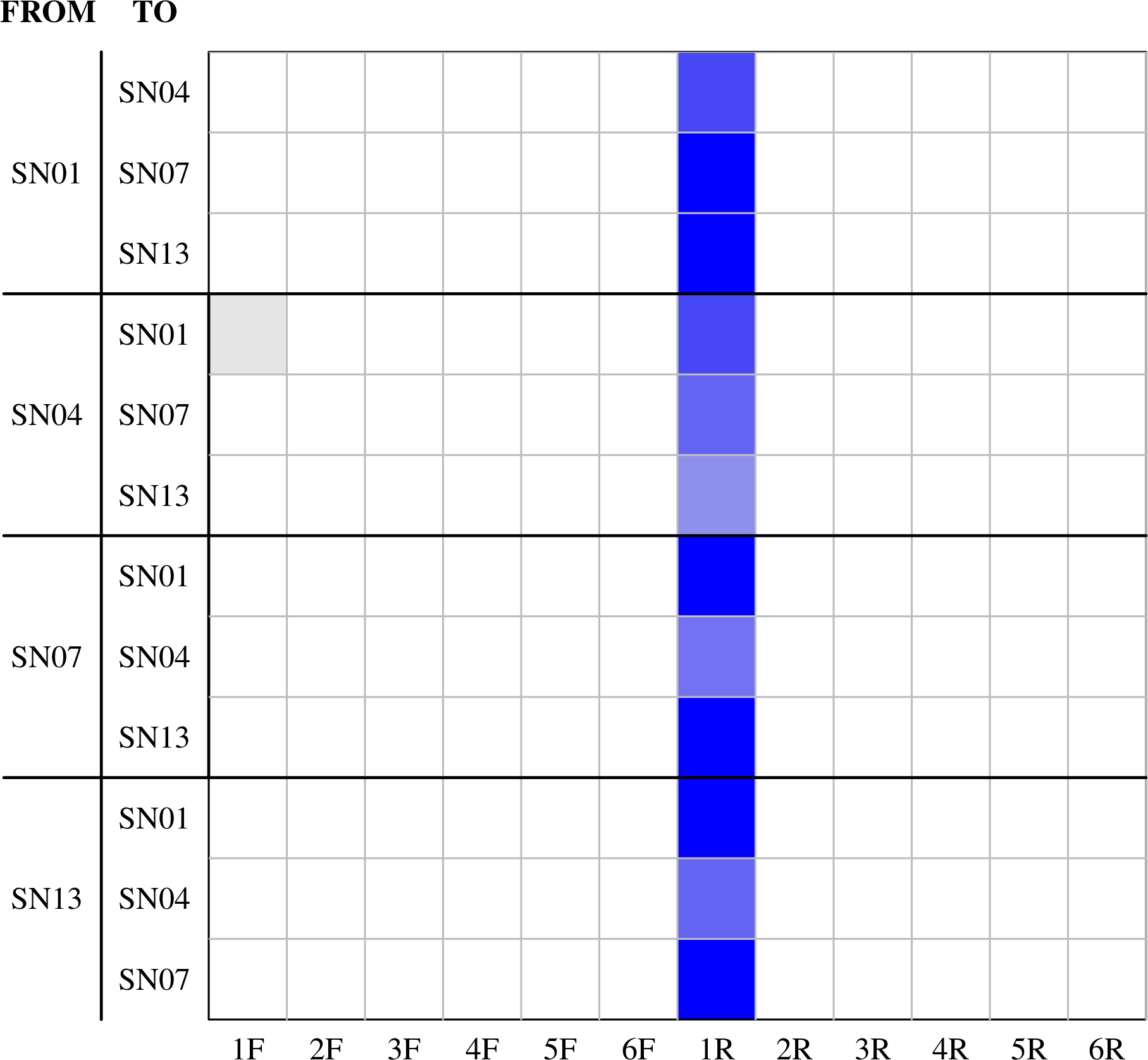}\label{fig:xsoc_vs_xsoc_1r}}\hspace{0.2mm}
    \subfloat[][]{\includegraphics[height=1in, width=1.5in]{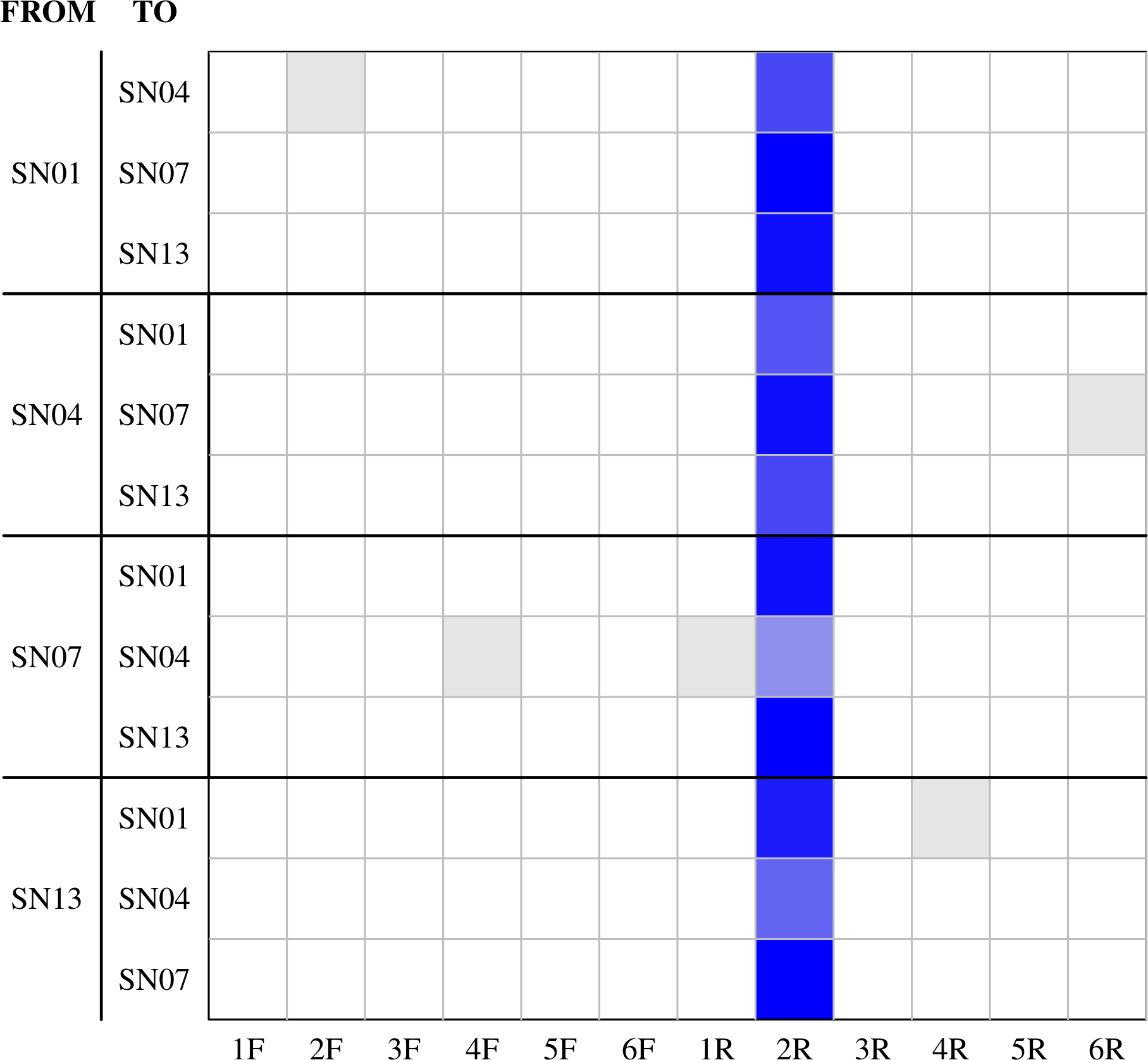}\label{fig:xsoc_vs_xsoc_2r}}\hspace{0.2mm}
    \subfloat[][]{\includegraphics[height=1in, width=1.5in]{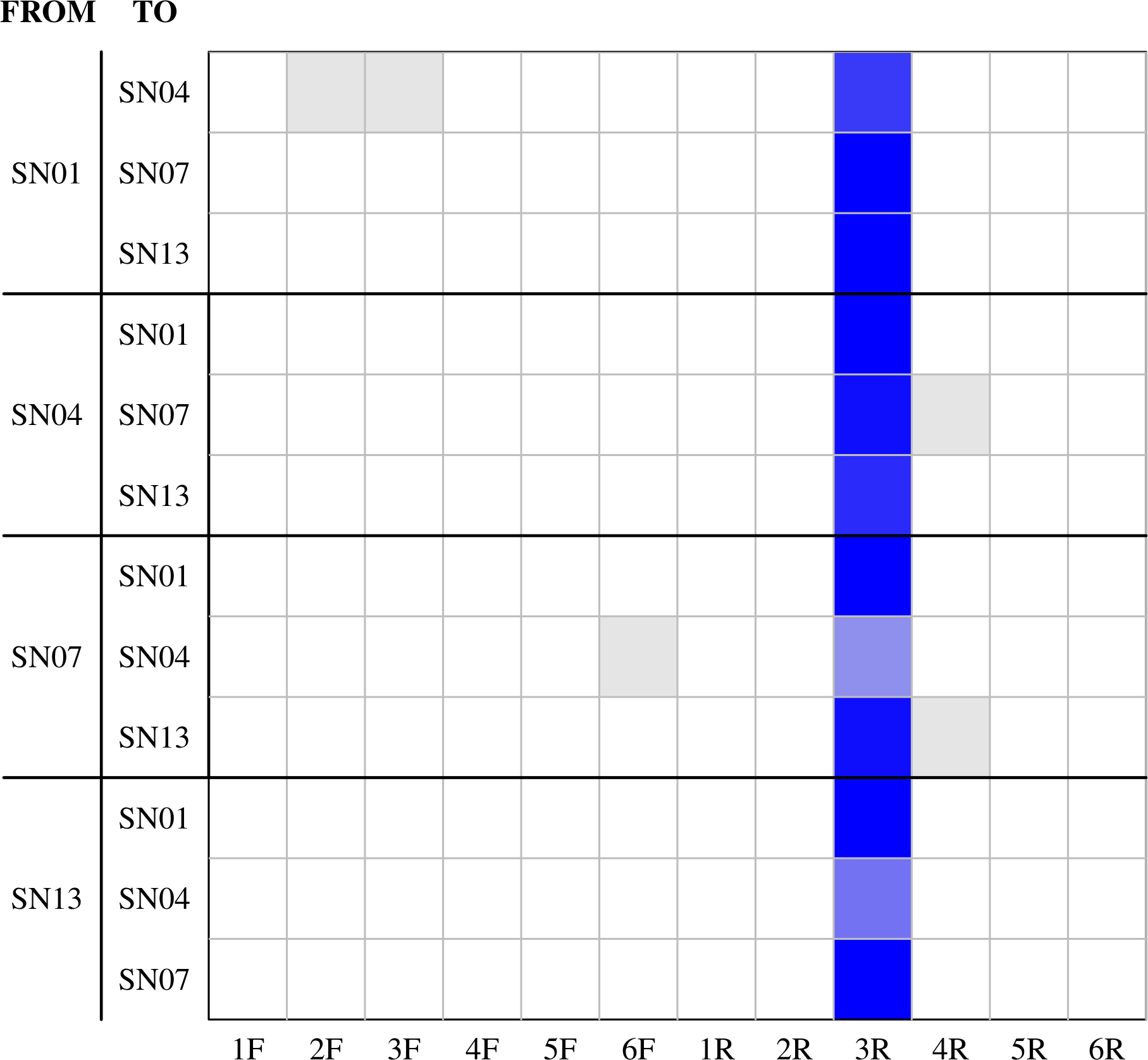}\label{fig:xsoc_vs_xsoc_3r}}\hspace{0.2mm}
    \subfloat[][]{\includegraphics[height=1in, width=1.5in]{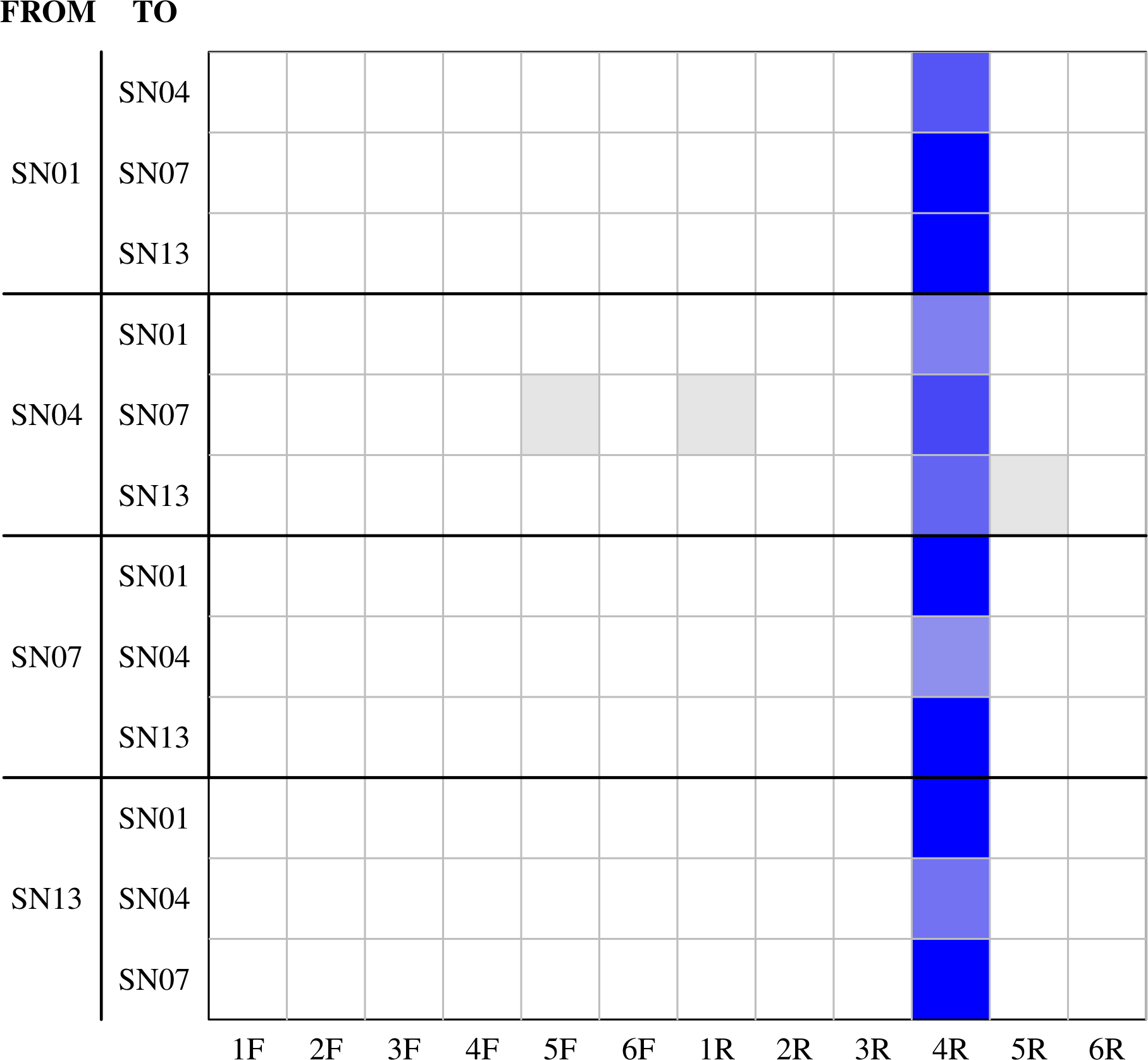}\label{fig:xsoc_vs_xsoc_4r}}\hspace{0.2mm}
    \subfloat[][]{\includegraphics[height=1in, width=1.5in]{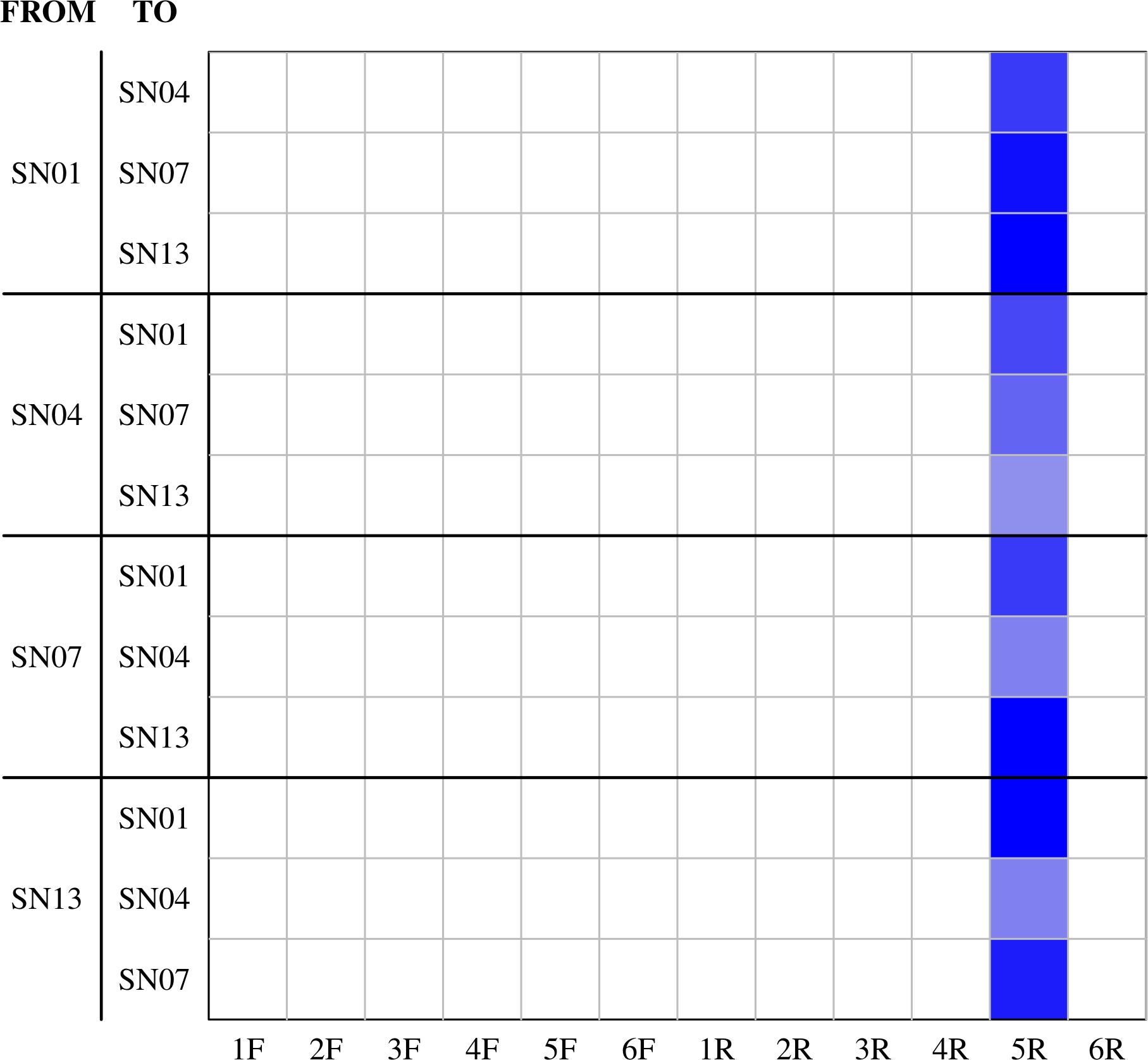}\label{fig:xsoc_vs_xsoc_5r}}\hspace{0.2mm}
    \subfloat[][]{\includegraphics[height=1in, width=1.5in]{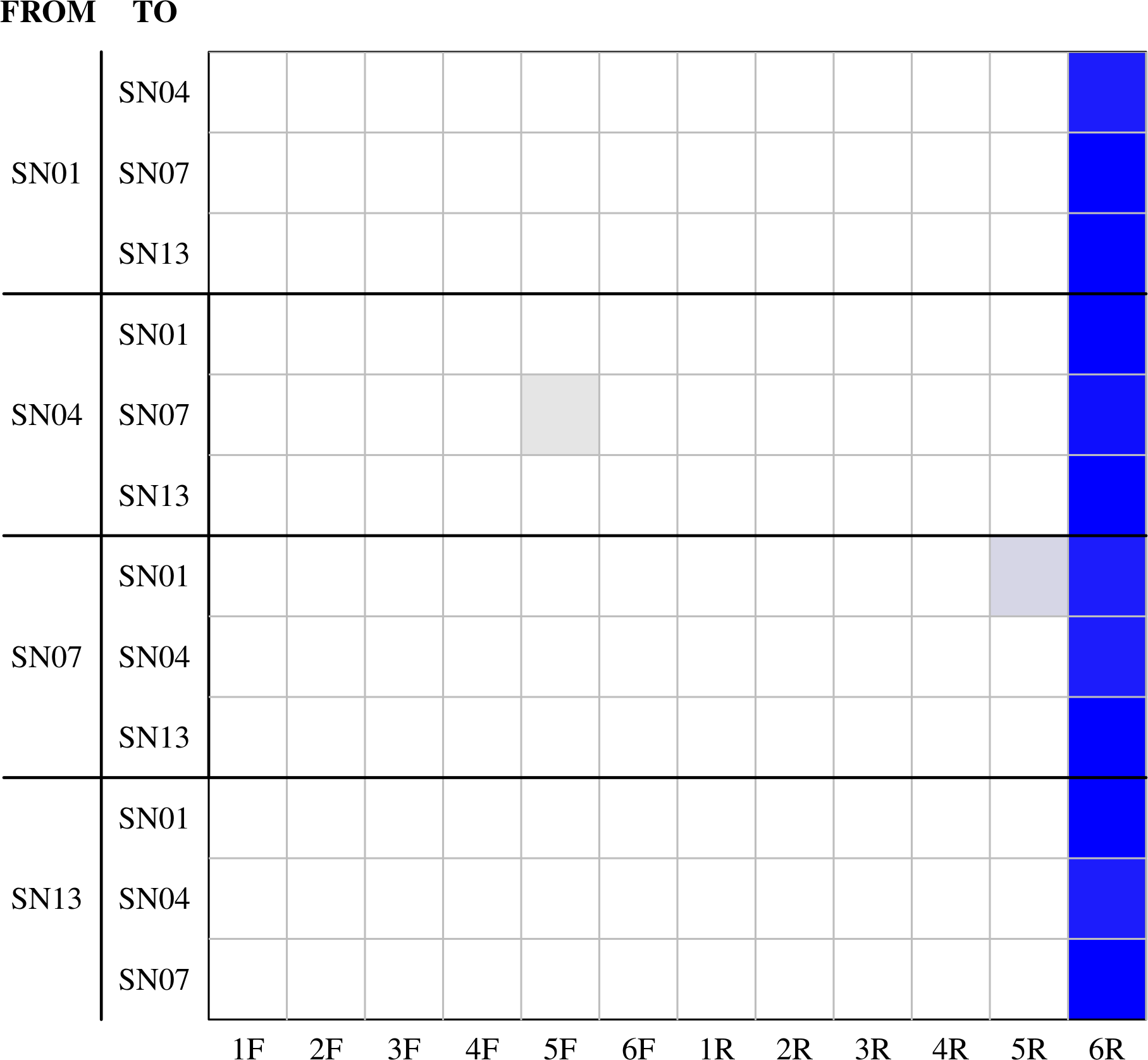}\label{fig:xsoc_vs_xsoc_6r}}\\
    \caption{\emph{Inter-layer user profiles linking} results. Each graph groups the results of a single source on all thirteen SNs, six for the front cameras (first and second rows) and six for the rear cameras (third and fourth rows). The number of images in each cell is identified through a white-to-blue scale, from 0 (white) to 17 (blue).}\label{fig:xsoc_vs_xsoc}
\end{figure}

The combination of all the selected SNs produced the results shown in Figure \ref{fig:xsoc_vs_xsoc}, front cameras are in the first two rows and rear cameras are in the second two rows. Each graph represents the \emph{inter-layer user profiles linking} performances for a given source camera. In particular, in each graph, each group of 3 rows from top to bottom represents the results using the fingerprint from a specific SNs (i.e., Facebook for the first three rows, Instagram for the second three rows, Telegram for the third three rows, and WhatsApp for the last three rows). The results show a slight worsening in comparison to \emph{intra-layer user profiles linking}, however, the number of misclassified images is very low. The problem is particularly significant when the Instagram images are involved in the comparison. In fact, the worst results occurred both when the fingerprint was defined with Instagram's images (i.e., the second group of three rows from the top) as well as when we tried to classify the Instagram's images (i.e., the first row in the first group of three rows and the second row in the third and fourth group of three rows). However, all the other SNs combinations that do not involve Instagram's images obtained good results, especially with front images.\\

The results show the effectiveness of the PRNU-based method in the task of defining a \emph{user unaware watermarking} for all the considered SNs. It is robust enough in spite of the use of the shared images degraded by SNs during the uploading and downloading process. Our results indicate that the proposed method can be successfully used both for the \emph{profile attribution} and for the \emph{intra/inter-layer user profiles linking}. Moreover, the method makes it possible to identify those profiles that belong to the same user, for instance like fake profiles.

\section{Conclusions}\label{sec:conc}
An increasingly large amount of data is loaded daily and shared on SNs and the majority of this flow of data includes images. Thus it is important to find solutions for authorship attribution and verification of these pictures to avoid problems like impersonating or fake profiles.

We trod multiple investigations in the domain of the image watermarking. 
All the experiments we conducted in the context of \emph{social networks watermarking} showed no evidence of watermarks. Facebook introduces some suspicious values in images' metadata that actually can not be considered a robust watermark in SNs context.

Next, we conducted a detailed review of how conventional image watermarking algorithms (\emph{user explicit watermarking}) behave on each selected SN. In particular, we examined whether they can resist the compression algorithms applied by social platforms. In this case, we found that none of the algorithms is able to apply a robust watermark that retains in all the 13 selected SNs. Moreover, in some cases, these watermarking algorithms visually altered the original image.

Finally, we showed that a \emph{user unaware watermark} based on PRNU is able to resist the compression algorithms of all the 13 SNs investigated in our study. The PRNU is a regular and systematic noise component characterizing each image captured by off-the-shelf cameras (including the smartphone's camera). We also proved that the method based on this unawarely trace allows to perform several challenging tasks, that is \emph{profile attribution}, \emph{intra} and \emph{inter-layer user profile linking}. Moreover, we showed how the \emph{user unaware watermark} can be exploited for \emph{fake profiles detection}, that is a corollary of the three previous tasks.

\section*{Acknowledgment}
We thank all the students of \emph{Data Base} and \emph{Data Base Complements} courses (AY 2015/2016) held by the prof. Montesi at the Department of Computer Science and Engineering, University of Bologna. They put a big effort in collecting plenty of images throughout the city during the winter period. We also thank Andrea Iann\`i for his efforts in \emph{user unaware watermarking} experiments.\\
\emph{Funding:} This work has been partially supported by H2020 project, SoBigData++ [grant number 871042].

\bibliography{Biblio}\label{sec:References}

\end{document}